\documentclass[11pt]{article}
\pdfoutput=1
\usepackage{amsmath,amssymb,epsfig,amsfonts}
\usepackage{graphicx}
\usepackage{subfig}
\usepackage[usenames, dvipsnames]{color}
\usepackage{hyperref}
\usepackage[dvipsnames]{xcolor}
\usepackage{cite}
\usepackage{multirow}
\usepackage{slashed}
\usepackage{verbatim} 
\usepackage{enumitem}
\usepackage{rotating}
\usepackage{xcolor}
\usepackage{multirow}
\usepackage{bm}
\usepackage[normalem]{ulem}
\usepackage{cleveref}
\usepackage{booktabs} 
\usepackage{tikz-cd}

\usepackage[fleqn,tbtags]{mathtools}

\usepackage{ytableau}

\usepackage{tikz}
\usepackage{diagbox}
\usetikzlibrary{positioning}
\usetikzlibrary{calc}
\usetikzlibrary{decorations.pathreplacing,calligraphy}

\usetikzlibrary{arrows}

\usepackage{xstring}
\usetikzlibrary{decorations.pathmorphing} 
\usetikzlibrary{decorations.markings} 
\usetikzlibrary{arrows} 
\usetikzlibrary{shapes} 
\usetikzlibrary{matrix} 
\usetikzlibrary{positioning} 
\usepackage[english]{babel} 
\usepackage[autostyle]{csquotes}

\usepackage{tcolorbox}

\usepackage{titlesec}

\titlespacing{\paragraph}{%
  0pt}{
  0.2\baselineskip}{
  0.5em}

\makeatletter

\newcommand{\Hom}{\text{Hom}}
\newcommand{\id}{\text{id}}
\newcommand{\Spin}{\text{Spin}}

\newcommand{\Pin}{\text{Pin}}

\newcommand{\Bock}{\text{Bock}}
\newcommand{\be}{\begin{equation}}
\newcommand{\ee}{\end{equation}}
\newcommand{\ba}{\begin{aligned}}
\newcommand{\ea}{\end{aligned}}

\newcommand{\SymTFT}{\text{SymTFT}}
\newcommand{\brho}{\bm{\rho}}

\newcommand{\cO}{\mathcal{O}}

\newcommand{\cS}{\mathcal{S}}
\newcommand{\cC}{\mathcal{C}}
\newcommand{\cT}{\mathcal{T}}

\newcommand{\bea}{\begin{eqnarray}}
\newcommand{\eea}{\end{eqnarray}}

\renewcommand{\Vec}{\mathsf{Vec}}
\newcommand{\Rep}{\mathsf{Rep}}

\newcommand{\TwoVec}{2\mathsf{Vec}}
\newcommand{\TwoRep}{2\mathsf{Rep}}
\renewcommand{\id}{\text{id}}

\newcommand{\cA}{\mathcal{A}}

\newcommand{\Z}{{\mathbb Z}}

\def\diag{\mathop{\mathrm{diag}}\nolimits}

\def\Tr{\mathop{\mathrm{Tr}}\nolimits}

\def\unit{{1\kern-.65ex {\rm l}}}
\def\1{{1\kern-.65ex {\rm l}}}

\def\bbZ{{\mathbb{Z}}}

\newcount\hour \newcount\minute
\hour=\time \divide \hour by 60
\minute=\time
\def\now{%
\ifnum \hour<13
  \ifnum \hour=0 \advance \hour by 12 \number\hour:\else \number\hour:\fi%
     \ifnum \minute<10 0\fi%
     \number\minute%
\ A.M.%
\else \advance \hour by -12 \number\hour:%
  \ifnum \minute<10 0\fi%
  \number\minute%
  \ P.M.%
\fi%
}

\makeatother

\begin{document}

\baselineskip=18pt  
\numberwithin{equation}{section}  
\allowdisplaybreaks  


%
%


\thispagestyle{empty}

\vspace*{1.2cm} 
\begin{center}

{
\Huge ICTP Lectures on 
\\ \bigskip
(Non-)Invertible Generalized Symmetries
}

 \vspace*{1.5cm}
Sakura Sch\"afer-Nameki\\

 \vspace*{0.7cm} 
{\it  Mathematical Institute, University of Oxford, \\
Andrew-Wiles Building,  Woodstock Road, Oxford, OX2 6GG, UK}\\

\vspace*{1cm}
\end{center}

\noindent
What comprises a global symmetry of a Quantum Field Theory (QFT) has  been vastly expanded in the past 10 years to include not only symmetries acting on higher-dimensional defects, but also most recently symmetries which do not have an inverse. The principle that enables this generalization is the identification of symmetries with topological defects in the QFT. In these lectures, we provide an introduction to  generalized symmetries, with a focus on non-invertible symmetries. We begin with a brief overview of invertible generalized symmetries, including higher-form and higher-group symmetries, and then move on to non-invertible symmetries. The main idea that underlies many constructions of non-invertible symmetries is that of stacking a QFT with topological QFTs (TQFTs) and then gauging a diagonal non-anomalous global symmetry. The TQFTs become topological defects in the gauged theory called (twisted) theta defects and comprise a large class of non-invertible symmetries including condensation defects, self-duality defects, and non-invertible symmetries of gauge theories with disconnected gauge groups. We will explain the general principle and provide numerous concrete examples. 
Following this extensive characterization of symmetry generators, we then discuss their  action on higher-charges, i.e. extended physical operators. As we will explain, even for invertible higher-form symmetries these are not only representations of the $p$-form symmetry group, but more generally what are called higher-representations. Finally, we  give an introduction to the Symmetry Topological Field Theory (SymTFT) and its utility in characterizing symmetries, their gauging and generalized charges.\\
{\it Lectures prepared for the ICTP Trieste Spring School, April 2023.}

\newpage

\tableofcontents


\section{Introduction}

Symmetries are fundamental throughout physics. In  quantum field theory (QFT),  global and local symmetries are indispensable, starting with the formulation of gauge theories, where symmetries organize the fields and interactions, and the spectrum; anomalies, which constrain the quantum theories further, and symmetry breaking, which is at the root of many phenomenological applications and predictions. Particularly powerful are global symmetries and their anomalies ('t Hooft anomalies), which are RG-invariants and thus provide insights into strongly-coupled regimes that are difficult to access by perturbative means. In this context, symmetries are conventionally understood to act  on point-like operators (as opposed to extended operators). Furthermore, symmetries are often understood to be group-like, i.e. the composition obeys a group multiplication law, and for each symmetry generator, there is an inverse. 

The past 10 years have seen several generalizations of the concept of global symmetries: first, the introduction of higher-form symmetries \cite{Gaiotto:2014kfa} has relaxed the restriction that symmetries act on point-like operators. Higher-form symmetries act on extended operators, such as line and surface operators. However they were introduced assuming that their composition satisfies a group-law, and so in particular every symmetry generator has an inverse.

The second generalization occurred only in the past  two years: here we relax the assumption that the composition is group-like, in particular, not every symmetry generator needs to have an inverse. Such symmetries have been called  non-invertible symmetries. Although known in lower dimensions, their existence in $d>2$ QFTs has led to a flurry of discoveries.

The purpose of the present lecture notes is to provide an introduction to this fast-developing topic, providing some basic background, and key examples, organized along the main lessons that we have learned so far. 
The lectures should enable the reader to subsequently pick up the ever-growing list of publications in this field and explore the subject themselves. Many of these developments in non-invertible symmetries point to 
exciting connections between high energy theory, condensed matter, quantum gravity and mathematics (in particular category theory), which should be subsequently accessible to the reader.

\subsection{Symmetries as Topological Operators}

\paragraph{Generalized Global Symmetries.}
In 2014, the seminal paper \cite{Gaiotto:2014kfa} proposed a generalization of the notion of global symmetries by introducing  higher-form symmetries which act on extended operators. 
For instance a 1-form symmetry $G^{(1)}$ acts on line operators such as Wilson and 't Hooft lines, a 2-form symmetries $G^{(2)}$ acts on surfaces operators and more generally a $p$-form symmetry  $G^{(p)}$ acts on $p$-dimensional extended operators. 
Ordinary (pre-2014) symmetries that act on local operators are a special instance, namely 0-form symmetries $G^{(0)}$.  
Higher-form symmetries have led to numerous important insights, e.g. the characterization of confinement, where the order parameter is the expectation value of a Wilson loop. This can be characterized in terms of an unbroken 1-form symmetry.

The central idea that underlies this and many subsequent extensions of the notion of a global symmetry is the following identification:

\begin{tcolorbox}[colback=white, colframe=black!50, rounded corners]
\centering {\text{Global Symmetries} $\longleftrightarrow$ \text{Topological Operators}}
\end{tcolorbox}
This identification is the reason for most of the studies of symmetries being focused on understanding topological operators in QFTs, their dimensionality and their composition (fusion). 

\begin{figure}
\centering
\begin{tikzpicture}
\draw [blue, thick] (0, 0) ellipse (2 and  2);
\draw [fill=blue, opacity = 0.2] (0, 0) ellipse (2 and  2);
\draw [dotted, bend left=30, color= blue, looseness=1.0] (-2,0) to (2,0);
\draw [bend right=30, color= blue, looseness=1.0] (-2,0) to (2,0);
\draw [red,fill=red] (0,0) ellipse (0.07 and 0.07);
\node[red, above] at (0,0) {$\cO_0({\bf p})$} ;
\node[blue, right] at (1.7, 1.7) {$D_{d-1}^{(g)}(M_{d-1})$};
\end{tikzpicture}
\caption{The co-dimension 1 (i.e. $d-1$-dimensional) topological operator $D_{d-1}^{(g)}(M_{d-1})$ is a generator of the 0-form symmetry group $G$. 
The charge of the local operator (physical, not necessarily topological) $\cO_0$ located at the point ${\bf p}$ is computed in terms of the linking of the manifold $M_{d-1}$ and the point ${\bf p}$. The charge is invariant under deformations of the manifold $M_{d-1}$, which do not change the linking, i.e. as long as the operator insertion is contained within the volume that $M_{d-1}$ bounds. 
\label{fig:Noethersphere}}
\end{figure}
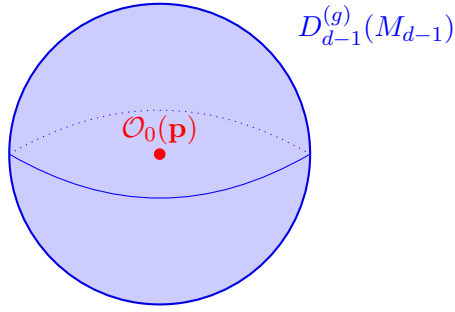

The main example to keep in mind is that of a standard global symmetry: we can measure the charge of a point-like operator by surrounding it with a codimension 1 sphere, as shown in figure \ref{fig:Noethersphere}. If the charge comes from a conserved current,  then we can also deform the sphere to any closed compact codimension 1 manifold, $M_{d-1}$ without changing the charge, as long as the operator is inserted at a point inside the volume that is bounded by $M_{d-1}$. 
This is precisely the hallmark of a topological operator. In this case it measures the charge, or put differently, the local operator transforms in some representation of the symmetry that this operator generates. We will refer to such codimension 1 operators that are topological as $D_{d-1}^{(g)}$, where $g$ is an element in the symmetry group $G$. Importantly we will abstract this notion of symmetry generator from the existence of a conserved current, which enables the generalization to e.g. finite group symmetries (and much more). 

0-form and higher-form symmetries (except in $d=2$\footnote{Throughout these lectures, $d$ denotes the spacetime dimension, which will be usually Euclidean.}) have until very recently (pre-2021) been groups $G$ -- Lie groups, or finite groups, which means  the following:
group-like symmetries are generated by topological operators, which compose according to the multiplications in a group $G$, and particles/operators/states transform in representations of $G$. 
In particular, a group $G$ is generated by $g_i$, including an element, the identity, $\id$, such that there is an associative map $\cdot: G \times G \to G$
\be\label{Group}
g_1 \cdot g_2 = g_3 \in G  
\ee
and $\id \cdot g  = g = g \cdot \id$, satisfying associativity 
\be
(g_1 \cdot g_2) \cdot g_3 = g_1 \cdot(g_2 \cdot g_3) \,. 
\ee
Furthermore, 
for every $g\in G$ there exists an inverse, i.e. a $g^{-1}\in G$ such that 
\be
g \cdot g^{-1} = \id  \,,
\ee
so that each symmetry generator has an inverse. 
Examples of such symmetries are gauge groups, e.g.  
$G$ a simple Lie group $SU(N)$, $\Spin(N)$, $E_n$, or global symmetries like flavor symmetries (0-form symmetry groups) acting on matter fields, charge conjugation $\Z_2$, R-symmetries, etc\footnote{There are also spacetime symmetries, such as the Poincar\'e group and supersymmetry, but we will focus here on internal symmetries of QFTs.}. Higher-form symmetries were introduced with this group composition law, and in fact most of the time form (finite or continuous) abelian groups. 

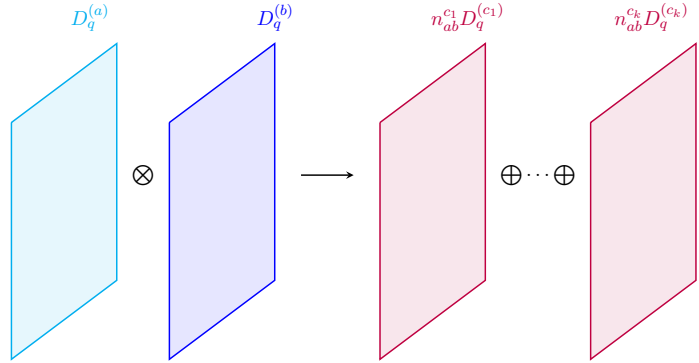
\begin{figure}
\centering
\scalebox{0.7}{\begin{tikzpicture}
\begin{scope}[shift={(11.5,0)}]
\draw [blue, fill=blue,opacity=0.1] (-2,0) -- (0,1.5) -- (0,-3) -- (-2,-4.5) -- (-2,0);
\draw [blue, thick] (-2,0) -- (0,1.5) -- (0,-3) -- (-2,-4.5) -- (-2,0);
\end{scope}
\begin{scope}[shift={(8.5,0)}]
\draw [cyan, fill=cyan,opacity=0.1] (-2,0) -- (0,1.5) -- (0,-3) -- (-2,-4.5) -- (-2,0);
\draw [cyan, thick] (-2,0) -- (0,1.5) -- (0,-3) -- (-2,-4.5) -- (-2,0);
\end{scope}
\begin{scope}[shift={(15.5,0)}]
\draw [purple, fill=purple,opacity=0.1] (-2,0) -- (0,1.5) -- (0,-3) -- (-2,-4.5) -- (-2,0);
\draw [purple, thick] (-2,0) -- (0,1.5) -- (0,-3) -- (-2,-4.5) -- (-2,0);
\end{scope}
\begin{scope}[shift={(19.5,0)}]
\draw [purple, fill=purple,opacity=0.1] (-2,0) -- (0,1.5) -- (0,-3) -- (-2,-4.5) -- (-2,0);
\draw [purple, thick] (-2,0) -- (0,1.5) -- (0,-3) -- (-2,-4.5) -- (-2,0);
\end{scope}
\draw [-stealth,thick](12,-1) -- (13,-1);
\node[left,cyan] at (8.5, 2) {$D_{q}^{(a)}$};
\node[left, blue] at (12,2) {$D_{q}^{(b)}$};
\node[left,purple] at (16,2) {$n_{ab}^{c_1} D_{q}^{(c_1)}$};
\node[black] at (16.5, -1) {$\bigoplus \cdots \bigoplus$}; 
\node[black] at (9, -1) {$\bigotimes$};
\node[left, purple] at (19.5,2) {$n_{ab}^{c_k}D_{q}^{(c_k)}$};
\end{tikzpicture}}
\caption{Non-Invertible fusion of two $q$-dimensional topological defects $D_q^{(a)}$ and $D_q^{(b)}$ into a sum of $q$-dimensional topological defects $D_q^{(c_k)}$. In the group-like case, $a, b \in G$, the right hand side would be  $D_q^{(ab)}$ with $ab \in G$. \label{fig:fusion}}
\end{figure}

These lectures are about symmetries which {\it do not} necessarily satisfy such group-like composition:
not every symmetry generator is required to have an inverse, and the composition can take a more general form than (\ref{Group}). 
This motivated the name {\bf non-invertible symmetries}. 
Schematically, this  symmetry $\cS$ has a set of generators, on which we define  a composition (fusion) $\cdot : \cS \otimes \cS \to \cS$, by 
\be
a \otimes b = \sum_{c \in \cS} n_{ab}^c\,c \,, \qquad a, b\in \cS \,,
\ee
where the right hand side is a sum over elements in $\cS$, with the coefficient $n_{ab}^c$ satisfying an (generalized) associativity constraint.  This requires more than a group-like structure,  and is akin to an algebra (there is a sum and a product), where furthermore the existence of an inverse is also not required. A cartoon of the composition is shown in figure \ref{fig:fusion}: bringing two symmetry generators together and fusing them, results in a sum over other symmetry generators.

As we will see, this is only a very crude approximation to the wealth of interesting structures that non-invertible symmetries provide and we will motivate physically, why the correct mathematical framework that replaces groups is in fact  fusion higher-categories\footnote{There is no reason to run at this point. We will not assume any prior knowledge of what a fusion category or even a higher-category is!}.

The last 2 years have shown surprising developments, which point to a wealth of non-invertible symmetries in $d>2$ dimensional QFTs. These are neither particularly exotic theories (e.g. pure gauge theories in 4d, or the Standard Model gauge theory), nor are these symmetries particularly rare. It seems to be an accident of history then that they only have been uncovered now. Their discovery opened up many interesting questions and is exciting opportunity for exploration, both of the  general structure, as well as  physical implications, of these new symmetries.

In these lectures, we will provide an introduction to various constructions of non-invertible symmetries in higher dimensions, focusing on 3d and 4d. Before discussing these recent constructions, an important laboratory are 2d QFTs, where many ideas that have higher-dimensional generalizations are already present, and very well understood.

\subsection{Non-Invertible Symmetries}

\paragraph{Non-Invertible Symmetries in 2d.}
In 2d such non-invertible, i.e. not group-like, symmetries are in fact very well known and studied. What replaces groups in this instances are fusion categories \cite{EGNO}, which have a long history in 2d QFTs, see \cite{Frohlich:2004ef,Fuchs:2007tx,Frohlich:2009gb,Bhardwaj:2017xup,Chang:2018iay,Thorngren:2019iar, Komargodski:2020mxz, Thorngren:2021yso} for some of the seminal papers. 
A particularly interesting set of examples arise in 2d rational conformal field theories (RCFTs). Rationality implies a finite set of conformal primaries, which in turn label the so-called Verlinde topological lines, which compose according to the  fusion rules of the RCFT. 

\paragraph{Ising model.}  
One of the most well-studied examples is the 2d Ising model, which is the $c=1/2$ rational CFT\footnote{E.g. you can realize it as a GKO coset model $SU(2)_2/U(1)$, with diagonal modular invariant.}. This has three conformal primaries: 
\be
\lambda_{h_i, \overline{h_i}} = \id _{0,0} \,, \ \epsilon_{1/2, 1/2}\,,\  \sigma_{1/16, 1/16} \,.
\ee
In turn, there are three topological line operators, 
usually denoted by $\id, \eta, D$ (with $\eta$ associated to $\epsilon$ and $D$ to $\sigma$) with fusion 
\be
\eta \otimes \eta = \id \,,\quad \eta \otimes D = D \otimes \eta = D \,,\quad  D \otimes D = \id \oplus \eta \,, 
\ee
where the latter exhibits the non-invertible fusion. 
There are numerous interesting implications of such symmetries in 2d, which we will see avatars of in higher-dimensions.

\begin{figure}
\centering
\begin{tikzpicture}
\begin{scope}[shift={(0,0)}]
\draw [blue, thick] (0,0)-- (0,4);
\draw [ PineGreen,fill= PineGreen] (-1,2) ellipse (0.07 and 0.07);
\node[above, PineGreen] at (-1,2) {$\sigma$};
\node[below, blue] at (0,0) {$D$};
\end{scope}
 \begin{scope}[shift={(3,0)}]
 \draw[blue, thick, rounded corners] (0,0) -- (0,2) -- (1.7,2) -- (1.7, 2.3) -- (0,2.3) -- (0,4);
 \draw [ PineGreen,fill= PineGreen] (1.5, 2.15) ellipse (0.07 and 0.07);
\node[above,  PineGreen] at (1.6, 2.3) {$\sigma$};
\node[below, blue] at (0,0) {$D$};
\node[above, blue] at (1,2.4) {$D$};
\node[below, blue] at (1,2) {$D$};
 \end{scope} 
 \begin{scope}[shift={(7,0)}]
 \draw[blue, thick] (0,0) -- (0,4);
  \draw[purple, thick] (0,2.2) -- (1.7,2.2);
 \draw [ PineGreen,fill= PineGreen] (1.7, 2.2) ellipse (0.07 and 0.07);
\node[above,  PineGreen] at (1.7, 2.2) {$\sigma'$};
\node[below, blue] at (0,0) {$D$};
\node[above, purple] at (0.8,2.2) {$\eta$};
 \end{scope} 
\end{tikzpicture}
\caption{Action of the Kramers-Wannier duality defect $D$ on the local operator $\sigma$. As we move this through the topological defect, via the figure in the middle, we can use the fusion $D^2 = \id \oplus \sigma$, which then results in the right most figure. 
The operator becomes a disorder operator $\sigma'$, i.e. an operator with the same conformal weights, but now attached to an $\eta$-line. \label{fig:Ising}}
\end{figure}
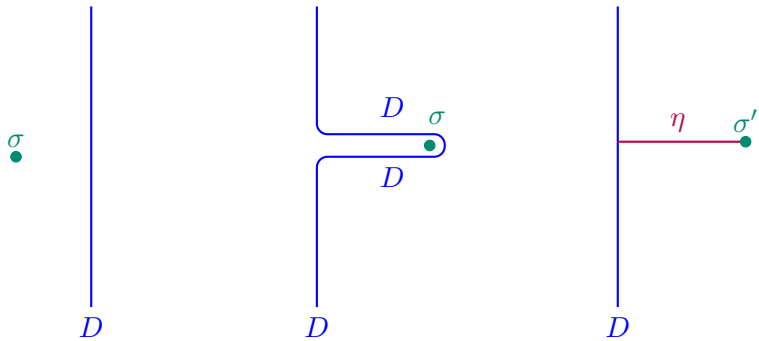

\paragraph{Action on Operators.}
One way to distinguish a non-invertible symmetry from an invertible one is by its action on operators (or charges). 
The non-invertible  line $D$, the so-called Kramers-Wannier duality defect, in the Ising model illustrates this as follows, see figure \ref{fig:Ising}: passing a local operator $\sigma$ through the topological line $D$, we can deform the line, and use the fusion to compute the effect of moving the operator through the line $D$ (middle of figure \ref{fig:Ising}). Applying the fusion $D\otimes D = \id \oplus \eta$ to the horizontal parallel lines implies that the operator $\sigma$ is now attached to a non-trivial line $\eta$. 
If we define a {\bf non-genuine $p$-dimensional operator} as a $p$-dimensional operator attached to the end of a $(p+1)$-dimensional operator, and a {\bf genuine} one is not the boundary of a higher-dimensional operator, then we can rephrase this observation as follows: 
a non-invertible symmetry can map a genuine local operator to a non-genuine one. 
In this case the operator $\sigma$ is genuine, but maps to the operator $\sigma'$, which is non-genuine as it is attached to the $\eta$-line, and corresponds to a disorder operator \cite{Frohlich:2004ef}. 

We will encounter a very similar situation when we consider the symmetries of pure 4d minimally supersymmetric Yang-Mills theory with gauge algebra $\mathfrak{su}(N)$. For the gauge group $SU(N)$ the theory has Wilson line operators, which all obey area law in the IR and there are $N$ confining vacua. Contrary to that, the $PSU(N)= SU(N)/\Z_N$ gauge theory, whose gauge configurations cannot be lifted to the simply-connected group $SU(N)$, will have a  non-invertible 0-form symmetry, very similar to the Kramers-Wannier duality defect. We will see that in this case the map from order to disorder is modeled by the map from confining to deconfining vacuum, where in one, the 't Hooft line is a genuine defect and in the other it acquires a fractional electric charge by the Witten effect \cite{Witten:1979ey} and becomes a non-genuine line operator, which is attached to a topological surface operator.

\paragraph{Non-Invertibles from Groups.}
A simpler class of non-invertible symmetries in 2d can be derived starting with an ordinary (invertible), though finite, group symmetry. 
A group-like 0-form symmetry $G^{(0)}$ in 2d is generated by topological lines that fuse according to the group composition (\ref{Group}). 
Gauging this 0-form symmetry, will result in a theory which again has a 0-form symmetry, however the topological lines are now Wilson lines, labeled by representations $\bm{R}$, and we will call this symmetry $\Rep(G)$, the finite-dimensional representations of $G$. The symmetry generators are topological lines, labeled by representations. However, their fusion can now be non-invertible, as it  obeys the Clebsch-Gordan decomposition of the tensor product of representations. 
 If $G$ is an abelian group, this is again invertible (and if $G$ is finite, then in fact the symmetries before and after gauging are isomorphic). However if $G$ is non-abelian $\Rep(G)$ is a non-invertible symmetry. We will study this relatively simple case at the start of the lectures, and then generalize it to higher-dimensions.

\paragraph{Non-Invertible Symmetries in $d\geq 3$.}
Starting in 2021, non-invertible symmetries were introduced in higher-dimensional QFTs, and numerous different constructions have since been developed 
\cite{Heidenreich:2021xpr,
Kaidi:2021xfk, Choi:2021kmx, Roumpedakis:2022aik,Bhardwaj:2022yxj, Choi:2022zal,Cordova:2022ieu,Choi:2022jqy,Kaidi:2022uux,Antinucci:2022eat,Bashmakov:2022jtl,Damia:2022bcd,Choi:2022rfe,Bhardwaj:2022lsg,Bartsch:2022mpm,Damia:2022rxw, Apruzzi:2022rei,Lin:2022xod,GarciaEtxebarria:2022vzq,Heckman:2022muc,Niro:2022ctq,Kaidi:2022cpf,Antinucci:2022vyk,Chen:2022cyw,Lin:2022dhv,Bashmakov:2022uek,Karasik:2022kkq,Cordova:2022fhg,GarciaEtxebarria:2022jky,Decoppet:2022dnz,Moradi:2022lqp,Runkel:2022fzi,Choi:2022fgx,Bhardwaj:2022kot, Bhardwaj:2022maz,Bartsch:2022ytj,Heckman:2022xgu,Antinucci:2022cdi,Apte:2022xtu,Delcamp:2023kew,Kaidi:2023maf,Li:2023mmw,Brennan:2023kpw,Etheredge:2023ler,Lin:2023uvm, Putrov:2023jqi, Carta:2023bqn,Koide:2023rqd, Zhang:2023wlu, Cao:2023doz, Dierigl:2023jdp, Inamura:2023qzl,Chen:2023qnv, Bashmakov:2023kwo, Choi:2023xjw}. One goal that we will set in these lectures is to give a road-map and some organizational structure to this vast and rapidly developing field. 
Broadly speaking, most constructions of non-invertible symmetries are in one way or another related to gauging of invertible symmetries. These were dubbed ``non-intrinsic non-invertible symmetries"\footnote{This double-negative naming was proposed in \cite{Kaidi:2022cpf}, and clearly a better name needs to be found. Mathematically these are called module categories over $n\Vec_G$ which also fails in the task of providing a suitably concise name.}, whereas those that are not are called ``intrinsic".

\paragraph{Stacking TQFTs.}
One of the recurring themes  throughout the construction of non-invertible symmetries is that of ``stacking" with TQFTs and subsequently gauging. The idea is very simple, and many constructions of non-invertible symmetries  have such a characterization.  
The main idea is to take a theory $\cT$ with a given global symmetry $\cS$ (e.g. an invertible 0-form symmetry, or higher-form symmetry) and before gauging $\cS$, we take a product (stack) with topological field theories (TQFTs) that also carry this symmetry $\cS$.
Gauging the diagonal symmetry
\be
(\cT \times \text{TQFT})/\cS
\ee
yields a new theory, where the TQFTs are no longer decoupled, but topological defects in the gauged theory, and usually obey non-invertible fusion! What precise TQFTs one has to consider can depend on the setup, but the underlying principle is the same: 
stacking with $\cS$-symmetric TQFTs before gauging $\cS$ results in topological defects, which have non-invertible symmetry. Topological defects obtained in this way will be called theta-defects.

Lets consider a concrete example, which will illustrate both the simplicity of the idea, as well as the ubiquity of such non-invertible defects: 
Consider a 3d theory $\cT$ with an invertible global symmetry $\Z_2^{(0)}$. The standard lore has it that gauging a 0-form symmetry yields a dual $1$-form symmetry (see section \ref{sec:Dual}), which is again given in terms of the group $\Z_2$. 
This is however not the full story: the symmetry of the gauged theory $\cT/\Z_2$ contains an additional topological surface defect $D_2^{(\Z_2)}$, which has non-invertible fusion: 
\be
D_2^{(\Z_2)} \otimes D_2^{(\Z_2)} = 2 D_2^{(\Z_2)} \,.
\ee
The defect $D_2^{(\Z_2)}$ arises precisely due to the stacking with $\Z_2$-symmetric 2d TQFTs \cite{Bhardwaj:2022lsg}. An equivalent description of this theta defect is as a condensation defect \cite{Gaiotto:2019xmp, Roumpedakis:2022aik}, i.e. obtained by gauging a higher-form symmetry on a subspace (in this case, the 1-form symmetry on the trivial surface defect). 

This demonstrates nicely the ubiquity of non-invertibles: even in the simple setting of gauging an  abelian finite 0-form symmetry, where the rules of \cite{Gaiotto:2014kfa} would have told us that the gauged theory simply has an invertible dual symmetry (in 3d, a dual 1-form symmetry, and in general $d$ a $(d-2)$-form symmetry), there are in addition non-invertible defects!

\paragraph{Layer-Structure of Topological Defects.}

\begin{figure}
\centering
\begin{tikzpicture}
\draw [fill=white,opacity=0.1] 
(0,0) -- (4,0) -- (4,4) -- (0,4) -- (0,0);
\draw [yellow, thick,  fill=yellow, opacity = 0.3] (0,2) -- (0,4) -- (4,4) -- (4,2) -- (0,2) ;
\draw [cyan, thick,  fill=cyan, opacity = 0.3] (0,0) -- (0,2) -- (4,2) -- (4,0) -- (0,0) ;
\draw [ultra thick,Green] (0,2) -- (2,2) ;
\draw [ultra thick,LimeGreen] (2,2) -- (4,2) ;
\draw [green,fill=green] (2,2) ellipse (0.07 and 0.07);
\node[Green] at (-0.7,2) {$D_{1}^{(a,b)}$};
\node[LimeGreen] at (4.7,2) {$D_{1}^{(a,b)'}$};
\node[black] at (2,1) {$D_{2}^{(a)}$};
\node[black] at (2,3.1) {$D_{2}^{(b)}$};
\node[green] at (2.3, 2.4) {$D_{0}^{(a,b),(a,b)'}$} ;
\begin{scope}[shift={(8,0)}]
\draw [fill=white,opacity=0.1] 
(0,0) -- (4,0) -- (4,4) -- (0,4) -- (0,0);
\draw [yellow, thick,  fill=yellow, opacity = 0.3] (0,2) -- (0,4) -- (4,4) -- (4,2) -- (0,2) ;
\draw [cyan, thick,  fill=cyan, opacity = 0.3] (0,0) -- (0,2) -- (4,2) -- (4,0) -- (0,0) ;
\draw [ultra thick,Green] (0,2) -- (2,2) ;
\draw [ultra thick,LimeGreen] (2,2) -- (4,2) ;
\draw [green,fill=green] (2,2) ellipse (0.07 and 0.07);
\node[Green] at (-1.2,2) {1-Morph$^{(a,b)}$};
\node[LimeGreen] at (5.2,2) {1-Morph$^{(a,b)'}$};
\node[black] at (2,1) {Object$^{(a)}$};
\node[black] at (2,3.1) {Object$^{(b)}$};
\node[green] at (2.5, 2.4) {2-Morph} ;
\end{scope}
\end{tikzpicture}
\caption{The layer structure of topological defects of dimensions $q=2, 1, 0$, denoted by $D_q^{(a)}$ is shown on the LHS. The two surfaces $D_2^{(a)}$ and $D_2^{(b)}$ can have topological line defects as an interfaces (or junctions) $D_1^{(a,b)}$ and $D_1^{(a,b)'}$, which in turn can have a 0-dimensional interface (junction) $D_0$. This is precisely the setting for a fusion 2-category (RHS): the objects are the surfaces, and 1-morphisms between objects are the line operators. In turn the morphisms between 1-morphisms, i.e. points, are called 2-morphisms.   \label{fig:UkrainianDoll}}
\end{figure}
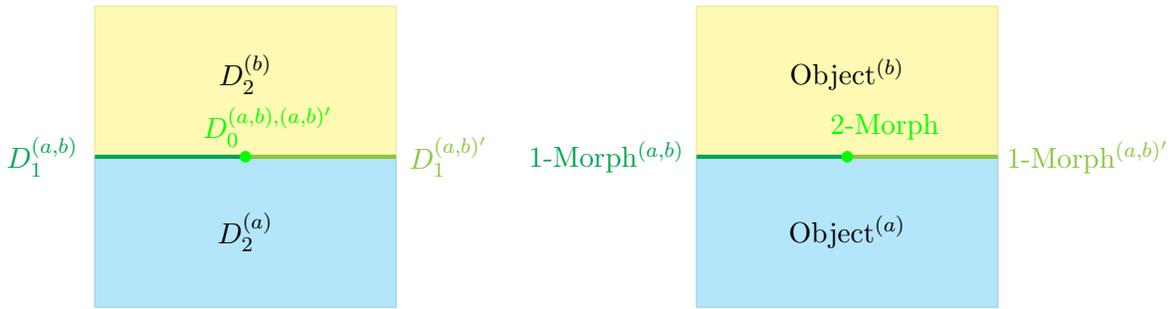

In higher dimensional theories, a symmetry comprises a set of topological defects, of various dimensionality and their fusion can in general be non-invertible. 
There is a natural layer structure that organizes the topological defects by dimension. 
In a $d$-dimensional theory, we have topological defects of dimension $d-1$ -- which will correspond to 0-form symmetry generators --  and lower-dimensional topological defects. The co-dimension $p+1$ topological defects $D_{d-p-1}$ generate a $p$-form symmetry. 
The most general symmetry will have non-invertible fusion with coefficients $N^{(p)}$ within each layer 
\be
D_{d-p-1}^{(a)} \otimes D_{d-p-1}^{(b)} = \bigoplus_{c} (N^{(p)})_{ab}^c\,  D_{d-p-1}^{(c)}  \,, \qquad p=0, \cdots, d-1\,.
\ee
This is depicted in figure \ref{fig:fusion}. 
Furthermore, topological defects of different dimension $q$ and $q'$ may have non-trivial ``interrelations". The simplest instance is that of a 2-group, when a 0-form and 1-form symmetry symmetry do not form a product, but an extension group. A $q$-dimensional defect can have a $(q-1)$-dimensional interface on its worldvolume, which in turn has a $(q-2)$-dimensional topological defect on its worldvolume etc. -- see figure \ref{fig:UkrainianDoll}. This layer structure together with a fusion within the set of fixed dimensional topological defects, is in effect the structure of a {\bf fusion higher-category}. Mathematically these have been initiated in \cite{DouglasReutter} for 2-categories, but overall this remains a very active field of research within mathematics.

\paragraph{Defusing Fusion Categories.}

At this point we should briefly comment on the elephant in the room: categories. 
One potential reason why categories may be of some disrepute with some physicists is that they are perceived as ``fancy wrappers with little content" (if one were to put it unkindly). There are many flavors of categories, and the ones we will be dealing with in the context of symmetries are {\bf fusion higher-categories}. 

We should essentially think of these as natural extensions of groups. Theoretical physics without group theory and its representation is unthinkable, and groups are far more than an organizationl principle, but are important computational tools. 

The extension of symmetries to higher-form and non-invertible symmetries naturally requires going beyond a formulation of symmetries by groups. 
The fact that topological operators, i.e. symmetries, in a $d$-dimensional QFT can have dimension $d-1$ to $0$, which will have a rules of composition, see figure \ref{fig:UkrainianDoll}, inevitably will lead us to characterizing symmetries in terms of fusion higher-categories (without having to start with a heavy-handed introduction to this technical subject). The categories emerge quite naturally from the physics.

These fusion higher-categories contain as a special case groups, but extend them in manifold ways. The important point is that like groups, these are not only conceptually important, but of  computational relevance: they help put an organizational structure to the symmetry defects of a theory, and enable their systematic study (and classification), and most importantly, they will tell us how generalized symmetries act on a physical theory, i.e. they provide the generalized version of representations.

Let us re-iterate that for these lectures no prior knowledge of higher-fusion categories is required. The approach we will take is to motivate the relevant mathematics by first describing the physical setups (symmetries, interfaces, composition of symmetry generators, charges, gauging) and to then identify them a posteriori with the appropriate mathematics (objects, morphisms, fusion, higher-representations, modules of higher-fusion categories). This hopefully will both take the edge off the potentially overwhelming category literature, but in turn also convince the reader that higher categories are naturally the mathematical language -- much like groups and their representations are for studying ordinary symmetries -- and will somewhat inevitably enter our vocabulary.

\subsection{Scope and Overview}

These lectures are intended as an introduction to the concept of generalized global symmetries and its extension into non-invertible symmetries assuming familiarity with QFT and standard considerations of symmetries. The goal is to provide an accessible approach to this topic, without overburdening the reader with generality. The focus will therefore be on introducing general ideas, with a focus on illustrating them with concrete examples.

\paragraph{Scope and Aim of these Lectures.}
The lectures are aimed at non-experts in the field of symmetries, and 
the main pre-requisit is knowledge of advanced QFT. Although categories will play a role, no prerequisite along those lines is required. 
The reader will however benefit from some mathematical background in algebraic topology, see e.g. Hatcher \cite{Hatcher} for a brilliant mathematics text book, but also \cite{Kapustin:2014gua, Benini:2018reh, Hsin:2020nts} for expositions within the generalized symmetry literature geared towards physicists. Some basic representation theory of (finite) groups, e.g. Fulton-Harris \cite{FultonHarris}, would not go amiss either.

\paragraph{Invertible Generalized Symmetries.}
Before getting to non-invertible symmetries, we will have to discuss  invertible, generalized symmetries. Although no comprehensive technical review of the field exists, there are some overviews and introductory texts \cite{Cordova:2022ruw, Gomes:2023ahz}. 
Section \ref{sec:Inv} will introduce some of the salient features of symmetries as topological operators, and the generalization from standard global symmetries (0-form symmetries) to higher-form, i.e. $p$-form  symmetry groups $G^{(p)}$, and higher-groups, i.e. $(p+1)$-groups $\mathbb{G}^{(p+1)}$. We consider invertible symmetries throughout section \ref{sec:Inv}. The concept of charges under higher-form symmetries will be introduced, and we will discuss coupling  to background fields and gauging of higher-form symmetries. We will briefly discuss 't Hooft anomalies and 2-group symmetries as well.

\paragraph{Non-Invertible Symmetries: General Constructions.}
In section \ref{sec:NonInv1}, we explore non-invertible symmetries: first discussing examples that occur ``naturally in QFT", which the reader may have encountered. This is includes 4d $O(2)$ gauge theory and 2d rational conformal field theories RCFTs, which exhibit non-invertible symmetries. 

We then turn to studying symmetries in 2d theories from the perspective of topological operators: many ideas that we will encounter in higher-dimensions already occur in 2d. In particular gauging a group-like 0-form symmetry which is non-abelian, results in a non-invertible symmetry. This idea will extend to higher dimensions in terms of  theta defects \cite{Bhardwaj:2022lsg} -- these are generalizations of theta-angles, and provide a universal origin for non-invertible symmetries. They give rise to e.g. condensation defects \cite{Gaiotto:2019xmp, Roumpedakis:2022aik}.

\paragraph{Non-Invertible Symmetries: 3d and 4d QFT Realizations.}

In section \ref{sec:QFTs} we consider realizations of these symmetries in QFTs, focusing on 3d and 4d. The idea of stacking and gauging will follow through the entire section. The first class of examples are theta or condensation defects in pure gauge theories. 
We then apply the same logic to theories with mixed 't Hooft anomalies which again give rise to non-invertible symmetries  following the construction in \cite{Kaidi:2021xfk} and for the ABJ anomaly in \cite{Choi:2022jqy, Cordova:2022ieu}. These have again a (twisted) theta-defect interpretation. 
We then return to the initial example of disconnected gauge groups and show that stacking with TQFTs, which have higher-group symmetries, yields the non-invertible symmetries (again as theta-defects). 
Finally, we discuss self-duality defects  in 4d \cite{Choi:2021kmx}  -- modelled along the lines of the Ising model in 2d. 

\paragraph{Generalized Charges.}

In section \ref{sec:Charges} we turn to the discussion of charged operators, which from a QFT perspective may be equally important to the symmetry generators that we discussed up until that point. In this section, we will in fact restrict to invertible symmetries. This may seem surprising. However, we show that even for group-like symmetries $G^{(p)}$ (and higher-groups) there is more than representation theory. I.e. the standard paradigm that a $p$-dimensional defect transforms in a representation of $G^{(p)}$ is by far not exhaustive  \cite{Bhardwaj:2023wzd, Bartsch:2023pzl}. 
We  discuss generalized charges for invertible symmetries -- and show, that they in fact also form a category, which is the higher representation category of the invertible symmetry $G^{(p)}$. So there is really no escaping from higher-categories, even if you stay within the realm of invertible symmetries! 

\paragraph{Symmetry TFT.}

What emerges from these recent developments in generalized symmetries, is that it is useful to separate the physical theory from the symmetry that it carries. The tool to do so is the Symmetry TFT (SymTFT) \cite{Gaiotto:2020iye, Apruzzi:2021nmk, Freed:2022qnc}. This is the topic of the last section \ref{sec:SymTFT}: the SymTFT is a $(d+1)$-dimensional topological field theory, which encodes the symmetries but also the generalized charges \cite{Bhardwaj:2023wzd, Bhardwaj:2023ayw}. 
As a bonus, it also connects very closely to constructions in string theory and holography. 

We conclude in section \ref{sec:Outlook} with some open questions.

\subsection{Conventions}

Throughout these lectures $d$ is the spacetime dimension and we will work usually in Euclidean signature (unless otherwise stated). Topological operators of dimension $q$ will be always denoted by 
\be
D_q^\alpha\,,
\ee
where $\alpha$ labels the specific operator (e.g. a group element or for non-invertible symmetries a suitable label set). 
Sometimes we will specify a manifold $M_q$ or a background field $B$ to stress that the operator is defined on this space coupled to the background field, by writing $D_q^\alpha (M_q, B)$.
Not necessarily topological operators, e.g. physical line or surface operators, of dimension $q$ will be labeled by 
\be
\cO_q \,.
\ee
Occasionally we will denote line operators by $L$. 
Background gauge fields will be denoted by upper case letters $B_1, B_2, B_3$ and lower case letters correspond to dynamical gauge fields $b_1, b_2, b_3$. 

A $p$-form symmetry will be labeled by $G^{(p)}$, though we may drop the superscript when referring to the underlying group $G$. These groups for $p=0$ can be non-abelian, and are abelian otherwise. Furthermore, $(p+1)$-groups, which in particular contain $p$-form symmetries and $0$-form symmetries, are denoted by $\mathbb{G}^{(p+1)}$. 


\section{Generalized Global Symmetries: Invertible Symmetries}
\label{sec:Inv}


\subsection{Symmetries as Topological Operators}

The fundamental insight of  \cite{Gaiotto:2014kfa} is that   global symmetries of a QFT should be identified with topological operators. 
 An ordinary (soon to be called 0-form) symmetry is generated by codimension 1 topological operator
\be
D_{d-1}^{(g)} \,,\qquad g\in G^{(0)} \,.
\ee
The generalization proposed in \cite{Gaiotto:2014kfa} and subsequent works on higher-form and higher-group symmetries is to consider operators of higher-codimension (which may not necessarily be independent, as in the higher-group case). 
However all these generalizations assume that the composition of such operators, when e.g. inserted into correlation functions, forms groups, and so every element of this generalized symmetry has an inverse. 
These {\it invertible or group-like} generalized symmetries are the topic of this first section. Some basic field theory introductions  can be found here \cite{TongGauge, Gomes:2023ahz}. The most comprehensive discussion of higher-form symmetries is \cite{Gaiotto:2014kfa}, and for higher-group symmetries \cite{Tachikawa:2017gyf}.


\subsection{Noether Currents and Global Symmetries}

Before discussing generalizations to higher-form symmetries, we should see how this is consistent with the known paradigms of global symmetries in QFT, e.g. arising from symmetries with conserved currents, as derived by Noether's theorem.

Let us recall how symmetries and conserved charges relate in standard field theory. Consider a symmetry $G$ of a field theory, which leaves invariant the action, $\delta S=0$.  Then Noether's theorem implies that there is a conserved current $\partial_\mu J^\mu= 0$ or in terms of differential forms
\be
 d\star\bm{J}  =0 \,,
\ee
where $\bm{J}= J^\mu dx_\mu$
 and a charge, which is obtained by integrating $\star {\bf J}$ over a codimension 1 manifold $M_{d-1}$
\be
Q(M_{d-1})=  \int_{M_{d-1}} \star \bm{J}\,,
\ee
which is conserved. Here $M_{d-1}$ is a $d-1$ dimensional spatial slice. 
In a QFT setting we can insert the current into correlators and derive Ward identities, e.g. for a single local operator $\cO$ inserted at a point ${\bf p}$ these are  
\be
\delta^{(d)}({\bf x} - {\bf p})\left\langle\delta  \cO({\bf p})  \right\rangle=  \partial_\mu\left\langle J^\mu({\bf x}) \cO({\bf p})  \right\rangle  \,.
\ee
Integrating this equation over a $d$-dimensional $N_d$ with boundary $M_{d-1}$ 
\be
\ba
\int_{N_d} d^d {\bf x }\delta^{(d)}({\bf x} - {\bf p})\left\langle\delta  \cO({\bf p})\right\rangle  
 =&\int_{N_d}   \langle d \star\bm{J}  \cO ({\bf p}) \rangle \cr
= & \int_{M_{d-1}} \langle \star \bm{J} \cO ({\bf p})\rangle  
= \langle Q(M_{d-1}) \cO({\bf p}) \rangle  
\,.
\ea
\ee
This is non-zero if the point ${\bf p}$, at which  $\cO$ is inserted,  is inside the volume $N_d$, or contained within the interior of $M_{d-1}$. Mathematically this means, the point ${\bf p}$ links with the codimension 1 space $M_{d-1}$, and we can write the charge in terms of the linking 
\be\label{QO}
\langle Q(M_{d-1}) \cO({\bf p}) \rangle = L(M_{d-1}, {\bf p}) \langle \delta \cO ({\bf p}) \rangle \,.
\ee 
There are a few things to note here:  
the current conservation ensures in fact that the charge is independent of $M_{d-1}$ as long as the linking is non-trivial with ${\bf p}$. Put differently the linking $L(M_{d-1}, {\bf p})$ is a topological property. 
We can deform $M_{d-1}$ whilst retaining the linking (independent of the precise shape, or metric properties, of $M_{d-1}$, as long as ${\bf p}$ is contained inside the deformation of $M_{d-1}$). 
From this perspective it is natural to  abstract this, and define a symmetry generator as such a topological operator of dimension $d-1$.

For a continuous symmetry  $G^{(0)}$, e.g. a $U(1)$, with a conserved current, we can define a 
topological operator $D_{d-1}^{(g)}(M_{d-1})$, which implements the action of the symmetry on the operators as follows (we are now firmly working in Euclidean signature). Here, $M_{d-1}$ is not generically a time-slice but any $(d-1)$-dimensional manifold in space-time. 
Parametrize  $g = e^{i \alpha} \in U(1)$, then we define 
\be
D_{d-1}^{(g= e^{i \alpha})}(M_{d-1}) = \exp\left(i \alpha Q (M_{d-1})  \right) \,.
\ee
We can then think of inserting this operator into correlators as in (\ref{QO}), which amounts to acting with the symmetry generator on the local operator. 
Inserting several of these, composes as in the group composition $g\cdot  h = gh$ i.e. 
\be
\left\langle D_{d-1}^{(g= e^{i \alpha})}(M_{d-1}) D_{d-1}^{(h= e^{i \beta})}(M_{d-1}) \cdots \right\rangle = 
\left\langle D_{d-1}^{(gh= e^{i (\alpha+ \beta)})}(M_{d-1})  \cdots \right\rangle \,.
\ee
More generally, we find that  inserting two such topological operators for elements $g, h\in G^{(0)}$ into correlation functions is equivalent to inserting the topological operator associated to their product $gh \in G^{(0)}$, i.e. 
\be
D_{d-1}^{(g)} \otimes D_{d-1}^{(h)} = D_{d-1}^{(gh)} \,.
\ee

For $U(1)$ the topological operator acts by a phase as in (\ref{QO}), but more generally\footnote{We will not write the correlation functions in the following.} 
\be
D_{d-1}^{(g)}(M_{d-1}) \cO_0^i = R_{ij}^g \cO_0^j \,, 
\ee
where $i$ labels the multiplet that this local operator transforms in, given in terms of a representation $R_{ij}^g$ of the group element $g$.  The subscript $0$ indicates that this is a point-like operator.

This operator $D_{d-1}^{(g)}(M_{d-1})$ is  topological as the associated current is conserved and thus by Gauss' theorem any small deformation of $M_{d-1}$ is immaterial, as long as the point where the local operator is inserted is not crossed, see figure \ref{fig:Noethersphere}.

In the following we will abstract from this setup with Noether currents to symmetries, which may not admit such a conserved current description, and simply identify symmetries with topological operators. Thus codimension 1  topological operators $D_{d-1}^{(g)}$ will be identified with generators of 0-form symmetries $G^{(0)}$, which can be either continuous groups or finite groups (and in the latter case do not have a usual conserved current formulation). They do not necessarily have to be abelian.


\subsection{Higher-Form Symmetries}
\label{sec:hfs}

The first generalization of global symmetries is to higher-form symmetries. The most straight-forward way to generalize the notion of a symmetry, which is generated by a topological operator $D_{d-1}^{(g)}$, $g\in G^{(0)}$, is to allow the topological operator to be of different dimension: 
\be
D_{d-(p+1)}^{(g)}\,,\qquad g\in G^{(p)} \,.
\ee
A $G^{(p)}$-form symmetry is generated by topological operators of codimension $p+1$, $D_{d-(p+1)}^{(g)}$,  $g\in G^{(p)}$, satisfying a group  law
\be
D_{d-(p+1)}^{(g)} \otimes D_{d-(p+1)}^{(h)}  = D_{d-(p+1)}^{(gh)} \,,\qquad g, h \in G^{(p)}  \,.
\ee

\paragraph{Charges.} 
Such a co-dimension $p+1$ operator naturally links with $p$-dimensional operators $\mathcal{O}_p$ -- although we will see that this is only a special case of higher-charges (see section \ref{sec:Charges}):
\be\label{squeeze}
\begin{tikzpicture}
\draw [thick](-4,0) -- (-2,0);
\node at (-4.5,0) {$\cO_{p}$};
\draw [ultra thick,white](-3.4,0) -- (-3.2,0);
\draw [thick,blue] (-3,0) ellipse (0.3 and 0.7);
\draw [ultra thick,white](-2.7,-0.1) -- (-2.7,0.1);
\draw [dashed,thick](-2.7396,0) -- (-2.6396,0);
\node[blue] at (-3,-1) {$D^{(g)}_{d-(p+1)}$};
\end{tikzpicture}
\qquad \qquad 
\begin{tikzpicture}
\draw [thick](-4,0) -- (-2,0);
\node at (-4.5,0) {$\cO_{p}$};
\draw [thick,blue] (-3,0) ellipse (0.05 and 0.1);
\node[blue] at (-3,-1) {$D^{(g)}_{d-(p+1)}$};
\end{tikzpicture}
\qquad \qquad 
\begin{tikzpicture}
\draw [thick](-4,0) -- (-2,0);
\node at (-4.5,0) {$\cO_{p}$};
\node[blue] at (-3,-1) {$q_g(\cO_{p}) \times \cO_p$};
\end{tikzpicture}
\ee
Let $\cO_p$ be a $p$-dimensional operator (it can be non-topological or topological, for instance for $p=1$ this could be a Wilson line), which does not end. 
Then a codimension $p+1$ topological operator $D_{d-(p+1)}^{(g)}$ can link with this. Then deforming this operator does not change the linking and squeezing as shown in (\ref{squeeze}) computes  the charge  $q_g (\cO_p)$ of $\cO_p$ under $g\in G^{(p)}$.

0-form symmetries can be non-abelian, i.e. the composition of codimension 1 defects depends on their ordering. For $p$-form symmetries, with $p\geq1$, two codimension $p+1$ defects $D_{d-p-1}$ can be continuously deformed to reverse their order in spacetime, so $G^{(p)}$ for $p\geq1$ have to be  abelian. 

As in the 0-form symmetry case, the linking is a topological invariant 
 (for a nice physics-motivated discussion of linking see \cite{horowitz1990quantum})
\be
L(M_{d-p-1}, W_p)  \,,
\ee
 where we have $D_{d-p-1}^{(g)}$ defined on a $(d-p-1)$-dimensional manifold $M_{d-p-1}$ and the charged operator $\cO_p$ has world-volume on a $p$-dimensional submanifold $W_{p}$. For the 0-form symmetry this was a codimension 1 surface and  a point. 

\paragraph{{Example}.}
The simplest example is 4d Maxwell theory. In this case the 1-form symmetry is continuous. We have a gauge field $A$ and field strength $F=dA$ with the conditions $dF= d\ast F=0$. Thus both $\ast F$ and $F$ are closed and we can construct topological surface defects 
\be
D_2^{\alpha, e} = e^{ i\alpha \int \ast F} \,,\qquad 
D_2^{\alpha, m} = e^{ i\alpha \int F} \,,\qquad \alpha \in [0, 2\pi) \,.
\ee
These form the  electric and magnetic $U(1)$ 1-form symmetries, respectively. They compose according to the $U(1)$ group multiplication, and there is an inverse to every element (by replacing $\alpha$ with $-\alpha$).
The charged line operators are Wilson lines $e^{iq_e \int A}$ and 't Hooft lines $e^{i q_m\int A^D}$, where $\ast F= dA^D$. 
The charge of the Wilson line can be measured by surrounding it with the 2d suface operator $D_2^{\alpha, e}$, and the charge is 
\be
D_2^{\alpha, e} (M_2) W_{q_e}(M_1) = e^{i \alpha q_e L(M_2, M_1)} W_{q_e} (M_1) \,,
\ee
where $L(M_2, M_1)$ is the linking of the surface and the line in 4d. 

\paragraph{1-Form Symmetries and Line Operators.}

Non-abelian gauge theories in 4d are another important set of examples.  
Let ${G}_{\text{gauge}}$ be the simply-connected group associated to a Lie algebra $\mathfrak{g}$. 
The 1-form symmetry $G^{(1)}$ for ${G}_{\text{gauge}}$-Yang-Mills theory is  the center of the gauge group 
\be
G^{(1)} = Z_{{G}_{\text{gauge}}}  \,,
\ee
where for a simply-connected Lie group $G$ we define 
\be
Z_{{G}}= \text{Center} ({G}) 
= \{g\in {G}: \ g h = h g \ \text{for all } h\in {G}\} \,.
\ee
The center for all the simple, simply-connected ADE Lie groups are shown in table \ref{tab:centers}. We will argue later on why this is the 1-form symmetry for this gauge theory. 

\begin{table}
$$
\begin{array}{|c|c|c|}\hline
G & Z_G & q(\bm{F}) \cr \hline \hline
SU(N) & \Z_N & 1 \mod N\cr \hline 
\Spin(4N) & \Z_2   \times \Z_2 & (1,1) \mod (2,2)\cr \hline 
\Spin(4N+2) & \Z_4  & 2 \mod 4\cr \hline 
\Spin(2N+1) & \Z_2 & 1 \mod 2\cr \hline 
E_6 & \Z_3  & 1 \mod 3\cr \hline
E_7 & \Z_2 & 1 \mod 2 \cr \hline 
E_8 & \Z_1 & 1 \mod 1 \cr \hline
\end{array}
$$
\caption{Simply-connected Lie groups $G$ and their centers $Z_G$, as well as the charge of the fundamental representation $\bm{F}$ under the generator(s) of the center. \label{tab:centers}}
\end{table}

For a pure gauge theory in 4d, with gauge algebra $\mathfrak{g}$, and particular choice of gauge group ${G}_{\text{gauge}}$, the set of line operators have been studied in depth in \cite{Kapustin:2005py, Aharony:2013hda}. Let us here illustrate this briefly. Let $\mathfrak{g}$ be simply-laced (all roots have the same length squared equal to $2$). Define $\Delta$ to be the set of all roots $\alpha$ of $\mathfrak{g}$, and $\Lambda_r$ the root lattice generated by $\Delta$. Let $\alpha^\vee$ be the co-roots (or Cartan subalgebra generators, which we can associate to roots again), i.e. $\alpha^\vee \cdot \beta \in \Z$ for all $\alpha, \beta \in \Delta$, and define $\Lambda_{cr}$ to be the lattice spanned by the coroots. This is the root lattice of the Goddard-Nuyts-Olive-dual algebra $\mathfrak{g}^*$ (though taking this dual is not relevant in the case of ADE types).
Then the dual lattices (i.e. the vectors that have integral inner product with all vectors in the lattice) define the weight $(w)$ and magnetic weight $(mw)$ lattices, respectively \cite{Goddard:1976qe}: 
\be
\Lambda_w= \Lambda_{cr}^* \,,\qquad 
\Lambda_{mw} = \Lambda_{r}^* \,.
\ee
Note that $\Lambda_r \subset \Lambda_{w}$ and $\Lambda_{cr} \subset \Lambda_{mw}$. A generic line operator is the world-line of a heavy probe particle, so carries some electric or magnetic charges or dyonic charge, and is labeled thus by an element $(\lambda_w, \lambda_{mw}) \in \Lambda_w \oplus \Lambda_{mw}$. We can furthermore restrict these charges to the Weyl chambers of $\mathfrak{g}$ and $\mathfrak{g}^*$, to remove redundancy in this labeling. 

For a pure gauge theory with gauge group ${G}_{\text{gauge}}$ we always have a line operator that is given in terms of the adjoint Wilson line with charge  $\lambda_w= \alpha$ and also $\lambda_{mw}=\alpha^\vee$, which can end. As the charges are additive, we can mod out by the root lattice in order to characterize the independent set of line operators, i.e. 
\be
\mathcal{L} = \Lambda_w/\Lambda_r \oplus \Lambda_{mw}/\Lambda_{cr} = Z_G \oplus Z_G  \,,\qquad \text{$Z_G$= center of $G$.}
\ee
The quotients by these sublattices are the center $Z_G$.

However not all lines in $\mathcal{L}$ are mutually local\footnote{The theory is not an absolute theory, but a relative theory, i.e. it is the boundary of a higher-dimensional theory,}. Let $L_{\lambda} = (\lambda_w, \lambda_{mw}) \in \mathcal{L}$. Inserting two arbitrary lines into a correlation function, may result in a sign ambiguity, as the lines  may not be mutually local: 
\be
L_\lambda L_\mu = L_\mu L_\lambda e^{2 \pi i \langle L_\lambda, L_\mu\rangle}\,,
\ee
where $\langle , \rangle$ is the Dirac pairing $\lambda_w \mu_{mw}  - \lambda_{mw} \mu_{w}$. 
To define a consistent theory, we choose  a maximal set of mutually local lines: this is a called a polarization $\Lambda\subset \mathcal{L}$, and gives rise to an absolute theory (i.e. a theory that is not a boundary condition to a higher-dimensional theory).  
Then the 1-form symmetry is the Pontryagin dual group $\widehat{\Lambda}$ to $\Lambda$
\be
G^{(1)} = \widehat{\Lambda} := \text{Hom} (\Lambda , U(1)) \,.
\ee
For all practical purposes, when $\Lambda= \Z_N$ (or sum of cyclic groups), the Pontryagin dual group is isomorphic to the group. We will argue later in section \ref{sec:Dual}, why this group appears here. 
For an abelian group, the Pontryagin dual group is the group of  the characters, i.e. indicate the irreducible representations. 
The resulting theory is an absolute theory with a definite 1-form symmetry. For now it suffices to require that an absolute theory should have at least a maximally, mutually local set of $p$-dimensional operators $\cO_p$. 

\paragraph{Global Form of the Gauge Group.} 
The spectrum of line operators is closely related with the global form of the gauge group. Let us consider a pure gauge theory with  gauge group  $G_{\text{gauge}}$, which is characterized in terms of a $G_{\text{gauge}}$-principal bundle with a connection, where $G_{\text{gauge}}$ is a Lie group with Lie algebra $\mathfrak{g}$. 
If $\widetilde{G}$ is the simply-connected gauge group, with center $Z_G$ as in table \ref{tab:centers},  Wilson lines of all charges under the center are allowed. If the gauge group is $\widetilde{G}/C$ where $C$ is a subgroup of the center, then the allowed bundles are restricted, and there are obstructions to lifting them to $\widetilde{G}$-bundles, which are given in terms of characteristic classes 
\be
w_2 \in H^2\left(M, C\right) \,.
\ee
Given a global form $G_{\text{gauge}}= \widetilde{G}/C$, only the weights invariant  under $C$ are allowed as Wilson line charges.  However there are now additional 't Hooft lines. An in depth analysis of all these possibilities can be found in ``Reading between the Lines" \cite{Aharony:2013hda}.

\paragraph{Example.}
Let the gauge algebra be $\mathfrak{su}(N)$. Then $\mathcal{L} = \mathbb{Z}_N \oplus \mathbb{Z}_N$ is generated by the fundamental Wilson line $W$ and 't Hooft line $W$, with Dirac pairing 
\be
\langle W, H \rangle = {1\over N} \,.
\ee
Two particularly simple choices of polarization  are:
\begin{itemize}
\item  $\Lambda= \langle W\rangle \cong \Z_N $: gauge group $G_{\text{gauge}}= SU(N)$
\item $\Lambda=  \langle H\rangle \cong \Z_N $: gauge group $G_{\text{gauge}}= PSU(N) = SU(N)/\Z_N$, i.e. $C= \Z_N$\,,
\end{itemize}
where the brackets indicate that the fundamental Wilson line $W$ or 't Hooft line $H$ span this polarization. 
These two theories have both $G^{(1)} \cong \Z_N$ (one is the electric, the other the magnetic, 1-form symmetry). For different choices of $C= \Z_k$ with $k|N$ other global forms of the gauge group exist (in addition to the choice of theta-angles).

\subsection{Screening of Higher-Form Symmetries}

Some line operators can end on local operators and in the process their higher-charge is screened (``trivialized"). More generally,  $p$-form symmetries can be screened by $p-1$ dimensional operators: whenever they are ``endable" in a $p-1$ dimensional operator the associated higher-form symmetry can be screened (trivialized). This arises from the equality of the  picture-sequence in figure \ref{fig:screen}.

\begin{figure}
\centering
\begin{tikzpicture}
\draw [thick](-4,0) -- (-2,0);
\node at (-4.5,0) {$\cO_{p}$};
\node[red] at (-2,-0.5) {$\cO_{p-1}$};
\draw [ultra thick,white](-3.4,0) -- (-3.2,0);
\draw [thick,blue] (-3,0) ellipse (0.3 and 0.7);
\draw [ultra thick,white](-2.7,-0.1) -- (-2.7,0.1);
\draw [red,fill=red] (-2,0) node (v1) {} ellipse (0.05 and 0.05);
\draw [dashed,thick](-2.7396,0) -- (-2.6396,0);
\node[blue] at (-3,-1) {$D^{(g)}_{d-(p+1)}$};
\begin{scope}[shift= {(5,0)}]
\node at (-6,0) {$=$};
\draw [thick](-4,0) -- (-2,0);
\node at (-4.5,0) {$\cO_{p}$};
\node[red] at (-2,-0.5) {$\cO_{p-1}$};
\draw [thick,blue] (-3,0) ellipse (0.05 and 0.1);
\draw [red,fill=red] (-2,0) node (v1) {} ellipse (0.05 and 0.05);
\node[blue] at (-3,-1) {$D^{(g)}_{d-(p+1)}$};
\end{scope}
\begin{scope}[shift= {(10,0)}]
\node at (-6,0) {$=$};
\draw [thick](-4,0) -- (-2,0);
\node at (-4.5,0) {$\cO_{p}$};
\node[red] at (-2,-0.5) {$\cO_{p-1}$};
\draw [red,fill=red] (-2,0) node (v1) {} ellipse (0.05 and 0.05);
\node[blue] at (-3,-1) {$q_g(\cO_{p}) \times \cO_p$};
\end{scope}
\begin{scope}[shift= {(0,-3)}]
\node[rotate= 90] at (-3,1) {$=$};
\draw [thick](-4,0) -- (-2.5,0);
\node at (-4.5,0) {$\cO_{p}$};
\node[red] at (-2.5,-0.5) {$\cO_{p-1}$};
\draw [thick,blue] (-1.5,0) ellipse (0.3 and 0.7);
\draw [red,fill=red] (-2.5,0) node (v1) {} ellipse (0.05 and 0.05);
\node[blue] at (-1.5,-1) {$D^{(g)}_{d-(p+1)}$};
\end{scope}
\begin{scope}[shift= {(5,-3)}]
\node at (-5.5,0) {$=$};
\draw [thick](-4,0) -- (-2,0);
\node at (-4.5,0) {$\cO_{p}$};
\node[red] at (-2,-0.5) {$\cO_{p-1}$};
\draw [red,fill=red] (-2,0) node (v1) {} ellipse (0.05 and 0.05);
\node[blue] at (-3,-1) {$1 \times \cO_p$};
\end{scope}
\end{tikzpicture}
\caption{Screening of a $p$-form symmetry:  $p$-dimensional defects  $\cO_{p}$  are charged under the $p$-form symmetry generators $D_{d-(p+1)}^{(g)}$. However, if $\cO_p$ can end on $\cO_{p-1}$, these $p$-form symmetries can get screened, by following the equality in the figure.  \label{fig:screen}}
\end{figure}
Here $D_{d-p-1}^{(g)}$ is a topological operators, which is a generator of the higher-form symmetry $G^{(p)}$. It links non-trivially with the charge, i.e. $p$-dimensional defect, $\cO_p$. If $\cO_p$ ends on a lower-dimensional defect $\cO_{p-1}$ (such as a line operator on a local operator, or a surface operator on a line), then we can compute the charge under the higher-form symmetry to be trivial:  by moving the symmetry generator off the end of the defect $\cO_p$. This effect is what is often referred to as ``screening" of higher-form symmetries. 
Equivalently, we now find that the position where we shrink the defect $D_{d-p-1}^{(g)}$ matters, and thus this ceased to be a topological defect, i.e. it is no longer a symmetry generator, and the symmetry is broken.

For the 1-form symmetries, line operators can be screened by local operators. The global form of the gauge group determines what matter representations of local operators are allowed. This in turn specifies which set of line operators are screened. Let us see this in a bit more detail:
the Wilson lines of a gauge theory transform in a representation $\bm{R}$ of the gauge group $G_{\text{gauge}}$
\be
W_{\bm{R}} = \Tr_{\bm{R}} P e^{i \int_\gamma A}  \,,
\ee
i.e. they are world-lines along the path $\gamma$ of heavy probe particles in the representation $\bm{R}$. 
A Wilson line $W_{\bm{R}}$ is endable on local operator $\cO_{\bm{R}}$ transforming in the same representation $\bm{R}$. This corresponds to an annihilation operator for the heavy probe particle that is creating the Wilson line. 
If a theory does not have a particular operator representation, the corresponding Wilson line does not end, and carries a non-trivial charge under the 1-form symmetry. 

Adding e.g. matter to pure  gauge theories with simply-connected groups can screen the center symmetry, depending on the charge of the matter, see table \ref{tab:centers} for the charge of the fundamental representations  under the centers of ADE gauge groups.

\paragraph{Example.} 
Consider the simplest non-abelian gauge algebra in 4d: $\mathfrak{su}(2)$. There are two global forms of the gauge group, the simply-connected $SU(2) = \Spin(3)$, and the non-simply-connected $SO(3)$. $SO(3)$ gauge bundles have an obstruction $w_2\in H^2 (M_4, \Z_2)$ to lifting the gauge bundles from $SO(3)$ to $SU(2)$. 
The irreducible representations $\bm{R}_j$ are labeled by their highest weight, which we normalize to be half-integral $j$, with dim($\bm{R}_j)= 2j+1$. The center in this case is $\Z_2$, and the charge under the center symmetry is 
\be
q (\bm{R}_j) = (-1)^{2j}  \,.
\ee
The two gauge groups $SU(2)$ and $SO(3)$ are related by gauging the center 
\be\label{SO3SU2}
SO(3) = SU(2)/\Z_2\,.
\ee
In particular, representations that are charged under the center (with half-integer $j$) will not be representations of $SO(3)$! 
In terms of Wilson lines, we have lines in representations $W_{\bm{R}_j}$, however for $SO(3)$ only integral $j$ are allowed, whereas for $SU(2)$ any half-integral value is allowed. 
For a pure gauge theory local operators  $\cO_{\text{Adj}}$ come in the $j=1$ adjoint representation and tensor products thereof. For pure $SU(2)$ gauge theories the fundamental Wilson line, i.e. $j=1/2$, cannot end, and is charged under the center. Adding fundamental matter, i.e. $\cO_{\bm{F}= \bm{R}_{j=1/2}}$, screens the fundamental Wilson line, and the resulting theory has no 1-form symmetry. 

For $SO(3)$, there is no fundamental Wilson line, and thus we do not get any non-trivial Wilson lines. However, there are 't Hooft lines which are characterized by a charge under $\pi_1 (SO(3)) = \mathbb{Z}_2$. Again, adding lines that correspond to world-lines of monopoles in the $j=1/2$ representation screens this 1-form symmetry. 

We will see momentarily that these 1-form symmetries of pure $SO(3)$ and $SU(2)$ are consistent with the relation (\ref{SO3SU2}), i.e. we can obtain $SO(3)$ by gauging the center (1-form) symmetry of $SU(2)$. We will see that gauging a 1-form symmetry in 4d yields a dual 1-form symmetry and this is precisely the one that the 't Hooft lines are charged under.

\paragraph{Perspective from the Charges.}
An equivalent way of thinking about the $p$-form symmetry is in terms of an equivalence relation on $p$-dimensional operators \cite{Bhardwaj:2021wif, Apruzzi:2021vcu, Bhardwaj:2022dyt}. The equivalence relation is furnished by means of local operators that can form junctions between lines, i.e. the charges. This perspective will be useful when considering higher-groups in the next subsection. 

Let $\Lambda_p$ be a set of mutually local $p$-dimensional defects $\cO_p$ (i.e. we picked a polarization).
There is an equivalence relation on $\Lambda_p$, which will determine the charges under the 1-form symmetry\footnote{These are not necessarily equivalent operators for other puposes.}: 
\be
\cO^{(1)}_p\sim \cO_p^{(2)} \quad \Leftrightarrow \quad \exists \ O_{p-1} \ \text{ at the junction between $\cO^{(1)}_p$ and $\cO_p^{(2)}$}\,.
\ee
In particular if a defect is endable, then it is by definition equivalent to the trivial defect. 
Let us define the identification by the equivalence relation to be 
\be
\widehat{G^{(p)}} := \Lambda_p/\sim  \,.
\ee
This is the set of inequivalent lines (which will be an abelian group) and is the Pontryagin dual group of $G^{(p)}$
\be
\widehat{G^{(p)}} = \text{Hom} (G^{(p)}, U(1)) \,.
\ee
If $\widehat{G}^{(p)}$ is finite and abelian, then $G^{(p)} \cong \widehat{G^{(p)}}$ as abelian groups.
Again we should think of this as the group of characters of the abelian group $G^{(p)}$.

\paragraph{Line Operators and 1-form symmetry.}
On line operators the equivalence relation is formulated  with junctions that are local operators:
\be\label{Freta}
\begin{tikzpicture}
\draw [thick](-4,0) -- (-2,0);
\node at (-4.5,0) {$\cO_1^{(1)}$};
\node at (0.5,0) {$\cO_1^{(2)}$};
\node[red] at (-2,-0.5) {$\cO_0^{(12)}$};
\draw [thick](0,0) -- (-2,0);
\draw [red,fill=red] (-2,0) node (v1) {} ellipse (0.05 and 0.05);
\node at (-8.5,0) {$\cO_1^{(1)} \sim \cO_1^{(2)} \quad \Longleftrightarrow \   \text{there exists}$};
\end{tikzpicture} 
\ee
E.g. in a pure $G_{\text{gauge}}$ (simply-connected) Yang Mills theory, we have fundamental Wilson lines. The only local operators are in the adjoint, which are uncharged under the center and so
\be
G^{(1)}= Z_{G_{\text{gauge}}}\,.
\ee
From this perspective it is also clear that adding matter, and thereby expanding the set of local operators that can be used as junctions will generically reduce the higher-form symmetry (depending on the charges under the center of the matter representation, see table \ref{tab:centers}).

\subsection{Dual Symmetries and Background Fields}
\label{sec:Dual}

Higher-form symmetries can be coupled to background fields. 
Lets first recall a slight extension of the standard 0-form symmetry background field story, by considering a finite, abelian 0-form symmetry. 
A background $A$ for a finite 0-form symmetry group $G^{(0)}$ is a specification of  holonomies $g\in G$ on each closed 1d submanifold in spacetime $g= e^{i \oint_{M_1} A}$.
Furthermore we want this to be a trivial holonomy if the 1d manifold is the boundary of a 2d manifold, i.e. $dA=0$. Finally, this should be invariant under gauge transformations $A\to A + d\lambda$. Such backgrounds mathematically take values in 
\be\label{AH1}
A \in H^1 (M, G^{(0)}) \,.
\ee

In analogy, we define for a higher-form symmetry $G^{(p)}$ the background field $B_{p+1}$, which is closed and assigns holonomies valued in $G^{(p)}$ to $(p+1)$-dimensional cycles. Again, we consider only gauge equivalence classes, under 
$B_{p+1} \to B_{p+1} + d\lambda_{p}$. The background fields for a $p$-form symmetry $G^{(p)}$ is then naturally a generalization of (\ref{AH1})
\be
B_{p+1} \in H^{p+1} (M_d, G^{(p)}) \,.
\ee
Given a $p$-form symmetry $G^{(p)}$, with background field $B_{p+1}$, there is dual $(d-p-2)$-form symmetry $\widehat{G}^{(d-p-2)}$ with background field $\widehat{B}_{d-p-1}$
The dual gauge field 
$\widehat{B}_{d-p-1}$ takes values again in
\be
\widehat{B}_{d-p-1} \in H^{d-p-1} (M_d, \widehat{G}^{(d-p-2)})\,.
\ee
For a $U(1)$ 1-form symmetry in 4d, we have seen this already in terms of the gauge field $A$ and its dual $A^D$ with $F = dA$ and $* F = dA^D$. 
The main examples we will encounter are: 0-form symmetries are dual to 0-form symmetries in 2d, likewise 1-form maps to 1-form symmetries under gauging in 4d. Finally, 0-form and 1-form symmetries are dual in 3d.

We can pass from a given $p$-form symmetry to its dual symmetry by gauging. 
For continuum fields this is simply the standard procedure of coupling to a dynamical gauge field. In the realm of finite symmetries -- as we will mostly encounter -- gauging reduces to 
 summing over all (flat) background field configurations.
 
For discrete symmetries we can then think  of the gauging as a kind of discrete Fourier transformation of the partition function:
 In terms of the partition function $Z(\cT, B)$ of a theory $\cT$ with global symmetry $G^{(p)}$ and background fields $B_{p+1}$, the partition function of the  gauged theory reads 
\be\label{gaugeZ}
Z \left(\cT/G^{(p)}, \widehat{B}_{d-p-1}\right) 
\sim \sum_{B_{p+1}\in H^{p+1} (M_d, G^{(p)})} 
e^{2 \pi i (\widehat{B}, B)}Z \left(\cT, B_{p+1}\right) \,.
\ee
The term $e^{2\pi i (\widehat{B}, B)}$ can be thought of as coupling the dual theory  to a ``current", which is the field $B$. 
The pairing between $B$ and $\widehat{B}$ is the weight in the ``Fourier transform"
\be
(,):\qquad H^{p+1} (M_d, G^{(p)})  \times H^{d-p-1} (M_d, \widehat{G}^{(d-p-2)}) \ \to \ \mathbb{R}/\mathbb{Z} \,,
\ee
which uses Poincar\'e duality $H^{d-p-1} \cong H_{p+1}$ (so we can evaluate $B$ on the Poincar'e dual cycle to $\hat{B}$) and the pairing between $G^{(p)}$ and its Pontryagin dual $\widehat{G}^{(d-p-2)}$, i.e. given 
\be
\widehat{g}  \in \widehat{G} = \Hom (G, U(1)) 
\ee
this defines a map from $G$ to $U(1)$
\be
\ba
\widehat{g}:\qquad  G \ &\to\  U(1) \cr 
 g\  & \mapsto \ \widehat{g} (g)  
 \,. 
 \ea
\ee
Gauging the dual symmetry results back in the orginal theory.

We can also see that the dual symmetry provides an explicit way to write the symmetry generators of the $p$-form symmetry namely 
\be
D_{d-p-1}= \exp \left(i \int_{M_{d-p-1}} \widehat{B}_{d-p-1} \right) \,, 
\ee
which link with $p$-dimensional operators $\cO_p$ in $d$-dimensions. 
Likewise the dual symmetry ${G}^{(d-p-2)}$ in the gauged theory can be written as 
\be
\widehat{D}_{p+1}= \exp \left(i \int_{M_{p+1}} {B}_{p+1} \right) \,, 
\ee
which again link with the defects $\cO_{d-p-2}$ that are charged under this $(d-p-2)$-form symmetry.

For instance consider the pure $SU(2)$ gauge theory in 4d. 
This has a 1-form symmetry $\Z_2^{(1)}$, with background fields $B_2 \in H^2 (M_4, \Z_2)$. Summing over these has a very simple interpretation as as sum over Stiefel-Whitney classes, which measure the obstruction to lifting $SO(3)$ bundles to $SU(2)\cong \Spin(3)$ bundles. We therefore see that gauging the 1-form symmetry results in a path integral over $SO(3)$ gauge bundles, which include the sum over Stiefel-Whitney classes. 
The gauged theory has a dual $\widehat{\Z}_2$ 1-form symmetry, and gauge group $SO(3)$, with background fields $\widehat{B}_2 \in H^2 (M_2, \widehat{\Z}_2)$. The 1-form symmetry generators, which are the Gukov-Witten operators, are $D_2^{(\widehat{B}_2)} = e^{i \int_{M_2} \widehat{B}_2}$ for the 1-form symmetry in $SU(2)$, whereas $D_2^{(B_2)}$ for the dual 1-form symmetry in $SO(3)$. 

At this point we should also make a comment about the formulation of background fields which are $\Z_N$ valued: we can either stay with continuous $U(1)$-valued fields $B_{p+1}$, and impose $N dB_{p+1}=0$, which implies they are $N$-torsion. Or we can use a cochain formulation, including a discussion of cup-products (generalizing the wedge product) and coboundary $\delta$ operation (which generalizes $d$). The latter is clearly  the more elegant version, but requires some reading of \cite{Hatcher}, or one of the physics references \cite{Kapustin:2014gua, Benini:2018reh, Hsin:2020nts}, which provide a quick introduction. 
For practical purposes, the map from continuous to discrete valued fields  
\be\label{boom}
B^{\text{continuum}} = {2\pi \over N} B^{\text{discrete}}\,.
\ee

\subsection{'t Hooft Anomalies}
\label{sec:Anomalies}

't Hooft anomalies of global symmetries are not inconsistencies, but in fact important tools to study RG-flows.
They are a feature of the theory, not a bug. They provide an obstruction to potentially gauging a symmetry (e.g. a pure 1-form symmetry anomaly would obstruct gauging that symmetry), however more importantly, they are RG-flow invariants and thus provide vital information that can be matched from the UV to the IR. 
In standard QFT textbooks 't Hooft anomalies for continuous global symmetries are computed from triangle diagrams, coupling the theory to flavor symmetry currents. When the symmetries are finite groups, we need to slightly abstract from this. 

Let us consider a theory for concreteness, with a 1-form symmetry. 
What are the 't Hooft anomalies? 
An anomaly for the 1-form symmetry arises if the correlation function of the theory $\cT$, coupled to the background field $B_2$ of the 1-form symmetry $Z(B_2)$ is \textit{not} a well-defined function of the cohomology class $[B_2]$ of $B_2$, i.e. if we perform a gauge transformations 
\be
B_2 \to B_2 + d \lambda_1 \,,
\ee
then the partition function is not invariant 
\be
Z(B_2+d\lambda_1)=\varphi(\lambda_1,B_2)\times Z(B_2) \,.
\ee
Here, $\varphi(\lambda_1,B_2)\in U(1)$ is a phase factor depending on $\lambda_1$ and $B_2$.

We can quantify the anomaly of $\mathcal{T}$ by constructing a so-called anomaly theory, which has the property that is cancels the anomalous transformation of $\mathcal{T}$ under background gauge transformations.
This theory lives in one higher-dimensions, and the process of cancelling the anomaly by $d+1$ dimensional theory, is usually referred to as anomaly inflow. 

Associate a $(d+1)$-dimensional anomaly theory  $\mathcal{A}_{d+1}(B_2)$ carrying a non-trivial 1-form symmetry ${G^{(1)}}$. 
The partition function $\mathcal{A}_{d+1}(B_2)$ on a compact manifold $M_{d+1}$ with a background $B_2$ is  well-defined function of $[B_2]$. However on  $\partial M_{d+1}=M_d \not= \emptyset$ it trarnsforms as follows
\be
\mathcal{A}_{d+1}(B_2+d\lambda_1)=\varphi^{-1}(\lambda_1,B_2)\times \mathcal{A}_{d+1}(B_2)  \,.
\ee
Regarding the $d$-dimensional theory $\cT$ as a boundary condition for the $(d+1)$-dimensional theory restores invariance under background gauge transformations 
\be
\widetilde{Z}(B_2)=Z(B_2)\mathcal{A}_{d+1}(B_2)\,,
\ee
which satisfies
\be
\widetilde{Z}(B_2+d\lambda_1)=\widetilde{Z}(B_2) \,.
\ee
Thus we have restored gauge-invariance under background gauge transformations of the 1-form symmetry, however at the price that the theory is now no-longer defined in $d$-dimensions only, but is coupled to a  $(d+1)$-dimensional theory. 

\paragraph{Examples: 3d.}
The types of 't Hooft anomalies that arise for 1-form symmetries in 3d are 
\be\label{B2ano}
\mathcal{A}_4=\text{exp}\left(2\pi i {1\over 2N}\int {\mathfrak{P}(B_2)}\right) \,,
\ee
where $\mathfrak{P}$ is the Pontryagin square\footnote{This is defined as a map 
$\mathfrak{P} \left(B_2\right): H^2\left(M_3, G^{(1)}\right) \rightarrow H^4\left(M_3, \mathbb{R} / \mathbb{Z}\right)
$, which depends on a bilinear pairing
$\eta: {G^{(1)}}\times{G^{(1)}}\to \mathbb{R}/\mathbb{Z}$
with quadratic refinement $\sigma: G^{(1)} \to \mathbb{R}/\mathbb{Z}$. See e.g. \cite{Hatcher}.}, which is in practice closely related to $B_2\cup B_2$ (or $B_2\wedge B_2$ in the continuum). For instance for $G^{(1)}= \Z_N$ we can state the Pontryagin square operation more explicitly: 
\be
\ba
N \text{ odd}: &\qquad   \mathfrak{P}(B_2 )=B_2 \cup B_2 \in H^4\left(M, \Z_N\right) \cr 
N \text{ even}: & \qquad \mathfrak{P}(B_2) \in H^4\left(M, \Z_{2N}\right) \,,\qquad \mathfrak{P}(B_2)= B \cup B \mod N \,.
\ea
\ee
The anomaly (\ref{B2ano}) obstructs gauging the 1-form symmetry. If you wonder how to see this explicitly, we will return to this question in  section \ref{sec:SymTFT} on the Symmetry TFT. 

We will encounter several types of 't Hooft anomalies in these lectures. The most important ones will be mixed 't Hooft anomalies between different higher-form symmetries. For instance in 4d $\mathcal{N}=1$ supersymmetric Yang-Mills with gauge group $SU(N)$, we have a chiral symmetry $\mathbb{Z}_{2N}$ and a 1-form symmetry $\mathbb{Z}_N$. These two enjoy a mixed 't Hooft anomaly 
\be
\mathcal{A}_5 = \exp \left(  {2 \pi i \over 2N} \int A_1 \cup \mathfrak{P} (B_2) \right) \,.
\ee
This type of anomaly will play an important role when constructing non-invertible symmetries, see section \ref{sec:KOZ}.

\subsection{Application: Confinement}

We will only discuss briefly one physical application of higher-form symmetries: confinement. A 1-form symmetry is a global symmetry, and acts as an order parameter for  confinement. A confining vacuum is one where the Wilson loop has area law, $\langle W \rangle \sim e^{- \text{Area}}$, whereas a deconfining vacuum the Wilson loop expectation value has perimeter law. In terms of the 1-form symmetry we can formulate this as follows: 
the 1-form symmetry is spontaneously unbroken/broken in a confining/deconfining vacuum. 

For 4d $\mathcal{N}=1$ $SU(N)$, the 1-form symmetry characterizes the IR physics. The theory has 
$N$ confining vacua. The 0-form R-symmetry gets broken from the UV to IR as follows:
\be
U(1)_R \quad \stackrel{\text{ABJ}}{\longrightarrow}\quad  \mathbb{Z}_{2N} \quad \stackrel{\chi_{\text{SB}}}{\longrightarrow} \quad  \mathbb{Z}_2 \,.
\ee
The ABJ anomaly breaks it to a finite subgroup $\Z_{2N}$, which then by chiral symmetry breaking gets further reduced to $\Z_2$. The broken generators $\Z_N$ map between the $N$ vacua. It is interesting to consider other global forms, see \cite{Aharony:2013hda} for general gauge groups. 
Let us consider here for simplicity the gauge algebra $\mathfrak{su}(2)$. There are two vacua: the vacuum where monopole ($m$) condenses, and  the one where the dyon ($d$) condenses. These can be derived from $\mathcal{N}=2$ after breaking to $\mathcal{N}=1$.  For $\mathfrak{su}(2)$ there are three polarizations, where the genuine line operators are the Wilson line $W$, the 't Hooft line $H$ or the dyonic line $W+H$, respectively. 
The preserved (i.e. spontaneously unbroken) 1-form symmetry for each polarization are then: 
\be
\begin{tabular}{|c||c|c|c|} \hline
$\Lambda$ & $G_{\text{gauge}} $ &$G_m^{(1)} $& $G_d^{(1)}$\cr \hline\hline
$W$ & $SU(2)$ &$\mathbb{Z}_2$ & $\mathbb{Z}_2$  \cr 
$H$ & $SO(3)_+$ &$\emptyset$ & $\mathbb{Z}_2$  \cr 
$H+W$ & $SO(3)_-$ &$\mathbb{Z}_2$ & $\emptyset$  \cr \hline
\end{tabular}
\ee
Here $G^{(1)}_{m, d}$ indicate the unbroken 1-form symmetry in the monopole/dyon condensing vacua. If there is an unbroken 1-form symmetry the vacuum is confining. 
In fact the symmetries of this class of theories (the $SO(3)$ gauge theory to be precise) go beyond these invertible 1-form symmetries, and we will return to this setup shortly.

\subsection{Higher-Group Symmetries}

Whenever a theory has a collection of higher-form symmetries $G^{(p)}$, $p= 0, 1,\cdots$ the combined symmetry group may be a product or  it forms what is called a higher-group symmetry. If a theory has 0-form and $p$-form symmetry which do not form a product, this is called a $(p+1)$-group symmetry: $\mathbb{G}^{(p+1)}$ \cite{Sharpe:2015mja, Tachikawa:2017gyf, Cordova:2018cvg, Benini:2018reh, Hsin:2020nts,  Apruzzi:2021mlh, Lee:2021crt}. 
We will focus here on  2-groups formed from 0-form and 1-form symmetry. As a warmup we should first however recall a few facts about ``1-groups". 

\paragraph{Group Theory Interlude.}
Let us consider finite abelian groups (imagine they act as 0-form symmetries), with 4 elements. There are two such groups 
\be
\text{Klein}=\mathbb{Z}_2\times \mathbb{Z}_2 \qquad \text{and}\qquad \mathbb{Z}_4 \,.
\ee
These are distinguished as follows. We can consider all groups which fit into the short exact sequence\footnote{A short exact sequence is a chain of maps, where at each entry, the kernel of the outgoing map is the image of the incoming map, e.g. $1\stackrel{a}{\to} G \stackrel{b}{\to} \cdots$ means that the map $b$ is injective, as $\text{image}(a)= 1 = \text{kernel} (b)$.}
\be
1 \to \Z_2 \to G \stackrel{\pi}{\to} \Z_2 \to 1\,,
\ee
i.e. $\Z_2$ embeds into $G$, and $G$ projects onto $\Z_2$. Such a short exact sequence is an example of a group extension, in this case $\Z_2$ by $\Z_2$. In particular central extensions are classified by group cohomology, which in this case is  
\be
H^2 (\mathbb{Z}_2 , \mathbb{Z}_2) = \mathbb{Z}_2 \,.
\ee
This indicates that there are two distinct extensions: one is trivial  (and the sequence splits), the other is a non-trivial central extension, and the resulting group is $\Z_4$. 
More generally for a finite group $G$ and abelian group $A=\Z_n$ these central extensions $E$ 
\be
1 \to A \to E \stackrel{\pi}{\to} G  \to 1\,,
\ee
are classified by 
\be
H^2 (G, A) \ni \omega: \quad G \times G \to A \,, 
\ee
which is a 2-cocycle. On pairs $(g, a) \in G \times A$ the multiplication in the central extension is defined as 
\be
(g_1, a_1) \cdot (g_2, a_2)  = (g_1 g_2 , a_1 +a_2 + \omega (g_1, g_2)) \,.
\ee
Associativity of this multiplication is equivalent to requiring $\delta\omega (g_1, g_2, g_3)=0$. 

The extension of two abelian groups does not have to be abelian. 
 The simplest instance is the semi-direct product. 
 A semi-direct product $G =H \ltimes N$, is given in terms of two subgroups $H$ and $N$, with $N$ normal subgroup ($N\lhd G$), 
 and a split extension 
 \be
1 \to N \to G \stackrel{\pi}{\to} H \to 1 \,,
 \ee
i.e.  there is a map $\sigma: H\to G$  such that $\pi \circ \sigma= \id_H$.
Then a semi-direct product is characterized by the action of $H$ on $N$ (automorphisms)
\be\label{semid}
H\to \text{Aut} (N) \,.
\ee
There is always the trivial action, which gives rise to the direct product. 
Examples of non-trivial semi-direct products are the permutation group on 3 elements, $S_3 = \Z_2\ltimes \Z_3$, which is a non-abelian semi-direct product of $\Z_2$ and $\Z_3$, as well as the Dihedral groups $D_{2n}= \Z_2\ltimes \Z_N$ (symmetries generated by rotations of an $n$-gon ($\Z_N$) and a reflection ($\Z_2$), which acts on the rotations).

\subsubsection{2-Group Symmetries}
\label{sec:TwoGp}

Higher-groups generalize this notion of extension, where formally (in a way we will make precise shortly) we consider an extension of a 0-form symmetry by a $p$-form symmetry. We will focus on 2-group symmetries, which are ``extensions" of $G^{(1)}$ and $G^{(0)}$. There will be also an analog of the semi-direct product. 

\begin{figure}
\centering
\begin{tikzpicture}
\draw [blue, fill=blue,opacity=0.2] 
(0,0) -- (0,4) -- (1.5,5) -- (1.5,1) -- (0,0);
 \draw [blue, thick] 
 (0,0) -- (0,4) -- (1.5,5) -- (1.5,1) -- (0,0);
 \draw [PineGreen, thick] (0.7,2.5) -- (3.5, 2.5) ;
 \draw [PineGreen, thick] (-2,2.5) -- (0.7, 2.5) ;
\node [PineGreen] at (-1.8, 2.8) {$D_2^{(C)}$} ;
\node[blue] at (0.8,-0.5) {$D^{(g)}_{3}, \, g\in \Z_2^{(0)}$};
\node [PineGreen, right] at (3.5, 2.5) {$D_2^{(C)}$} ;
 \draw [black,fill=black] (0.7,2.5) ellipse (0.03 and 0.03);
 \draw [orange, thick]  (0.7,2.5) -- (3, 1.5) ;
 \node[orange, right] at (3, 1.5) {$D_2^{(V)}$}; 
\end{tikzpicture}
\caption{Split 2-group: The 0-form symmetry $\Z_2^{(0)}$ generated by the 3d topological defect $D_3^{(g)}$ in 4d maps the two 1-form symmetry generators into each other: $D_2^{(C)}\leftrightarrow D_2^{(S)}$. This can be equivalently written as shown in the figure: as $D_2^{(C)}$ intersects $D_3^{(g)}$ it emits a $D_2^{(V)}$ defect (where $V$ parametrizes the diagonal $\Z_2^{(1)}$). 
\label{fig:Toes}}
\end{figure}
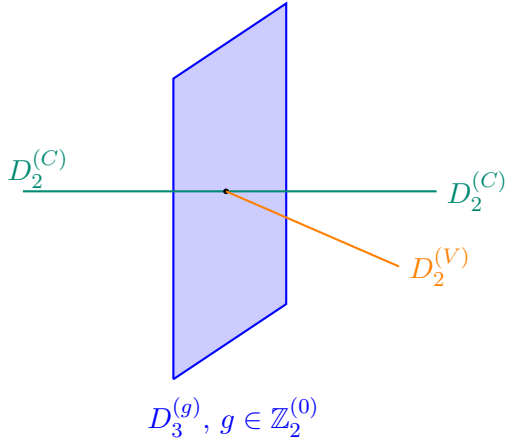

The main characteristic equation for a 2-group symmetry is 
\be
\delta B_2 \not =0 \,,
\ee
i.e. the  background $B_2$ of the 1-form symmetry $G^{(1)}$ is not closed. Instead it is given in terms of two possible contributions: a non-trivial Postnikov class (similar to a non-trivial extension) or a so-called split 2-group (similar to a semi-direct product). 
As we will encounter later on split 2-groups let us start with this: 

\paragraph{Split 2-groups.}
Consider the situation where we have a finite $G^{(0)}$ which acts on the 1-form symmetry as an outer automorphism. E.g. $G^{(1)} = \Z_N \times Z_N$ and $G^{(0)}= \Z_2$ acts as exchange of the two $\Z_N$ factors, i.e. we have a map similar to (\ref{semid})
\be
G^{(0)} \to \text{Aut} (G^{(1)}) \,.
\ee
An example is the theory with gauge group $\Spin(4N)$ whose center is $G^{(1)} = \Z_2^{(S)} \times \Z_2^{(C)}$, see table \ref{tab:centers}, and the outer automorphism exchanges these -- see figure \ref{fig:Toes}. 
In terms of the backgrounds we can state this as follows: denote the background for the diagonal $\Z_2^{(1), \diag} =\Z_2^{(V)}$ be $B_2$. Then 
\be\label{S2G}
\delta B_2 = A_1 C_2 \,,
\ee
with $A_1$ the background field for the 0-form symmetry, and $C_2$ the background for $\Z_2^{(C)}$. The variation of the background $B_2$ depends thus on the 0-form symmetry. This is a type of 2-group, which we will refer to as split 2-group. We will encounter them again as a useful building block for generating non-invertible symmetries.

\paragraph{2-Groups with Postnikov class.} 
The second type of 2-group symmetry is slightly more complicated. We will discuss it here for completeness, but will not require it in the subsequent discussions, and the reader should feel free to skip the remainder of this section. Let us start however with a very instructive example \cite{Lee:2021crt}:

\paragraph{Example: Spin$(4N+2)$ with Vector Matter.} 
Consider a 4d Spin$(4N+2)$ gauge group, whose center is {$\mathbb{Z}_4^{(1)}$}. Adding $2N_F$ matter fields in the vector representations results in an $\mathfrak{su}(2N_F)$ flavor symmetry algebra. This theory will now have a 2-group symmetry as shown in  \cite{Lee:2021crt}.
The Wilson line in $\bm{V}$ is now screened by the matter and the remaining 1-form symmetry is  $G^{(1)} = \mathbb{Z}_2^{(1)}$, generated by the spinor Wilson line $\bm{S}$.
However, 
the spinor line satisfies $\bm{S} \otimes \bm{S}  = \bm{ V}$, i.e. the Wilson line in the vector representations, which gets screened by an operator that is  
charged under the flavor center $\mathbb{Z}_2^{F} \subset \mathbb{Z}_{2N_F} \subset SU(2N_F)$. I.e. this gives rise to a ``flavor Wilson line". The 1-form and 0-form center symmetries fit into the extension 
\be
1 \rightarrow \mathbb{Z}_2^{(1)}   \rightarrow \mathbb{Z}_4 \rightarrow  \mathbb{Z}_2^{F} \rightarrow 1 \,.
\ee
In terms of the background fields for the symmetries, this can be stated as follows:
the $G^{(1)}$-background $B_2$, which is a cochain $B_2 \in C^2 (M, G^{(1)})$ and the obstruction {${\bf w}_2$ to lifting the flavor symmetry from $\mathcal{F} = SU(2N_F)/\mathbb{Z}_2^F$ to $SU(2N_F)$}  are related by: 
\be
\delta B_2 = B_1^* \text{Bock} ({\bf w}_2)\,,
\ee
where
\be
\Bock: \qquad  H^2 (B\mathcal{F}, \mathbb{Z}_2^F)\  \rightarrow \ H^3 (B\mathcal{F}, \mathbb{Z}_2^{(1)}) \,.
\ee
$B\mathcal{F}$ is the classifying space for $\mathcal{F}$ flavor group bundles, and $B_1: M \to B\mathcal{F}$, which is used to pull-back the class $\Bock ({\bf w}_2)$, which is the so-called Postnikov class for this 2-group. 
This dependence of the variation of the 1-form symmetry background on the background of the 0-form symmetry is referred to as a Postnikov 2-group symmetry.

Similarly one can see that 5d $SU(2)_0$ theory, which is a strongly coupled UV SCFT, has a 2-group symmetry \cite{Apruzzi:2021vcu}, which again is formed between the 1-form symmetry, i.e. the center $\Z_2^{(1)}$, and the non-simply-connected flavor symmetry group $\mathcal{F}=SO(3)= SU(2)/C$, so that $C$ and $\Z_2^{(1)}$ form a 2-group. Another class of theories, where a full classification of 2-groups is known are 6d SCFTs in \cite{Apruzzi:2021mlh}.

\paragraph{Perspective from the Line Operators and Screening.}

An alternative description of 2-groups is in terms of the screening picture \cite{Bhardwaj:2021wif, Apruzzi:2021vcu, Bhardwaj:2022dyt}. Recall that we defined  (the Pontryagin dual of) the 1-form symmetry as the quotient of all lines operators modulo the equivalence relation (\ref{Freta}). 

We can refine this equivalence relation as follows. Consider 
screening by local operators that can be charged under a 0-form (flavor) symmetry group 
\be
\mathcal{F} = {F\over C}\,.
\ee
Here $F$ is the simply-connected group,  and the global form of the flavor symmetry group is the quotient by $C$, which is a subgroup of the center of $F$. 
By coupling the system to flavor symmetry backgrounds we keep track of this flavor charge in terms of ``flavor symmetry Wilson lines". 
We then define a new equivalence relation on the set of  line operators $L$ and flavor Wilson lines $R$:
\be
\begin{tikzpicture}
\node at (-6.5,0) {$\left(L_1^{(1)},R_1^{(1)}\right) \sim \left(L_1^{(2)},R_1^{(2)}\right) \quad \Longleftrightarrow \quad \exists $};
\draw [thick] (-2,0) -- (2,0) ;
\draw [thick](-2,0) -- (-2,0);
\node[left] at (-2,0) {$L_1^{(1)}$};
\node[right] at (2,0) {$L_1^{(2)}$};
\node[red, below] at (0,0) {$\cO_0^{(12)}$};
\draw [blue, thick](0,0) -- (2,1.5);
\node[right] at (2,1.5) {$R_1^{(1)}\otimes (R_1^{(2)})^\ast$};
\draw [red,fill=red] (0,0)  ellipse (0.05 and 0.05);
\end{tikzpicture}
\ee
We can define the following equivalence class, which will be the analog of the extension group:  
\be
 \widehat{\mathcal{E}} = \{(L, R)\}/\sim  \,.
\ee
We can project from $\widehat{\mathcal{E}}$ to $\widehat{G}^{(1)}$,  and likewise $\widehat{C}$ injects into it, so we have a short exact sequence\footnote{We take the Pontryagin dual groups, which are all isomorphic as we assume all groups $G^{(1)}$, $C$, $\mathcal{E}$ here to be  finite.}
\be\label{2GpE}
{0\to G^{(1)} \to \mathcal{E} \to C \to 0 } \,.
\ee
This implies a long exact sequence in cohomology, where these finite groups form the coefficient groups, with the connecting map given by the Bockstein homomorphism (Bock) \cite{Hatcher}
\be
\begin{tikzcd}
\cdots \rightarrow  H^2 (M,G^{(1)}) \arrow[r] & H^2 (M, \mathcal{E}) \arrow[r] 
\arrow[d, phantom, ""{coordinate, name=Z}] &  H^2 (M, C) \arrow[dll,
"\text{Bock}",
rounded corners,
to path={ -- ([xshift=2ex]\tikztostart.east)
|- (Z) [near end]\tikztonodes
-| ([xshift=-3ex]\tikztotarget.west)
-- (\tikztotarget)}] \\ 
 H^3 (M, G^{(1)}) \arrow[r] & H^3 (M, \mathcal{E}) \arrow[r] &
 H^3 (M, C) \rightarrow \cdots 
\end{tikzcd}
\ee
If the sequence (\ref{2GpE}) does not split then there is a non-trivial Bockstein homomorphism
\be
\Bock: H^2( -,C) \to H^3(- , G^{(1)}) \,,
\ee
and there is a 2-group if
\be
{\delta B_2 = \Bock (w_2)} \,,
\ee
with {$w_2 \in H^2 (B\mathcal{F}, C)$ is the obstruction to lifting $\mathcal{F}$- to $F$-bundles.}

\paragraph{Summary: 2-groups with Postnikov class. }
To summarize, a 2-group with Postnikov class is specified by the following data
\be
\big(G^{(1)},\mathcal{F},\Theta\big) \,,
\ee
where $G^{(1)}$ is a finite 1-form symmetry with background $B_2$, $\mathcal{F}$ is the flavor symmetry group (0-form, but not necessarily finite) and 
the {Postnikov class}
\be
\Theta\in H^3(B\mathcal{F},G^{(1)})\,,
\ee
which specifies the relationship between the background fields
\be
{\delta B_2 = B_1^\ast\Theta  \,,\qquad B_1: \ M_{d}\to B\mathcal{F}} \,.
\ee

\subsubsection{2-Groups and 't Hooft Anomalies}

There is a close relationship between 2-group symmetries and 't Hooft anomalies -- via gauging. For an in depth discussion of this see \cite{Tachikawa:2017gyf}.
Consider a 2-group (including both Postnikov class and split 2-group)
\be\label{deltaB}
\delta B_2 = \Bock (w_2) + A_1 C_2 \,.
\ee
Gauging $B_2$ in (\ref{deltaB}), i.e. the 1-form symmetry (which assumes that there is no 't Hooft anomaly for $B_2$), we obtain a mixed anomaly 
\be
\int_{M_{d+1}}  \widehat{B}_{d-2} \cup \left( \Bock (w_2) + A_1 C_2\right) \,,
\ee
where $\widehat{B}_{d-2}$ is the background for the dual to the 1-form symmetry, i.e. a $(d-3)$-form symmetry.  Thus, gauging the 1-form symmetry in a  2-group symmetry results in a mixed anomaly for the  dual of the 1-form symmetry. For more examples of this type see \cite{Tachikawa:2017gyf}.


\section{Non-Invertible Symmetries from Stacking TQFTs}
\label{sec:NonInv1}

The starting point for our  symmetry exploration was the identification of symmetries as topological operators in a QFT. So far, in the lightning review of invertible symmetries of the last section, every symmetry generator had an inverse, and composition followed a group law (or higher-group). 

The core of these lectures will venture beyond this to  symmetries which do not follow a group-like composition and where not every symmetry generator has an inverse. 
On a crude level, we are replacing groups by algebra -- but we will see that algebras are not going to be sufficient in order to capture all the structure of symmetries. And we will be naturally led into a much more exciting world, that of categories. But more of that later. 

\subsection{First Encounter: Disconnected Gauge Groups}
\label{sec:O2Take1}

The simplest instance where non-invertible symmetries are apparent starts again with 4d abelian $U(1)$ gauge theory. We have seen in section \ref{sec:hfs} that this has topological surface operators, which generate the 1-form symmetries $U(1)_e^{(1)}$ and $U(1)_m^{(1)}$. In terms of the field strength we can write down the topological surface operators as 
\be
D_2^{\alpha, e} = e^{ i\alpha \int \ast F} \,,\qquad 
D_2^{\alpha, m} = e^{ i\alpha \int  F} \,,\qquad \alpha \in [0, 2\pi) \,.
\ee
This theory also has a 0-form symmetry, charge conjugation, which is $\Z_2^{(0)}$ and acts on the gauge field 
\be
A_\mu \to - A_\mu \,,
\ee
and thus acts on the Wilson lines $W_a \to W_{-a}$, but also on the 1-form symmetry generators 
\be
D_2^{\alpha} \to D_2^{-\alpha} \,.
\ee

As long as the charge conjugation symmetry has no anomaly, we can gauge it. 
Gauging the charge conjugation symmetry results in the theory with gauge group 
\be
G_{\text{gauge}} = U(1) \rtimes \Z_2 = O(2) \,.
\ee
Let us determine now the symmetry generators in the gauged theory, by considering invariant topological defects. 
Clearly the operators 
\be
D_2^{\alpha=0} \quad \text{and }\quad D_2^{\alpha = \pi}
\ee
are invariant.
However for the remaining values of $\alpha$ we have to take the invariant linear combinations 
\be
D_2^{\alpha, +}= D_2^{\alpha} \oplus D_2^{-\alpha} \,,\qquad \alpha\in (0,\pi) \,.
\ee
This linear combination of topological defects is the cause of the non-invertible fusion (composition) for many constructions. 
We can compute the fusion  of these defects in the original $U(1)$ theory, and find
\be\label{O2NI}
U(1):\qquad 
\ba
D_2^{\alpha, +} \otimes D_2^{0} &= D_2^{\alpha, +} \cr 
D_2^{\alpha, +} \otimes D_2^{\pi} & = D_2^{\alpha + \pi} \oplus D_2^{-\alpha + \pi} = D_2^{\alpha+ \pi, +} \cr 
D_2^{\alpha, + }\otimes D_2^{\beta, +} &= D_2^{\alpha + \beta} \oplus D_2^{-\alpha-\beta} \oplus D_2^{\alpha-\beta} \oplus D_2^{-\alpha + \beta}  = \underline{D_2^{\alpha+ \beta, +} \oplus D_2^{\alpha - \beta, +}}\,,\quad \alpha\not= \beta\,.
\ea
\ee
On the RHS, we have again written everything in terms of the invariant combinations under the 0-form symmetry. Thus after gauging, we expect these to be (at least some part of) the fusion. The last row indicates that this is indeed a non-invertible symmetry. 
The reason why this fusion is non-invertible in the sense that the RHS is a sum of two terms, can be traced back to the fact that $D_2^{\alpha, +}$ is a linear combination of two defects. In the $U(1)$ theory this is not an irreducible (or more correctly simple) defect, as it is the sum of two irreducible ones $D_2^{\alpha}$ and $D_2^{-\alpha}$, which each have invertible fusion. 
However, in the gauged theory i.e. $O(2)$ Yang-Mills, $D_2^{\alpha, +}$ is an irreducible topological defect (neither $D_2^{\pm\alpha}$ are well-defined on their own) and the above fusion becomes the fusion of an irreducible symmetry generator. 
This idea percolates through many constructions of non-invertible symmetries, as we shall see.

The most peculiar fusion arises for defects with $\alpha=\beta$ or $\beta= \pi-\alpha$. The paper \cite{Heidenreich:2021xpr} proposed the following: 
\be\label{weirdos}\ba
D_2^{\alpha, + }\otimes D_2^{\alpha, +} & = D_2^{0} \oplus W_{\det} \oplus D_2^{2\alpha, +} \cr
D_2^{\alpha, + }\otimes D_2^{\pi - \alpha, +} & = 
D_2^{\pi} \oplus D_2^{\pi}(W_{\det}) \oplus D_2^{2\alpha -\pi, +} \,.
\ea
\ee
Here $W_{\det}$ is the Wilson line in the determinant representation (for $O(2)$ there is a representation which maps the group element to its determinant), and $D_2^\pi (W_{\det})$ means the surface operator with the line immersed (the precise location is immaterial as the defects are topological). 
This fusion has various peculiarities: 
the first one in (\ref{weirdos}) is stating that the fusion of surfaces results in a line. 
The second one is even more peculiar, in that setting $\alpha=\pi/2$ we end up with a fusion that is  dubious
\be
``\, D_2^{\pi/2, + }\otimes D_2^{\pi/2, +}  = 
D_2^{\pi} \oplus D_2^{\pi}(W_{\det}) \oplus D_2^{0}  \, " \,.
\ee
It does not preserve the quantum dimensions. 
The quantum dimension is defined as  the number of distinct local operators that exist on each defect. If this is 1, the defect is called simple. This should usually be preserved under fusion. 
Here $D_2^{\alpha, +}$ has quantum dimension 2 as it is the sum of two simple (i.e. quantum dimension 1) operators, so that the left hand side has quantum dimension 4, whereas on the right hand side we have 3. 
Clearly this begs for some clarification, at the very least. 
We will return to this example in section \ref{sec:O2}, once we have developed a better understanding of non-invertible symmetries, and acquired some more  technology to determine their fusion rules. 

This first example of constructing  non-invertible symmetries using charge conjugation gauging generalizes to many non-abelian gauge theories. We can gauge of outer automorphisms of the gauge algebra or charge conjugation and this results in non-invertible symmetries. For instance the gauge theories with disconnected gauge groups 
\be
G_{\text{gauge}} = \Pin^+ (2N) \,,\  \widetilde{SU(N)}\,,
\ee
have non-invertible symmetries for all dimensions $d\geq 3$\cite{Bhardwaj:2022yxj}.
For example for the Lie algebras $\mathfrak{so}(4N)$ the simply-connected gauge group is $\Spin(4N)$. This has an outer automorphism acting on the Dynkin diagram. Quotienting $\Spin(4N)$ by this automorphism, gives rise to $\Pin^+(4N)$. Likewise $\widetilde{SU(N)}$ is the $SU(N)$ pure gauge theory with charge conjugation gauged \cite{Bourget:2018ond, Bourget:2018ond}. 

The principle for obtaining non-invertible symmetries is always that the invariant combinations in the theory before gauging will have fusion that are already indicative of the non-invertible structure as in the last line of (\ref{O2NI}).

\subsection{Second Encounter: Verlinde Lines in 2d Rational CFTs}

Another important example of non-invertible symmetries arises in 2d, for rational CFTs, e.g. the coset models $SU(2)_k/U(1)$. A rational CFT (RCFT) is characterized by a finite number of conformal primaries $\lambda_i$ with conformal dimensions $h_i$, with fusion 
\be
\lambda_i \otimes \lambda_j = \sum_{k} N_{ij}^k \lambda_k \,.
\ee
The torus partition function of these RCFTs (for simplicity we take it to be diagonal modular invariant) 
\be
Z(\tau, \bar\tau) = \sum \chi_i(\tau) \chi_i(\bar\tau) \,.
\ee
Here $\tau$ is the modular parameter (complex structure of the torus) and $\chi_i$ the characters with conformal weight $h_i$.
There is an $S$-matrix which provides the map
\be
\chi_i (-1/\tau) = \sum_{j} S_{ij} \chi_j(\tau) \,.
\ee
The fusion coefficients can then be computed from the Verlinde formula 
\be
N_{ij}^k = \sum_{l} {S_{il} S_{jl} S_{kl}^* \over S_{0l}} \,,
\ee
where $0$ denotes the vacuum. 
Any such RCFT now has a set of topological lines which act as 0-form symmetries. They commute with the Virasoro algebra and are defined by the action on the conformal primary states $| i\rangle $ as 
\be
D_1^{(i)} |j\rangle = {S_{ij} \over S_{0j}} |j\rangle \,.
\ee
The fusion of the topological lines $D_1^{(i)}$ is then precisely given by $N_{ij}^k$ 
\be
D_1^{(i)} \otimes D_1^{(j)} = \bigoplus_k N_{ij}^k D_1^{(k)} \,.
\ee
We have seen the example of the Ising model in the introduction. 
The goal of these lectures is to now construct in a systematic way, symmetries in $d>2$ that have such  generalized fusion rules.

\subsection{Non-Invertible Symmetries in $d=2$}
\label{sec:2d}

One realm where non-invertible symmetries have been well-known and studied are 2d QFTs. In this case the topological operators are limited to lines $D_1$ and point-operators $D_0$. An excellent exposition for physicists can be found here \cite{Bhardwaj:2017xup, EGNO, Thorngren:2019iar, Thorngren:2021yso}.

\subsubsection{Topological lines in 2d}

\paragraph{Example: Group-like Symmetries.}

The simplest fusion category symmetry is an invertible 0-form symmetry given by a finite group $G^{(0)}$. This is simply a standard invertible, global symmetry. 
We will now recast this slightly, and formulate this symmetry in terms of topological lines, and their fusion, as well as junctions between such lines. In this language, a 0-form symmetry $G^{(0)}$ in 2d takes the following formulation -- as a fusion category. It is given in terms of a set of objects, morphisms and a fusion product: 
\begin{itemize}
\item Objects: the objects are topological lines, $D_1^{(g)}$, $g\in G$.  
\item Morphisms: 
A topological local operator $D_0$ that acts as a homomorphism between two topological lines $\Hom(D_1^{(g)}, D_1^{(h)})$ is called a morphism. 
\item Simple objects have the property that the only local operator is the identity one
\be
\Hom(D_1^{(g)}, D_1^{(g)}) = \id \,. 
\ee
\item Fusion: The fusion for the 0-form symmetry is 
\be\label{fusD1g}
D_1^{(g)} \otimes D_1^{(h)} = D_1^{(gh)} \,,\qquad g , h \in G \,.
\ee
We can also fuse two lines and obtain a third line using a local operator $\Hom (D_1^{(g)} \otimes D_1^{(h)} , D_1^{(gh)})$: 
\be
\begin{tikzpicture}
    \draw [thick] (0,0) -- (0,1) ; 
    \draw [thick] (-1, -1) -- (0,0) -- (1,-1) ; 
    \node[right] at (0,0) {$D_0$};
    \node[below] at (-1,-1) {$D_1^{(g)}$};
    \node[below] at (1,-1) {$D_1^{(h)}$};
    \node[above] at (0,1) {$D_1^{(gh)}$};
    \end{tikzpicture}
\ee
\end{itemize}
This structure is modeling a set of codimension 1 topological operators, which fuse according to a group multiplication. This is the fusion category of $G$-graded vector spaces (each defect is a vector space, and maps between these are the morphisms)
\be
\Vec_{G} \,.
\ee
A small modification of this allows for a co-cycle (which will in fact act as an 't Hooft anomaly for this symmetry): let 
\be
\omega: \quad G^3 \to U(1)
\ee
be a 3-cocycle satisfying
\be\label{cocycome}
\omega\left(g_1 g_2, g_3, g_4\right) \omega\left(g_1, g_2, g_3 g_4\right)
=\omega\left(g_1, g_2, g_3\right) \omega\left(g_1, g_2 g_3, g_4\right) \omega\left(g_2, g_3, g_4\right)\,.
\ee
The fusion category 
\be
\Vec_G^{\omega} 
\ee
is then defined the same way as $\Vec_G$ with the difference that there is now a non-trivial associator, i.e. 
\be\label{assoc2d}
\begin{tikzpicture}[x=0.7cm,y=0.7cm] 
\begin{scope}[shift= {(0,0)}]
    \draw [thick] (0,0) -- (3,3) ;
    \draw [thick] (1,1) -- (2,0) ;
    \draw [thick] (2,2) -- (4,0) ;
    \node[below] at (0,0) {$D_1^{(g_1)}$};
    \node[below] at (2,0) {$D_1^{(g_2)}$};
    \node[below] at (4,0) {$D_1^{(g_3)}$};
    \node[above] at (3,3) {$D_1^{(g_1 g_2 g_3)}$};
\end{scope}
\begin{scope}[shift= {(10,0)}]
\node at (-3,2) {$= \omega(g_1, g_2, g_3)$} ;
    \draw [thick] (3,0) -- (0,3) ;
    \draw [thick] (-1,0) -- (1,2) ;
    \draw [thick] (1,0) -- (2,1) ;
    \node[below] at (-1,0) {$D_1^{(g_1)}$};
    \node[below] at (1,0) {$D_1^{(g_2)}$};
    \node[below] at (3,0) {$D_1^{(g_3)}$};
    \node[above] at (0,3) {$D_1^{(g_1 g_2 g_3)}$};
\end{scope}
    \end{tikzpicture}
\ee
Requiring that the fusion of four topological lines is consistent (a diagram called the pentagon identity) imposes (\ref{cocycome}), and thus  $\delta \omega=0$. 
Such co-cycles are thus indeed valued in $H^3 (G, U(1))$. 
As we will see later, they correspond to 't Hooft anomalies $\mathcal{A}_3 = \int_{M_3} \omega$, where $\partial M_3 = M_2$. 
These can obstruct gauging of certain (sub)groups of the 0-form symmetry group. 
Note that for a general fusion category symmetry, the associativity is captured by the F-symbols. We confine ourselves here to the special case of group-like symmetries to start with.

\subsubsection{Gauging 0-form Symmetries}

Gauging (non-anomalous) symmetries is a way of generating new symmetries. Even in higher dimensions, gauging invertible symmetries is an important way to construct  non-invertible symmetries. We assume that the symmetries that we gauge do not have any 't Hooft anomaly. 

The first instance of this can be seen in 2d. Let us start with a theory $\cT$ that has symmetry $\cS= \Vec_G$, i.e. a standard, invertible, 0-form symmetry $G$.
Gauging $G$ means introducing a dynamical $G$ gauge field. What are the topological operators after gauging?
\begin{itemize}
\item For abelian $G$ we can describe this as summing over flat gauge configurations $B_1$. We denote the dynamical $G$ gauge field by $b_1$ and this can be used to form Wilson lines in representations $\bm{R}$ of $G$
\be
{D_1^{(\bm{R})} = \Tr_{\bm{R}} e^{\int_{M_1}  b_1}} \,.
\ee
These are the topological lines generating the dual symmetry $\widehat{G}$.
For non-abelian $G$ we can still identify the dual symmetries as Wilson lines for $G$, i.e. 1d cycles which carry labels of representations $\bm{R}$ of $G$.
\item These Wilson lines fuse according to the representations  of $G$, {$\Rep (G)$}: 
\be
 D_1^{(\bm{R}_1)} \otimes D_1^{(\bm{R}_2)} = \bigoplus_{\bm{R}_3} N_{\bm{R}_3}^{{\bm{R}_1} {\bm{R}_2}}D_1^{(\bm{R}_3)} \,.
\ee
Here the coefficients $N_{\bm{R}_3}^{{\bm{R}_1} {\bm{R}_2}}$ are the Clebsch-Gordan coefficients in the decomposition of the tensor product of representations 
\be
\bm{R}_1 \otimes \bm{R}_2 = \bigoplus N_{\bm{R}_3}^{{\bm{R}_1} {\bm{R}_2}} \bm{R}_3 \,.
\ee
\item The identity element is the trivial representation. 
\item Intertwiners of $G$-representations (i.e. linear maps acting between representation $\alpha_{1,2}: \bm{R}_1 \to \bm{R}_2$ which commute with the action of the group) are junctions $D_0$ (i.e. topological local operators) between topological lines labeled by representations. 
\end{itemize}
Topological lines with the above properties form the fusion category $\Rep(G)$. So starting with $\cS= \Vec_G$, gauging $G$ results in the theory $\cT/G$ with symmetry $\cS' =\Rep(G)$. Both of these symmetries are 0-form symmetries, but the latter is non-invertible, when $G$ non-abelian. 

\paragraph{Example: Abelian $G$.}
For a finite abelian groups we argued in section \ref{sec:Inv} that the dual symmetry is again isomorphic to the original symmetry, i.e. an abelian finite group. Indeed, for $G$ finite abelian, 
$\Rep(G)$ is encoded in the characters. Lets assume for simplicity there is one cyclic factor $G= \Z_N$, with  $g=e^{2\pi i/N}$ its generator.
Then all irreducible representations are complex 1d, and $g$ acts on $v_k$ the basis vector of the 1d representation as 
\be
g:  \qquad v_k \to e^{2 \pi i k/N} v_k \,,\qquad k=0, \cdots, N-1 \,.
\ee
The representation is determined in terms of its character 
\be
\chi_{V_k}(g) = \Tr_{V_k} (g) = e^{2 \pi i k/N} \in U(1)\,.
\ee
More abstractly, we can say: the representation is encoded in the characters, i.e. group homomorphisms to $U(1)$, which is the Pontryagin dual group $\widehat{G}$ 
\be
\chi: \quad G \to U(1) \,,\quad \chi (g) \chi (h) = \chi (gh)\,.
\ee
Then $\chi (g)^N= \chi (g^N) = \chi (\id) = 1$. So we find  
\be
{\Rep(\mathbb{Z}_N)    \equiv \text{Hom} (\mathbb{Z}_N, U(1)) \cong \mathbb{Z}_N} \,.
\ee
So the Pontryagin dual group of a cyclic group is isomorphic to the group itself. And the dual symmetry is 
\be
\Rep(\mathbb{Z}_N) \cong \Vec (\mathbb{Z}_N) \,.
\ee
Consistent with the arguments in section \ref{sec:Inv}. 

\paragraph{Example: Non-Abelian $G$.} 
The situation is quite different for non-abelian finite group 0-form symmetries in 2d, where the representations can be higher-dimensional and their tensor product decomposition has generically multiplet factors.
The symmetry $\cS= \Rep (G)$ for $G$ non-abelian is non-invertible. 

Let us consider the simplest non-abelian finite group $G= S_3 = \mathbb{Z}_3 \rtimes \mathbb{Z}_2$ generated by 
\be
S_3 =\left\langle \id, a, b \, |\ a^3 =\id, \ b^2 = \id, \ bab=a^2 \right\rangle  \,.
\ee
The irreducible representations are the trivial representation $\bm{1}_+$, the sign representation $\bm{1}_-$ (which maps every permutation to $(-1)^{|\text{transpositions}|}$ i.e. the sign of the permutation) and the 2d representation $V$ (which can be constructed by letting $S_3$ permute the basis vectors $e_i$ of $\mathbb{R}^3$, subject to  the constraint that they lie on the plane $(x_1, x_2, -x_1-x_2)$). The simple topological lines are 
\be 
\Rep (S_3)^{\text{1d top-ops}}= \left\{ D_1^{(\bm{R})}:\quad \bm{R}=\bm{1}_+, \ \bm{1}_-, \  V \right\} \,.
\ee
The fusion $N_{k}^{ij}$ of the topological lines $D_{1}^{\bm{R}_i}$ is  
\be\ba
\bm{1}_{+} \otimes \bm{1}_{\pm} &= \bm{1}_{\pm}\cr 
\bm{1}_{-} \otimes \bm{1}_- &= \bm{1}_+ \cr 
V \otimes \bm{1}_{\pm} & = V \cr 
V \otimes V & = \bm{1}_+ \oplus \bm{1}_- \oplus V \,.
\ea
\ee
The last line clearly is a  non-invertible fusion. 
This holds more generally for any non-abelian finite $G$, the fusion category formed by the representations, $\Rep(G)$, gives rise to  a non-invertible 0-form symmetry.

\subsubsection{Gauging Take 2: Non-Invertibles from Stacking TQFTs}

This idea of generating non-invertible symmetries by gauging extends  to higher-dimensions. To prepare for this, it is however useful to think of the gauging in 2d from a slightly different, but equivalent,  perspective \cite{Bhardwaj:2022lsg}.
Consider a 2d theory $\mathcal{T}$ with 0-form finite group symmetry $G$, i.e. $\Vec_G$.

\begin{figure}
\centering
\begin{tikzpicture}
\begin{scope}[shift={(0,0)}]
\draw [fill=blue,opacity=0.1] 
(0,0) -- (4,0) -- (4,4) -- (0,4) -- (0,0);
\draw [thick] (0,0) -- (4,0) -- (4,4) -- (0,4) -- (0,0);
\draw [thick,red] (2,0) -- (2,4) ;
\node[red] at (2, -0.5) {1d $G$-$\text{TQFT}$}; 
\draw [thick, ->] (5,1.7) -- (7,1.7);
 \node[blue] at (1,2) {$\mathcal{T}$};
\node[black] at (6,2) {$\text{gauge }G$};
\end{scope}
\begin{scope}[shift={(8,0)}]
\draw [fill=blue,opacity=0.1] 
(0,0) -- (4,0) -- (4,4) -- (0,4) -- (0,0);
\draw [thick] (0,0) -- (4,0) -- (4,4) -- (0,4) -- (0,0);
\draw [thick,red] (2,0) -- (2,4) ;
\node[red] at (2, -0.5) {Topological line defect $D_1$}; 
 \node[blue] at (1,2) {$\mathcal{T}/G$};
\end{scope}
\end{tikzpicture}
\caption{2d theory $\cT$ with $G$ 0-form symmetry. We stack a 1d $G$-TQFT before gauging the diagonal $G$ symmetry. The 1d TQFT becomes a topological defect $D_1$ in the gauged theory $\cT/G$. This is the simplest instance of a theta-defect, which generically is non-invertible.\label{fig:Stack1d}}
\end{figure}
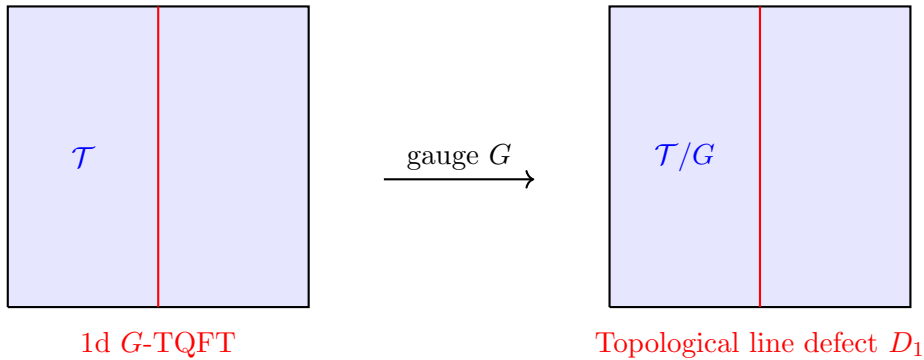

The most general gauging of $G$ can be thought of as follows:
we stack a 1d TQFT with $G$-symmetry $\text{TQFT}_1^{G}$, 
 and gauge the diagonal $G$. The TQFT becomes a 1d topological defect in the gauged theory, which will be part of the dual symmetry. 
So our setup is as follows:
\be
\mathcal{T} \times \text{TQFT}_1^{G}  \quad \stackrel{\text{gauge }G }{\longrightarrow} \quad (\mathcal{T} \times \text{TQFT}_1^{G})/G\,.
\ee
with the 1d TQFT becomes a topological defect in the gauged theory $\mathcal{T}/G$ as shown in figure \ref{fig:Stack1d}. 

What are the properties of these defects, in particular, how are 1d $G$-symmetric TQFTs characterized? These are specified in terms of the following data:  
\begin{itemize}
\item The number of vacua $\bm{v}_i$, $i= 1, \cdots N$. This corresponds to a finite dimensional vector space. 
\item An action of $G$ on these vacua, i.e. this makes $V$ into a $G$-representation.
\end{itemize}
These fuse precisely according to $\Rep(G)$, and in the gauged theory, become the generators of the dual symmetry $\Rep(G)$. 
In the theory after gauging these TQFTs become 1d topological defects, i.e. symmetry generators of a 0-form symmetry in 2d. 
This perspective will be hugely important for the higher-dimensional case of gauging group-like symmetries. The defects obtained in this way by stacking $G$-TQFTs will be called {\bf theta-defects} in the following \cite{Bhardwaj:2022lsg}.

\subsubsection{Gauging Take 3: Making Lines Invisible}
\label{sec:bimod2d}

Although by now we have a fairly good understanding of the gauging of finite 0-form symmetries in 2d, there is another perspective discussed in 2d in \cite{Bhardwaj:2017xup}, which generalizes to higher dimensions:  physically we study the topological lines in the gauged theory, describing the latter as obtained by ``summing over a mesh of topological lines $D_1^{(g)}$"\footnote{Mathematically it goes under the name of bimodules of algebras in the fusion category.}.
This also has a higher-dimensional avatar \cite{Bhardwaj:2022lsg, Bartsch:2022mpm,Bhardwaj:2022maz, Bartsch:2022ytj}, but may be less physically intuitive than the theta defects -- although these descriptions are equivalent \cite{Bhardwaj:2022maz, Bhardwaj:2022kot}.

Because of its extendability to higher-dimensions (in terms of bimodules of algebras in higher-fusion categories), some readers may find it useful to see this in lower dimensions first. We will not require this formulation in the following though (except perhaps for the description of condensation defects in section \ref{sec:Cond}).

\paragraph{Invisibility of lines.}
Consider, again, a 2d theory $\cT$ with a 0-form symmetry group $G^{(0)}$, generated by topological lines $D_1^{(g)}$, $g\in G$. We would like to gauge $G^{(0)}$ and determine what topological lines the resulting gauged theory $\cT/G$ has. 

First of all, let us recall that gauging of a finite group can be formulated in terms of the partition function as in (\ref{gaugeZ}) as 
summing over all gauge configurations $B_1\in H^1 (M_2, G)$. 

Equivalently, using Poincar\'e duality, we can map this to a sum over insertions of topological lines $D_1^{(g)}$ for some $g\in G$ (i.e. 0-form symmetry generators) on a mesh: starting with a triangulation of the 2d spacetime $M_2$ consider the dual graph, which produces a trivalent graph on $M_2$. Inserting topological lines of the original theory on this mesh and summing over all configurations corresponds to gauging. 

There is one additional constraint: we cannot simply insert arbitrary lines $D_1^{(g)}$. One way to see this can be problematic is that the partition function of the gauged theory should be invariant under changes of triangulations. A set of lines that ensures that this independence on the triangulation is guaranteed is usually called an algebra object $A = \bigoplus_{g} m_g D_1^{(g)}$ in the fusion category. Here will gauge the full symmetry $G$, for which the algebra is 
\be
A= \bigoplus_g D_1^{(g)}\,.
\ee
The gauged theory $\cT/G$ on $M_2$ has then the following description, as the original theory $\cT$ with  a sum over insertions of lines in $A$ on a trivalent mesh, that is dual to the triangulation of $M_2$: 
\be\label{TheMesh}
\begin{tikzpicture}
\begin{scope}[shift={(11.5,9)}]
\draw [thick] (0,3) -- (4,3) -- (4,6) -- (0,6) -- (0,3);
\node[blue] at (3,4.7) {$A$} ;
\end{scope}
\draw [blue,thick](12,12) -- (12,12.5) -- (13,13.5) -- (12,13.5) -- (11.5,13) (12.5,14.5) -- (12,13.5);
\draw [blue,thick](12,12.5) -- (13,12.5) -- (13.5,12);
\draw [blue,thick](14,13.5) -- (13,12.5);
\draw [blue,thick](12.5,14.5) -- (13,15);
\draw [blue,thick](15,14) -- (14.5,15);
\draw [blue,thick](13,13.5) -- (13.5,14.5);
\draw [blue,thick](13.5,14.5) -- (14,13.5);
\draw [blue,thick](12.5,14.5) -- (11.5,14.5);
\draw [blue,thick](14,12) -- (14,13.5);
\draw [blue,thick](14.5,12) -- (14.5,13) -- (15.5,12.5) (14.5,13) -- (15,14);
\draw [blue,thick](15,14) -- (13.5,14.5);
\end{tikzpicture}
\ee

Again, we ask the question: what is the symmetry of the gauged theory $\mathcal{T}/G$? 
We only allow the gauged theory to have lines, which are gauge-invariant,  i.e. configurations of lines that can moved freely through the mesh. Lets denote the topological lines in $\cT/G$ by $L_1$. Then consistency with the mesh of lines we have to require the following conditions: 
 \begin{itemize}
\item $D_1^{(g)}$ lines (or more precisely the lines in the algebra) can end from the left on $L_1$ subject to:
\be\label{innout}
\begin{tikzpicture}
\draw [thick, red] (0,0) -- (0,3);
\draw [thick,blue] (-1,0) -- (0,1) ;
\draw [thick,blue] (-2,0) -- (0,2) ;
\draw [black,fill=black] (0,1) ellipse (0.05 and 0.05);
\node [right] at (0,1) {$\cO_L$};
\node [right] at (0,2) {$\cO_L$};
\draw [black,fill=black] (0,2) ellipse (0.05 and 0.05);
\node[blue, below] at (-2, 0) {{$A$}};
\node[blue, below] at (-1,0) {{$A$}};
\node[red, below] at (0, 0) {$L_1$};
\node at (2,1) {=};
\begin{scope}[shift={(5,0)}]
\draw [thick, red] (0,0) -- (0,3);
\draw [thick,blue] (-1,0) -- (-1,1) ;
\draw [thick,blue] (-2,0) -- (0,2) ;
\node [right] at (0,2) {$\cO_L$};
\draw [black,fill=black] (0,2) ellipse (0.05 and 0.05);
\node[blue, below] at (-2, 0) {{$A$}};
\node[blue, below] at (-1,0) {{$A$}};
\node[red, below] at (0, 0) {$L_1$};
\end{scope}
\end{tikzpicture}
\ee
This is consistency of the fusion in the mesh with fusion of operators on the line.  
\item Similarly, we can couple to $G$-backgrounds from the right and obtain operators $\cO_R$, satifying the mirrored conditions. 
\item Compatibility of left and right coupling: $A$-lines can end from the left and right and these two operations need to be compatible:
\be\label{bimodu}
\begin{tikzpicture}
\draw [thick, red] (0,0) -- (0,3);
\draw [thick,blue] (1,0) -- (0,1) ;
\draw [thick,blue] (-2,0) -- (0,2) ;
\draw [black,fill=black] (0,1) ellipse (0.05 and 0.05);
\draw [black,fill=black] (0,2) ellipse (0.05 and 0.05);
\node [left] at (0,1) {$\cO_R$};
\node [right] at (0,2) {$\cO_L$};
\node[blue, below] at (-2, 0) {{$A$}};
\node[blue, below] at (1,0) {{$A$}};
\node[red, below] at (0, 0) {$L_1$};
\node at (2,1) {=};
\begin{scope}[shift={(6,0)}]
\draw [thick, red] (0,0) -- (0,3);
\draw [thick,blue] (-1,0) -- (0,1) ;
\draw [thick,blue] (2,0) -- (0,2) ;
\draw [black,fill=black] (0,1) ellipse (0.05 and 0.05);
\draw [black,fill=black] (0,2) ellipse (0.05 and 0.05);
\node [left] at (0,2) {$\cO_R$};
\node [right] at (0,1) {$\cO_L$};
\node[blue, below] at (2, 0) {{$A$}};
\node[blue, below] at (-1,0) {{$A$}};
\node[red, below] at (0, 0) {$L_1$};
\end{scope}
\end{tikzpicture}
\ee
\item Compatibility of fusions: Consider two lines $L_1$ and $L_1'$ in the gauged theory. Then fusing an $A$-line with one before fusing with the other line, should be compatible in the following sense: 
\be
\begin{tikzpicture}
\draw [thick, red] (0,2) -- (0,4);
\draw [thick, red] (-2,0) -- (0,2) -- (2,0);
\draw [thick,blue] (0,0) -- (-1,1) ;
\draw [black,fill=black] (-1,1) ellipse (0.05 and 0.05);
\draw [black,fill=black] (0,2) ellipse (0.05 and 0.05);
\node [left] at (0,1) {$\cO_R$};
\node[red, below] at (-2, 0) {{$L_1$}};
\node[blue, below] at (0,0) {{$A$}};
\node[red, below] at (2, 0) {$L_1'$};
\node at (3,2) {=};
\begin{scope}[shift={(6,0)}]
\draw [thick, red] (0,2) -- (0,4);
\draw [thick, red] (-2,0) -- (0,2) -- (2,0);
\draw [thick,blue] (0,0) -- (1,1) ;
\draw [black,fill=black] (1,1) ellipse (0.05 and 0.05);
\draw [black,fill=black] (0,2) ellipse (0.05 and 0.05);
\node [right] at (1,1) {$\cO_L$};
\node[red, below] at (-2, 0) {{$L_1$}};
\node[blue, below] at (0,0) {{$A$}};
\node[red, below] at (2, 0) {$L_1'$};
\end{scope}
\end{tikzpicture}
\ee
\end{itemize}
These compatibility conditions tell us what lines $L_1$ are allowed in the gauged theory. 

Mathematically, in order to gauge we need to pick an algebra $A$ in the symmetry category $\Vec_G$, and solve the above  {\it bimodule conditions}. For $A= \oplus_{g\in G} D_1^{(g)}$ in $\Vec_G$ these bimodules will be precisely the representations of $G$
\be
\text{Bimod}_{\Vec_G} (A) = \Rep(G) = \text{representations of $G$}\,,
\ee
confirming what we already showed with alternative methods. 
Lets see the representations emerge from the bimodule construction. 

\paragraph{Bimodules for $\Vec_G$.}
We now solve the bimodule conditions for $\Vec_G$.
The algebra is $A= \bigoplus_{g\in G} D_1^{(g)}$, and we would like to solve for bimodules $L_1$. To simplify notation, recall that $\Vec_G$ is simply $G$-graded vector spaces. Let us make an ansatz for $L_1$ in terms of a general linear combination of $g$-graded vector spaces: 
\be
L_1 \ \longleftrightarrow\  V = \bigoplus n_g D_1^{(g)}  = \bigoplus V_g \,,
\ee
where $V_g$ are $n_g$ dimensional vector spaces. This introduces a $g$-grading, which can be used to write the left and right actions 
\be
\cO_L: \quad A \otimes V  \to V \,,\qquad 
\cO_R:  \quad  V \otimes A \to V 
\ee
in their decomposition into $g$-graded components
\be
\cO_L^{g, h}: \quad g \otimes V_h \to V_{gh} \,,\qquad 
\cO_R^{h,g}: \quad V_h \otimes g \to V_{hg} \,.
\ee
Pictorially we can represent this as  
\be
\begin{tikzpicture}
\begin{scope}[shift={(0,0)}]
\draw [thick, red] (0,0) -- (0,2);
\draw [thick,blue] (-1,0) -- (0,1) ;
\draw [black,fill=black] (0,1) ellipse (0.05 and 0.05);
\node [right] at (0,1) {$\cO_L^{g,h}$};
\node[blue, below] at (-1,0) {{$g$}};
\node[red, below] at (0, 0) {$V_h$};
\node[red, above] at (0, 2) {$V_{gh}$};
\end{scope}
\begin{scope}[shift={(4,0)}]
\draw [thick, red] (0,0) -- (0,2);
\draw [thick,blue] (1,0) -- (0,1) ;
\draw [black,fill=black] (0,1) ellipse (0.05 and 0.05);
\node [left] at (0,1) {$\cO_R^{g,h}$};
\node[blue, below] at (1,0) {{$g$}};
\node[red, below] at (0, 0) {$V_h$};
\node[red, above] at (0, 2) {$V_{hg}$};
\end{scope}
\end{tikzpicture}
\ee
We can now evaluate the conditions on being a left and a right module, as well as the compatibility condition. 
In particular (\ref{innout}) implies 
\be
\cO_L^{h k, g} = \cO_L^{h, k g} \cO_L^{ k, g}\,,
\ee
and similarly for $\cO_R$. Note that $\cO_L^{1, g} = \id$  and likewise $\cO_R^{g,1} = \id $, where $1$ is the identity in the group. So picking $h k= 1$ this means that 
\be
\cO_L^{h, h^{-1} g} = (\cO_L^{h^{-1}, g})^{-1}\,,
\ee
and since these are invertible, it means the dimensiosn of the vector spaces $V_g$ have to be all the same: $n_g=n$ for all $g$. 

The bimodule condition (\ref{bimodu}) can be written as 
\be\label{leri}
\cO_L^{h, g k} \cO_R^{g, k}=\cO_R^{h g, k} \cO_L^{h, g}
 \,.\ee
 Putting these points together we can then write every single element in terms of elements of the form $\cO_L^{g, 1}$ and $\cO_R^{1,g}$, 
 \be\label{simpli}
 \ba
\cO_L^{g, h} = \cO_{L}^{gh, 1} (\cO_L^{h, 1})^{-1}\cr
\cO_R^{g,h} = \cO_{R}^{1,gh} (\cO_R^{1,g})^{-1} \
\,. \ea\ee
The bimodule compatibility of left and right actions (\ref{leri}) can be now rewritten in terms of the expressions (\ref{simpli}) and then recombine the terms into 
$\rho(g) = (\cO_R^{1,g})^{-1} \cO_L^{g,1}$. Then the condition for (\ref{leri}) is satisfied if $\rho$ is a representation, i.e. $\rho (g) \rho (h) = \rho(gh)$,
i.e. these form  $G$-representation, and the bimodules are precisely $\Rep(G)$. 

From this perspective it is also clear now, that the non-trivial associator $\omega$ in (\ref{assoc2d}) is an obstruction to gauging. It means in particular that changing triangulations introduces a phase $\omega$, which corresponds to an F-move on the dual triangulation, and thus obstructs gauging. The anomaly can be non-trivial also for a sub-group only and examples of that are discussed in \cite{Bhardwaj:2017xup}.

All three perspectives on gauging in 2d have certain advantages. However the approach using theta-defects -- i.e. stacking with $G$-TQFTs before gauging  -- seems to have wider scope (and generalizes to related constructions, so-called twisted theta defects), and also is perhaps physically the most well-motivated. So we will focus on this perspective in the following.


\subsection{Theta-Defects: Universal Non-Invertible Symmetries}

Lets consider gauging of finite global symmetries in higher dimensions. We have seen in 2d that gauging a non-abelian 0-form symmetry $G$ results in a non-invertible symmetry after gauging $\Rep(G)$. 
The same reasonings as in 2d can be applied in higher dimensions in order to gauge finite, non-anomalous symmetries. Contrary to 2d however, we will see that even gauging an {\it abelian} 0-form symmetry in e.g. 3d, will give rise to non-invertible symmetries! 
Again, there are several approaches to gauging: theta-defects and bi-modules, and we will focus here on the former.  

\paragraph{Theta-Defects.}

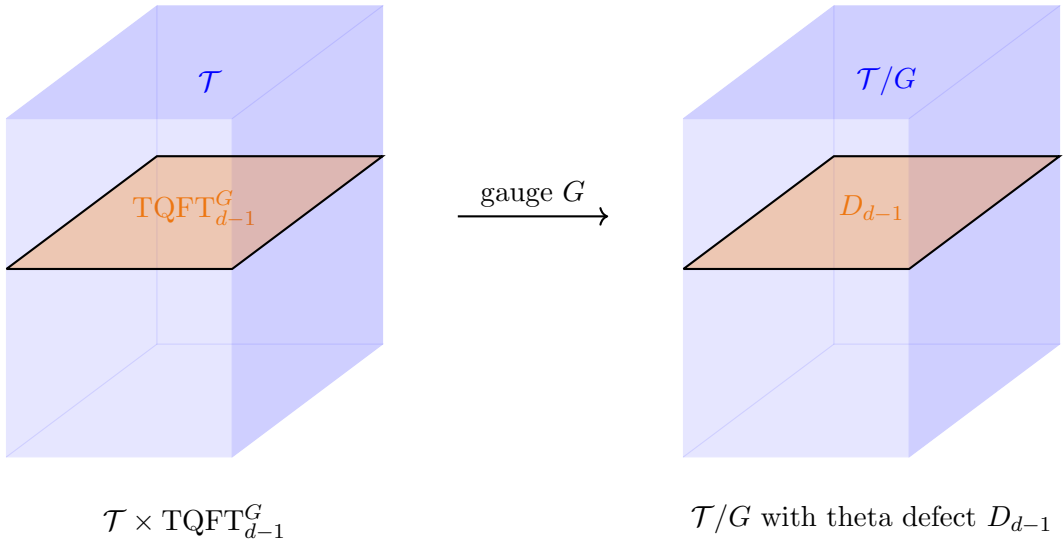
\begin{figure}
\centering
{\begin{tikzpicture}
\begin{scope}[shift={(0,0)}]
 \node[blue] at (0.7,0.5) {$\mathcal{T}$};
 \node[orange] at (0.5,-1.2) {$\text{TQFT}^G_{d-1}$};
\draw [blue, fill=blue,opacity=0.1] (0,1.5) -- (0,-3) -- (3,-3) -- (3,1.5) -- (0,1.5);
\draw [blue, fill=blue,opacity=0.1] (-2,0) -- (0,1.5) -- (0,-3) -- (-2,-4.5) -- (-2,0);
  \draw [blue, fill=blue,opacity=0.1] (-2,0) -- (1,0) -- (3,1.5) -- (0,1.5) -- (-2,0);
      \draw [blue, fill=blue,opacity=0.1] (-2,-4.5) -- (1,-4.5) -- (3,-3) -- (0,-3) -- (-2,-4.5);
      \draw [blue, fill=blue,opacity=0.1] (1,0) -- (3,1.5) -- (3,-3) -- (1,-4.5) -- (1,0);
    \draw [orange, fill=orange,opacity=0.3] (-2,-2) -- (1,-2) -- (3,-0.5) -- (0,-0.5) -- (-2,-2); 
    \draw [black, thick] (-2,-2) -- (1,-2) -- (3,-0.5) -- (0,-0.5) -- (-2,-2);
    \node[below] at (0.5, -5) {$\cT \times \text{TQFT}^G_{d-1}$};
 \end{scope}
\begin{scope}[shift={(-1,-3)}]
\draw [thick, ->] (5,1.7) -- (7,1.7);
\node[black] at (6,2) {$\text{gauge }G$};
\end{scope}
\begin{scope}[shift={(9,0)}]
 \node[blue] at (0.7,0.5) {$\mathcal{T}/G$};
 \node[orange] at (0.5,-1.2) {$D_{d-1}$};
\draw [blue, fill=blue,opacity=0.1] (0,1.5) -- (0,-3) -- (3,-3) -- (3,1.5) -- (0,1.5);
\draw [blue, fill=blue,opacity=0.1] (-2,0) -- (0,1.5) -- (0,-3) -- (-2,-4.5) -- (-2,0);
  \draw [blue, fill=blue,opacity=0.1] (-2,0) -- (1,0) -- (3,1.5) -- (0,1.5) -- (-2,0);
      \draw [blue, fill=blue,opacity=0.1] (-2,-4.5) -- (1,-4.5) -- (3,-3) -- (0,-3) -- (-2,-4.5);
      \draw [blue, fill=blue,opacity=0.1] (1,0) -- (3,1.5) -- (3,-3) -- (1,-4.5) -- (1,0);
    \draw [orange, fill=orange,opacity=0.3] (-2,-2) -- (1,-2) -- (3,-0.5) -- (0,-0.5) -- (-2,-2); 
    \draw [black, thick] (-2,-2) -- (1,-2) -- (3,-0.5) -- (0,-0.5) -- (-2,-2);
       \node[below] at (0.5, -5) {$\cT/G$ with theta defect $D_{d-1}$};
 \end{scope}
\end{tikzpicture}}
\caption{We start with the theory $\cT$ with $G$ 0-form symmetry, then take a product (``stack") the $G$-TQFT on top  and gauge $G$. In the gauged theory, the TQFT becomes a topological defect $D_{d-1}$. This is the theta-defect.  \label{fig:theta}}
\end{figure}

Consider a theory $\cT$ with a non-anomalous global $0$-form symmetry $G^{(0)}$. 
Before gauging $G^{(0)}$ we have the freedom to stack, i.e. take a product, with a 
$(d-1)$-dimensional $G$-symmetric TQFT. This is depicted in figure \ref{fig:theta}. 
Upon gauging the diagonal $G$-symmetry
\be
 \begin{tikzcd}
\mathcal{T} \times \text{TQFT}_{d-1}^{G^{(0)}}  \quad    \rar["\text{gauge }G^{(0)}"]&  \quad 
 (\mathcal{T} \times \text{TQFT}_{d-1}^{G^{(0)}})/G^{(0)}   \end{tikzcd}
 \,,
\ee
the TQFT becomes part of the theory as a topological defect, i.e. a symmetry generator. 
Fusion of these TQFTs is generally non-invertible (even for abelian $G$). This construction is universal, in the sense that it does not depend on the specifics of the theory $\cT$ (excpet for the existence of a non-anomalous $G^{(0)}$), and the resulting defects are called {\bf theta defects}. 
This construction has extensions to gauging of non-anomalous $G^{(p)}$ $p$-form symmetries and associated SPT-phases, which can be found in \cite{Bhardwaj:2022lsg}.

To carry out this construction requires knowing all $G$-symmetric TQFTs.

\paragraph{SPT-Phases.}
One class of $G$-TQFTs are called $G$-symmetry protected topological (SPT) phases. 
A $G$-SPT is a  topological theory, i.e. gapped,  and it has a global symmetry $G$. It has a unique vacuum (on a closed manifold). SPT phases of cohomology type are classified by $H^d(BG, U(1))$. For a classification of the topological actions for 0-, 1-form and 2-groups see \cite{Kapustin:2013uxa} and more general results also in cond-mat that discuss the beyond group-cohomology cases of SPTs \cite{Vishwanath:2012tq, Gaiotto:2017zba}. In addition to SPTs that preserve the full symmetry $G$, there can also be TQFTs that have spontaneous symmetry breaking from $G$ to a subgroup $H$, and these in turn can be dressed by an SPT $H^d(H, U(1))$. We will refer to this combination of SPT and SPT with spontaneous symmetry breaking as SPT+SSB type TQFTs. 
These are present in any dimension\footnote{There can be theory and dimension specific TQFTs that preserve $G$, e.g. 3d TQFTs go beyond the above SPT+SSB and can have non-trivial so-called topological order, which are classified by modular tensor categories.}.

\paragraph{Theta-angles.} 
Theta-defects are in some sense a natural generalization of theta angles, which can be thought of as obtained via stacking with a spacetime-filling TQFT. This results in a  defect generating a $(-1)$-form symmetry (i.e. a parameter). 
Consider a 4d $U(1)$ gauge theory, which can have a theta-angle that is a  topological term added to the Yang-Mills term
\be
 \theta\int  F\wedge F \,.
\ee
This $U(1)$ gauge theory can be thought of as arising in the following -- slightly unusual -- fashion:
consider a 4d trivial theory with a trivial $U(1)$ {\it global} symmetry, and background field $A$. We also allow for a $U(1)$-symmetry protected topological phase 
\be
\mathcal{L}_T = \mathcal{L}_{\text{trivial}} + \text{SPT} \,,\qquad \text{SPT}=  \theta \int F\wedge F\,.
\ee
Gauging the $U(1)$, we obtain a dynamical gauge field with Yang-Mills term and the SPT-phase has becomes the theta-angle 
\be
\mathcal{L}_{T/U(1)} = {1\over g^2} \int F \wedge \star F + \theta\int  F\wedge F \,.
\ee
This can be generalized to any theory with $U(1)$ global symmetry that can be gauged: 
{Stacking with $U(1)$-SPT and gauging adds a $\theta$-angle.} Theta-defects generalize this principle to $G$-SPT phases, where $G$ can be now also finite (non-)abelian, or a higher-form/higher-group symmetry. 



\subsection{Theta-Defects in 3d}
\label{sec:Theta3d}

To start with we consider a 3d QFT $\cT$, with non-anomalous 0-form symmetry $G^{(0)}$ (finite, but not necessarily abelian). We would like to gauge this symmetry and determine what the dual symmetry is. 
First we need to recall what the symmetries of the original theory $\cT$ are: 
\begin{itemize}
\item 0-form symmetry generated by topological surfaces $D_2^{(g)}$ with group-like fusion $D_2^{(g)}\otimes D_2^{(h)} = D_2^{(gh)}$.
\item If the theory only has a 0-form symmetry, there are no non-trivial topological lines. However, it is useful to still include the  trivial (identity) lines on each topological surface:  $D_1^{(g)}$ on the surfaces $D_2^{(g)}$. This may seem at this point a redundancy, but we will see that it is useful to keep track of the topological defects in all dimensions.
\item Topological local operators $D_0^{(g)}$ on each line (again these are trivial in the present case).
\item  In principle there could be interfaces between two topological surfaces and lines, but they are trivial in this case. 
\end{itemize}
This structure is usually referred to (similar to $\Vec_G$ in 2d), as
\be
\TwoVec_G \,.
\ee
At this point this is simply a concise labeling of the above structure: an invertible 0-form symmetry in 3d (or in $d>3$ dimension, the symmetry generated by topological surfaces, i.e. $p= d-3$ form symmetries). The $2$ indicates that this is a fusion 2-category, and we have 3 layers: topological surfaces, lines and points (so fusion categories are 1-categories, with only two layers: lines and points).

\subsubsection{Non-Invertibles from Stacking 2d $G$-TQFTs}
\label{sec:Rep}

We can now gauge the 0-form symmetry and construct the associated theta-defects. For this we should be stacking with 2d TQFTs with $G^{(0)}$ 0-form symmetry. Let us assume for simplicity that $G^{(0)}$ is {\bf abelian} (the non-abelian case is discussed in \cite{Bhardwaj:2022lsg}, and the non-abelian structure changes mainly the morphisms). Such 2d TQFTs are of the SPT+SSB type: 
\begin{enumerate}
\item Symmetry preserving TQFTs:\\ 
these preserve 
$G^{(0)}$ completely, and have a unique vacuum. Such TQFTs are  the aforementioned $G^{(0)}$-SPT phases.

This can be understood as 2d TQFTs, which have only trivial lines, with trivial operator junctions, that have to satisfy a cocyle condition (as in our 2d analysis earlier) and are classified by  
\be
\alpha_G \in H^2 \left(G^{(0)}, U(1)\right) \,.
\ee
A useful reference that computes these cohomology groups for many $G^{(0)}$ is \cite{Chen:2011pg}. The simplest example where this is non-trivial is $G^{(0)}= \Z_2 \times \Z_2$, for which there are $\Z_2$ many SPT phases. 
\item Spontaneous symmetry breaking: \\
The second option is that the 0-form symmetry is broken spontaneously to a subgroup $H$ of $G^{(0)}$: $H<G^{(0)}$. These TQFTs have a vacuum degeneracy. 
They are characterized by the unbroken symmetry $H$, which is a subgroup of $G^{(0)}$, and an SPT phase for this group $H$, which takes values in $\alpha_H\in H^2 (H, U(1))$. The vacuum degeneracy is $|G/H|$, and all vacua carry the same SPT phase.
\end{enumerate}
After gauging $G$, these TQFTs will be identified with topological surface defects, for which we would like to define a fusion product. This descends from a fusion (composition) on the TQFTs. 
For each TQFT with $G^{(0)}$-symmetry, we have two pieces of data: the unbroken subgroup $H$ and the SPT $\alpha_H$, i.e. we can label each TQFT by 
\be
\bm{T}_{H, \alpha_H}  \,.
\ee 
To define $\bm{T}_{H, \alpha_H} \otimes \bm{T}_{K, \alpha_K}$ note that the TQFTs have vacua $v_i$, $i=1,\cdots, |G/H|$ and $w_a$, $a= 1, \cdots, |G/K|$, respectively. The fusion then has vacua labeled by 
\be
u_{(i, a)} \,,
\ee
and the symmetry that is unbroken in these vacua is 
\be
U= H\cap K \,.
\ee 
The spontaneously broken part is $G/U= G/(H\cap K)$. 
Each vacuum $u_{(i, a)}$ is in an orbit of the broken symmetry $G/U$, of size $|G/U|$, and the number of such orbits is 
\be\label{marsbar}
n_{H, K} = {|G/H| \, |G/K| \over \left|{G\over (H\cap K)}\right|} \,.
\ee
In addition there are SPT phases for the unbroken symmetry, i.e. $\alpha \in H^2 (U, U(1))$. 
We get the fusion 
\be\label{TTT}
\bm{T}_{H, \alpha_H} \otimes \bm{T}_{K, \alpha_K} = n_{H, K} \bm{T}_{H\cap K, \alpha_{H\cap K}} \,.
\ee
Before we discuss the 1d topological defects after gauging, lets first consider an example to be more concrete. 

\paragraph{Example: Gauging a $\Z_2^{(0)}$ in 3d.}
Consider a non-anomalous
 {$G^{(0)}= \mathbb{Z}_2$ in a 3d theory $\cT$ (i.e. the invertibe group-like symmetry $2\Vec_{\Z_2}$).
 The symmetries of the gauged theory $\cT/\Z_2$ are determined by first considering the theta-defects: 
  there are two choices for unbroken subgroups of $G$, $H= 1$ or $\mathbb{Z}_2$}, and $\alpha$ is trivial for both of these (as the cohomology groups $H^2 (H, U(1))=0$). The topological surfaces and their fusion follow from the general analysis as follows: 
\begin{enumerate}
\item \underline{Topological surface defects:}
\begin{itemize}
\item $H= \mathbb{Z}_2$: \\
This is the (trivial) SPT phase, i.e. a TQFT with 1 vacuum $|0\rangle$, which gives rise to a trivial topological defect (identity).
The associated topological defect will be denoted by  
\be
 {D_2^{(\id )}} := D_2^{(1)} \,.
\ee
\item  $H=1$: \\
This is the SSB phase, and the TQFT has $|G/H|=2$ vacua, denoted by  $|\pm\rangle$. The resulting defect is non-trivial and will be denoted by 
\be
 D_2^{(\mathbb{Z}_2)} \,.
\ee  
\end{itemize}
\item \underline{Fusion:} \\
The fusion of the topological defects follows from the product of vacua of the TQFT and subsequent decomposition into orbits of the remnant symmetry. 
\begin{itemize}
\item $|0\rangle \otimes |0\rangle$: there is a single vacuum and thus  
\be
D_2^{(\id )}\otimes D_2^{(\id)} = D_2^{(\id)} \,.
\ee 
\item $|0\rangle \otimes (|+ \rangle \oplus |-\rangle)$: this gives rise to two vacua, which however form a single $\mathbb{Z}_2$-orbit: 
\be
D_2^{(\id )}\otimes D_2^{(\mathbb{Z}_2)} = D_2^{(\mathbb{Z}_2)} \,.
\ee
\item $|\pm\rangle \otimes |\mp\rangle$: 
The vacua are 
\be
| +, +\rangle \oplus |+, -\rangle \oplus |-, +\rangle\oplus |-, -\rangle 
\ee
and the $\Z_2$ decomposes them into two orbits ($++$ and $--$ into one orbit and $+-$ and $-+$ into another). In terms of defects we then get 
\be
\boxed{
D_2^{(\mathbb{Z}_2)}\otimes D_2^{(\mathbb{Z}_2)} = 2 D_2^{(\mathbb{Z}_2)} 
}
\ee
which is precisely the statement that $n_{1, 1}=2$ in (\ref{marsbar}). It is this last fusion that is obviously non-invertible! 
\end{itemize}
\end{enumerate}
In addition to the topological surfaces there are also  topological lines, in particular the dual line to the 0-form symmetry that we gauged. Denoting this by 
\be
D_1^{(-)}\,,
\ee
it generates an additional invertible 1-form symmetry (this is the ``dual" 1-form symmetry that we would expect from general higher-form symmetry considerations, and in terms of the gauge field we can write it as $e^{ i\int A_1}$). We can think of this line as an interface between $D_2^{(\id)}$ and itself, which corresponds to a genuine line in the 3d theory (a general discussion of the lines/interfaces will follow after this example). There are other interfaces between the various surface defects (see \cite{Bhardwaj:2022lsg}) and we will provide a general discussion of them below. The symmetry of the theory $\cT/\Z_2$ in this example is referred to as 
\be
\TwoRep (\Z_2) \,,
\ee
which comprises the topological surfaces, lines, and also local operators, as well as their fusion.

\paragraph{Interfaces between TQFTs.} 
For completeness we also discuss the topological lines and interfaces after gauging a 0-form symmetry. 
In addition to 2d TQFTs we also have $G$-symmetric topological interfaces between the 2d $G$-TQFTs. 
The interface between $\bm{T}^{(H,\alpha_H)}$ and $\bm{T}_{G, 0}$ is easy to see: these are $G$-symmetric boundary conditions of $\bm{T}^{(H,\alpha_H)}$. The 1d theory on the boundary has to carry $H$ symmetry and an anomaly $\alpha_H$, which corresponds to quantum mechanics with projective $H$-representations. Maps between two such representations are intertwiners of $H$-representations. So the interface is characterized by the fusion category of projective $H$-representations, which is denoted as
\be
\bm{I}^{(H, \alpha), (G, 0)} = \Rep^{\alpha_H}(H) \,.
\ee
Similarly generalizing this to any interface gives \cite{Bhardwaj:2022lsg}
\be
\bm{I}^{(H, \alpha_H), (K, \alpha_K)} = n_{H, K}\times  \Rep^{\alpha_{H\cap K}}(H\cap K)\,.
\ee

\paragraph{TQFTs to Symmetries.}
The theta-defects are identified with the TQFTs  as follows 
\be
\bm{T}_{H, \alpha_H} \ \longleftrightarrow\  D_2^{(G/H, \alpha_H)}  \,,
\ee
and each topological interface is a topological line defect (these form junctions between two surfaces)
\be
\bm{I}^{(H, \alpha_H), (K, \alpha_K)} \ \longleftrightarrow\  D_1^{(G/H, \alpha_H) (G/K, \alpha_K)} \,.
\ee
The combination of these topological defects and their composition/fusion is called the category of 2-representations 
\be
\cS_{\cT/G}= 2\Rep(G) \,.
\ee
Much like in 2d, gauging the invertible symmetry $G^{(0)}$ results in a reprsentation category.  However unlike in 2d, we saw that already gauging an abelian group $G=\Z_2$ results in non-invertible symmetries, namely the theta defect $D_2^{\Z_2}$!

\subsubsection{Condensation Defects are Theta Defects}
\label{sec:Cond}

The theta-defects $D_2^{(G/H), \alpha_H}$ have also an alternative description in terms of 
condensation defects \cite{Gaiotto:2019xmp, Roumpedakis:2022aik}. In this case they are condensation defects for the dual symmetry to the 0-form symmetry\footnote{This holds for abelian $G$ -- the situation for for non-abelian $G$ is more complicated, and we refer the reader to \cite{Bhardwaj:2022lsg}).}, which in 3d is a 1-form symmetry $\widehat{G}^{(1)}$. 

Let us consider this in the specific example: $G^{(0)}= \Z_2$. The theory $\cT/\Z_2$ has a 1-form symmetry $\Z_2$ generated by topological lines $D_1^{(-)}$. This fuses according to the group law of $\Z_2$. 
From this 1-form symmetry, we can construct topological surface operators 
\be
C\left(D_2^{(\id)}, D_1^{(-)}\right)= \text{gauge $\Z_2^{(1)}$ on the trivial $D_2^{(\id)}$ surface defect} \,,
\ee
i.e. we sum over all configurations of networks of  $D_1^{(g)}$ lines on the trivial surface $D_2^{(\id)}$  -- precisely as in the discussion around (\ref{TheMesh}). If $D_2^{(\id)}$ is defined on a 2d manifold $M_2$ then 
\be
C\left(D_2^{(\id)}, D_1^{(-)}\right) \sim \sum_{\Sigma_1\in H_1 (M_2, \mathbb{Z}_2)} D_1^{(-)}(\Sigma_1) D_2^{(\id)}\,.
\ee
This is precisely the defect 
\be
D_2^{(\Z_2)} = C\left(D_2^{(\id)}, D_1^{(-)}\right) \,.
\ee
More generally, condensation defects are obtained by gauging an algebra in the set of topological lines (see section \ref{sec:bimod2d}) on a topological defect $D_2^{(\alpha)}$
\be
C\left(D_2^{(\alpha)}, A\right) = \text{Condensation defect for the algebra $A$ on $D_2^{(\alpha)}$}\,.
\ee
The algebras in this case are precisely given in terms of subgroups $H< G$. 

\paragraph{Summary.} Let us take stock of what we have seen so far. A theory $\cT$ with 0-form symmetry $G^{(0)}$ in 3d has symmetry structure $2\Vec_{G^{(0)}}$: this means a 0-form symmetry generated by surface defects $D_2^{(g)}$, $g\in G$, with invertible, group-like fusion. 

If there is no 't Hooft anomaly we can gauge the symmetry to a theory $\cT/G$. The symmetry of the gauged theory is then: 
\begin{tcolorbox}[colback=white, colframe=black!50, rounded corners]
\be
\ba
&\cT \text{ with symmetry $G^{(0)}$} \cr
&\qquad\quad  2\Vec_G 
\ea 
\quad 
\begin{tikzcd}
 \quad    \rar["\text{gauge }G^{(0)}"]&  \quad 
  \end{tikzcd}
\quad 
\ba
&\cT/G^{(0)} \text{ with symmetry} \cr
&\qquad\quad  2\Rep(G)
\ea 
\ee
\end{tcolorbox}

The symmetry has topological surface operators, which are labeled by SPT+ SSB (spontaneous symmetry beaking) $D_2^{(G/H, \alpha_H)}$, and have an interpretation in terms of condensation defects. For both abelian or non-abelian groups, the fusion of topological surfaces is non-invertible (see (\ref{TTT})).  

The topological surfaces have interfaces, $D_1$. 
One special such line operator is given by the dual symmetry, which is generated by topological lines $D_1^{(a)}$ as follows:
\begin{itemize}
\item The simplest setting is that $G$ is abelian. 
Then the topological lines in the gauged theory $\cT/G$ will form again group-like fusion $D_1^{(g)}$. There are however also surface defects, $D_2^{(G/H, \alpha_H)}$, where $H$ is a subgroup and $\alpha_H$ is a cocycle in $H^2 (H, U(1))$. These are precisely the condensation defects for the topological lines $D_1^{(g)}$. Note that even though the lines are invertible, their associated condensation surfaces are not, see  (\ref{TTT}). 

\item For non-abelian $G$, the dual symmetry is not invertible, and the topological lines $D_1^{(a)}$ are labeled by representations of $G$, and form $\Rep(G)$ -- recall the 2d case in section \ref{sec:2d}. 
Again the topological surfaces are $D_2^{(G/H, \alpha_H)}$, which also have an interpretation (though less obvious) with condensation defects of the lines in $\Rep(G)$ \cite{Bhardwaj:2022lsg}.  
\end{itemize}
The generic interface will have the structure  $\Rep^{\alpha_H}(H)$ which also determines the local operators.
Although we focused on 3d theories in order to describe theta-defects that are surfaces, these symmetries can also appear as sub-symmetries of a larger symmetry whenever the theory has topological surface defects.

\paragraph{Twisted Theta Defects.}

The theta-defects that we discussed so far are universal, i.e. exist for any theory which has a non-anomalous higher-form or higher-group symmetry that we gauge. A generalization is to {\bf twisted theta defects} -- which are theory-dependent \cite{Bhardwaj:2022kot}. 

An example is the following setup: Consider a theory with a global symmetry. However which has a topological defect $D_p^{(a)}$, which cannot be made $G^{(0)}$-symmetric (e.g. due to an anomaly). Stacking with a topological field theory, that cancels the anomalous variation under $G^{(0)}$-transformations renders the combined system $G^{(0)}$-symmetric. This TQFT becomes a theta-defect, twisted by $D_p^{(a)}$. We will see an explicit example later on.

The nested structure of topological defects and fusion at each level is what determines the symmetries $2\Vec_G$ and $2\Rep(G)$. These are examples of so-called fusion 2-categories. Lets take a quick look at these structures.

\subsection{Higher-Categorical Structure}

The theta-defects, i.e. 2d $G^{(0)}$-TQFTs and their fusion (as well as an analog analysis of the lines) as seen in the last section, mathematically form a fusion 2-category $\TwoRep(G)$. This generalizes the symmetry $\Rep(G)$ that was studied in 2d. Both of these were constructed using theta-defects, i.e. by stacking $G$-symmetric TQFTs before gauging, and result in potentially non-invertible symmetries. In the generalization to higher-categories, we allow for higher-dimensional topologogical defects. In an $n$-category, the top dimension (usually referred to as ``objects") is formed by $n$-dimensional topological defects. Lower-dimensional defects, which can occur as interfaces between higher-dimensional topological defects are (higher) morphisms. 

Lets consider fusion 2-categories -- i.e. symmetries of 3d theories: 
\begin{enumerate}
\item Topological surfaces $D_2$: Objects \\
These generate the 0-form symmetry.
\item Topological lines/interfaces $D_1$, or $I_1^{(a,b)}$: 1-Morphisms.\\
These generate the 1-form symmetry. 

\item Topological local operators $D_0$: 2-Morphisms\\
These generate the 2-form symmetry. 
\end{enumerate}
None of these need to have invertible fusion.
See figure \ref{fig:UkrainianDoll}.
At each level, there is a fusion. Some key works within mathematics studying these categories can be found here \cite{DouglasReutter, Kong:2019brm, Johnson-Freyd:2020ivj, decoppet2022drinfeld}\footnote{In particular of interest is the last reference here which provides a classification: essentially every 2-fusion category is related via gauging to a (Deligne) product of an invertible symmetry (i.e. $2\Vec_G$) and a modular tensor category. }

The simplest case is a pure 0-form group symmetry, where only the top layer is populated with group-like fusion, and all morphisms are trivial. This is in fact the symmetry structure of the 3d theory at the start of section \ref{sec:Theta3d}. This is denoted by the category (genearlizing $\Vec_G$ in 2d)
\be
2\Vec_G^\omega \,.
\ee
This can have an anomaly (cocycle) $\omega \in H^4 (G, U(1))$. 
Gauging $G$ for $\omega=0$ results in the 2-category 
\be
2\Rep(G) \,.
\ee
More interesting situations arise in non-trivial cocycle situations, e.g. $G= \Z_2 \times \Z_2$ with non-trivial co-cycle results in $\Z_4$ after gauging, see \cite{Bhardwaj:2022maz}.

The generalization to any dimension $d$ is at least formally straight-forward: We stack with $d-1$-dimensional $G^{(0)}$-TQFTs and gauge. These universal defects will be called {\it theta defects}. The resulting symmetry structure always has a description in the category 
\be
\mathcal{S}=(d-1)\Rep (G) \,.
\ee
There are important subtleties for $d\geq 4$, where there  are TQFTs which have non-trivial topological order. E.g. $3\Rep(G)$ is not only determined by  SPT phases and SSB, but there is also the possibility of stacking with $G$-symmetric modular tensor categories, which characterize non-trivial 3d topological order, and may not even admit gapped boundary conditions.


\section{Non-Invertible Symmetries in $d\geq 3$ QFTs}
\label{sec:QFTs}

We now turn to realizing non-invertible symmetries in physical QFTs, in particular gauge theories in various dimensions. Our main focus will be on 3d and 4d gauge theories, but the sky's the limit, and many interesting applications exist for  $d>4$ theories as well. 

At this point in time, there are several approaches for constructing non-invertible symmetries in gauge theories in $d\geq 3$. 
The known constructions can be roughly speaking categorized as follows (citing the first instances of these constructions, which include a comprehensive discussion of defects and fusions)
\begin{enumerate}
\item Duality defects \cite{Choi:2021kmx, Kaidi:2021xfk}
\item Gauging outer automorphisms or non-normal subgroups \cite{Bhardwaj:2022yxj}
\item Theta-Defects or Condensation Defects \cite{Bhardwaj:2022lsg, Roumpedakis:2022aik}
\item Gauging mixed 't Hooft anomalies \cite{Kaidi:2021xfk, Bhardwaj:2022yxj}
\end{enumerate}
There is non-trivial overlap between these constructions. We will also see, that most of these -- with the exception of some duality and triality defects --  can be thought of using the same philosophy as applied in the construction of theta-defects: before gauging an invertible symmetry, we stack with a TQFT, which in the gauged theory becomes a topological defect (usually with non-invertible fusion). We will highlight this perspective  in the discussion of these non-invertible symmetries.

\subsection{Non-Invertible Symmetries in 3d QFTs}

\subsubsection{Theta-Defects as Symmetries}

In the last section we have seen that theta-defects give rise to universal non-invertible symmetries. What physical theories carry such symmetries? 
The symmetry $2\Rep(\Z_2)$ is realized the in a pure $SU(2)$ gauge theory in 3d. Let us start with gauge group $SO(3)$, which has only a 0-form symmetry $\Z_2^{(0)}$, generated by $D_2^{(g)}$, $g\in \Z_2$. Gauging this results in
\be
  \begin{tikzcd}
  SO(3) \quad    \rar["\text{gauge } \Z_2^{(0)}"]&  \quad 
     SU(2)   \end{tikzcd}
    \,.
\ee
In the $SU(2)$ gauge theory, we have the dual set of lines $D_1^{(g)}$, $g\in\Z_2$, which generate the 1-form symmetry group $\Z_2^{(1)}$. But there are also theta or condensation defects $D_2^{(G/H)} = D_2^{(\id)},\ D_2^{(\Z_2)}$, with the non-invertible fusion 
\be
D_2^{(\Z_2)} \otimes D_2^{(\Z_2)} = 2 D_2^{(\Z_2)}\,.
\ee
In summary we have: 
\begin{itemize}
\item Symmetry of 3d $SO(3)$ pure gauge theory: $2\Vec_{\Z_2}$.
\item Symmetry of 3d $SU(2)$ pure gauge theory: $2\Rep (\Z_2)$.
\end{itemize}

There is a straight forward generalization to the pair of $PSU(N)$ and $SU(N)$ gauge theories in 3d. Replacing $\Z_2$ by $\Z_N$: we start with 0-form $2\Vec_{\Z_N}$ of $PSU(N)$. Gauging this, produces the dual 1-from symmetry generators $D_1^{(g)}$, $g\in \Z_N$ and theta defects $D_2^{(G/H)}$, where $H$ are the subgroups of $\Z_N$. The symmetry of the $SU(N)$ theory is thus $2\Rep (\Z_N)$.

For $N>2$ this is however only a subset of the full symmetry:  in addition to $\Z_N^{(0)}$ there is also an outer automorphism that acts on the gauge algebra $\Z_2^{(0), \text{out}}$. Consider e.g. the pure 3d gauge theory with gauge group
\be
G_{\text{gauge}} = PSU(3)\,. 
\ee
This has a non-abelian 0-form symmetry 
\be
G^{(0)} = \Z_3 \rtimes \Z_2  = S_3\,,
\ee
so that after gauging this, we get to 
\be
\cC_{{PSU(3)\over S_3} = \widetilde{SU(3)}}: \qquad 2\Rep (S_3)  \,.
\ee
Here $\widetilde{SU(3)}$ is the dis-connected gauge group, obtained from $SU(3)$ by gauging charge-conjugation, which is the outer-automorphism $\Z_2^{(0), \text{out}}$.

\subsubsection{Orthogonal Gauge Groups}

A very rich class of non-invertible symmetries can be found in the pure gauge theories in 3d with gauge algebra $\mathfrak{so}(4N)$ (similarly $\mathfrak{so}(4N+2)$, see \cite{Bhardwaj:2022maz}). There are numerous global forms of the gauge group, and each will reveal an interesting global symmetry structure. 

From earlier considerations we have learned that it is easiest to start with a theory that has only 0-form symmetry. For the orthogonal groups this is the gauge group
\be
G_{\text{gauge}} = PSO(4N) \,.
\ee
This has only 0-form symmetries: 
\be
\mathcal{C}_{PSO(4N)}= 2\Vec (D_8) \,.
\ee
This means we only have topological surfaces in 3d, generating a non-abelian dihedral group $D_8$: 
\be
D_2^{(g)}\,,\qquad g\in D_8 =\left( \Z_2^{S} \times \Z_2^{C} \right) \rtimes \Z_2^{o}  \,.
\ee
There are also topological lines $D_1^{(g)}$ on $D_2^{(g)}$, which are all trivial lines that fuse according to the group law. 
The 0-form symmetry $\Z_2^{o}$ acts by exchanging two of the 0-form symmetry generators 
\be
D_2^{S} \leftrightarrow D_2^C \,.
\ee
The action on the Dynkin diagram is 
\be\label{SpinOut}
\begin{tikzpicture}
\begin{scope}[shift={(3,-8)}]
\draw [thick](4,0) -- (5,1);
\draw [thick](4,0) -- (5,-1);
\draw [thick](0,0) -- (1.5,0);
\draw [thick,fill=black] (0,0) ellipse (0.1 and 0.1);
\draw [thick,fill=black] (1,0) ellipse (0.1 and 0.1);
\draw [thick,fill=black] (3,0) ellipse (0.1 and 0.1);
\draw [thick,fill=black] (4,0) ellipse (0.1 and 0.1);
\node at (2,0) {$\cdots$};
\draw [thick] (2.5,0) -- (4,0);
\draw [thick,fill=black] (5,1) ellipse (0.1 and 0.1);
\draw [thick,fill=black] (5,-1) ellipse (0.1 and 0.1);
\node[above] at (5,1.2) {$C$};
\node[below] at (5, -1.2) {$S$};
\node at (-3,0) {{$\mathfrak{so}(4N)$}:};
\draw[ thick, <->] (5,-0.7) -- (5, 0.7) ;
\node[right] at (5.2,0) {$\Z_2^{o}$};
\end{scope}
\end{tikzpicture}
\ee
For concreteness the group multiplication table is as follows:
\be
g=\id,  S, C, V, o: \quad g^2 = \id \,,\  SC= V\,,  \ oSo = C \,.
\ee
There is a whole zoo of possibilities of gauging subgroups of $D_8$, which result in different global forms of the gauge group, and also different global symmetries.

\paragraph{Gauging $G=D_8$.}
By gauging the full 0-form symmetry, we can apply our earlier analysis on theta-defects and stack $D_8$-symmetric 2d TQFTs. This will give a symmetry 
\be
\mathcal{C}_{\Pin^+(4N)} = 2\Rep (D_8) \,.
\ee
To characterize the symmetry of this theory we  should apply the general theta-defect analysis to $2\Vec_{D_8}$.  
This is a  nice exercise and the result can be compared with section 5.5 in \cite{Bhardwaj:2022maz}.

\begin{figure}\centering
\begin{tikzpicture}
\node[draw,rectangle,thick,align=center](Spin) at (0,0) {{Spin$(4N)$} \\ 2Vec($(\bbZ_2^{(1)} \times \bbZ_2^{(1)})\rtimes \bbZ_2^{(0)}$) \\ 2-group};
\node[draw,rectangle,thick,align=center](PSO) at (8,0) {{PSO$(4N)$} \\ 2Vec($D_8^{(0)}$) \\ group};
\node[draw,rectangle,thick,align=center](Pin) at (0,-6) {{Pin$^+(4N)$} \\ 2Rep($D_8$) \\ non-invertibles};
\node[draw,rectangle,thick,align=center](PO) at (8,-6) {{PO$(4N)$} \\ 2Rep($(\bbZ_2^{(1)} \times \bbZ_2^{(1)})\rtimes \bbZ_2^{(0)}$) \\ non-invertibles};
\draw[thick,->] (-0.15,-1) -- (-0.15,-5);
\draw[thick,<-] (0.15,-1) -- (0.15,-5);
\draw[thick,->] (7.85,-1) -- (7.85,-5);
\draw[thick,<-] (8.15,-1) -- (8.15,-5);
\draw[thick,->] (2.5,0.15) -- (6.75,0.15);
\draw[thick,<-] (2.5,-0.15) -- (6.75,-0.15);
\draw[thick,->] (Spin) to (PO);
\draw[thick,<-] (Pin) to (PSO);
\node at (4.5,0.75) {$\Z_2^{(1)} \times \bbZ_2^{(1)}$};
\node at (4.5,-0.75) {$\Z_2^{(0)} \times \bbZ_2^{(0)}$};
\node at (-0.75,-3) {$\Z_2^{(0)}$};
\node at (0.75,-3) {$\Z_2^{(1)}$};
\node at (7.25,-3) {$\Z_2^{(0)}$};
\node at (8.75,-3) {$\Z_2^{(1)}$};
\node at (4.7, -2) {$D_8^{(0)}$};
\node at (4.7, -4) {$\mathbb{G}^{(2)}$};
\end{tikzpicture}
\caption{A sub-web of the 3d gauge theories with $\mathfrak{so}(4N)$ gauge algebra, which can be obtained by gauging various $0$- and 1-form symmetries and the 2-group 
$\mathbb{G}^{(2)}=(\Z_2^{(1)}\times \Z_2^{(1)} )\rtimes \Z_2^{(0)}$.
We list the gauge group in each box, then the symmetry category, and finally a description -- (2-)group symmetries are invertible, whereas the remaining two are non-invertible. The groups adjacent to the arrows indicate the symmetries that are gauged to get from one end to the other. Note that the inverse operations to the diagonals are non-invertible fusion category gaugings. 
\label{fig:miniweb}}
\end{figure}
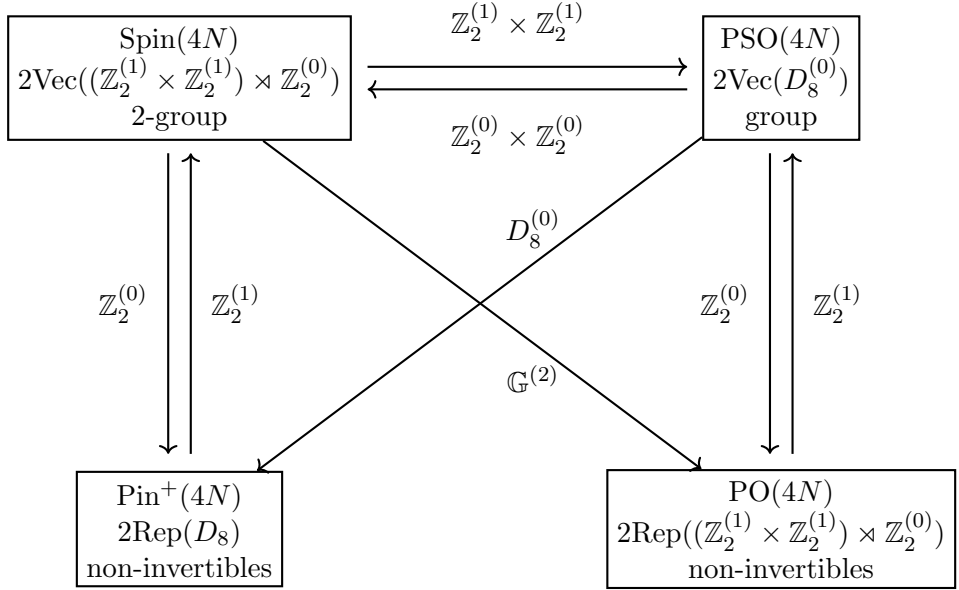

\paragraph{Gauging Subgroups of $D_8$.}
We can also gauge subgroups $G$ of $D_8$, and stack accordingly only with $G$-TQFTs, and obtain a whole non-invertible symmetry web, see \cite{Bhardwaj:2022maz}. An interesting  subset of symmetries are shown in figure \ref{fig:miniweb}, which illustrate the relation between invertible 0-, 1-form and 2-groups, non-invertible symmetries via gauging. The symmetry of $PSO(4N)$ is an invertible $G^{(0)}=D_8$. Gauging this full 0-form symmetry results in the theory $\Pin^+(4N)$, which has  non-invertible symmetry $2\Rep(D_8)$. This theory is alternatively the outer-automorphism gauged version of $\Spin(4N)$
\be
\Pin^+(4N) = \Spin(4N)/\Z_2^{o}\,.
\ee
$\Spin(4N)$ has in addition to the outer autmorphism $\Z_2^{o}$ 0-form symmetry, also a 1-form symmetry $G^{(1)}=\Z_2 \times \Z_2$. As the outer automorphism acts on the 1-form symmetry by exchanging the two $\Z_2$s this is precisely the symmetry that we denoted by a split 2-group (recall section \ref{sec:TwoGp}): 
\be
\mathbb{G}^{(2)}=\left(\Z_2^{(1)}\times \Z_2^{(1)} \right)\rtimes \Z_2^{(0)} \,.
\ee
This is an invertible symmetry, but the 0-form and 1-form symmetries have a non-trivial relation (\ref{S2G}). Gauging the full 2-group (which again can be done using theta-defects \cite{Bhardwaj:2022lsg} by stacking 2-group-protected TQFTs) results in the theory $PO(4N)$, whose symmetry is 2-representations of the 2-group $\mathbb{G}^{(2)}$. For the interested reader, there are several more details to be found in \cite{Bhardwaj:2022maz}, including a complete set of global forms that can occur for $\mathfrak{so}(4N)$ theories in 3d. Alternatively one can also generalize the bimodule picture we discussed in 2d to higher-dimensional theories  \cite{Bartsch:2022mpm, Bhardwaj:2022maz}.

\subsection{Non-Invertible Symmetries of 4d QFTs}
\label{sec:KOZ}

\subsubsection{General Remarks}

A symmetry for a 4d QFT will be comprised of 3d topological defects $D_3$, surfaces $D_2$, lines $D_1$ and points $D_0$. Generally speaking these form a fusion 3-category. 
Such structures are in fully generality still being explored -- both in physics and mathematics.  

However we can again apply the theta construction, and obtain non-invertible symmetries. Gauging a $p$-form symmetry, requires stacking with TQFTs that $p$-form symmetric. 
For example, to gauge a 1-form symmetry, we have to stack 2d TQFTs that are $G^{(1)}$-SPTs, etc. However a crucial difference occurs when considering 3d topological defects and gauging the associated 0-form symmetry. 

A 3d TQFT that is a $G^{(0)}$-SPT+SSB type TQFT is only a very special 3d TQFT. Applying the theta-defect construction works of course, but we can also stack  3d TQFT with non-trivial topological order. 
Thus $3\Rep(G)$ is much more than SPT+SSB!

Moreover 3d TQFTs may not admit gapped boundary conditions, i.e. the 3d TQFT put on a 3-manifold with boundary $\Sigma_2$ gives rise to a 2d CFT. Examples are $U(1)_M$ CS-theories, which have a boundary that is a chiral boson in 2d. 
This can be used for a theta-defect construction if the 0-form symmetry has a  mixed 't Hooft anomaly. We will provide an example in section \ref{sec:TwistTheta}.

\subsubsection{Twisted Thetas and Gauging in the Presence of Mixed Anomalies}
\label{sec:TwistTheta}

For this section we assume that we are on a spin-manifold, and consider a spin-QFT.
So far we have considered gauging without any 't Hooft anomalies for the symmetries under consideration. 
An interesting class of non-invertibles arises when considering theories with mixed anomalies, as pioneered in \cite{Kaidi:2021xfk}. The classic example by now starts in 4d with a  mixed 0-form/1-form symmetry anomaly that we encountered already in section \ref{sec:Anomalies}
\be\label{mixedano}
{\mathcal{A} = 
-{2 \pi   \over N} \int A_1 \cup  {\mathfrak{P} (B_2)\over 2} } \,.
\ee
Here $B_2\in H^2 (M, \Z_N)$ and $A_1\in H^1 (M, \Z_{2N})$. There is a mixed anomaly between the 0-form $\Z_N\subset \Z_{2N}$ and 1-form symmetry $\Z_N^{(1)}$. 
This occurs for instance in 4d pure super-Yang Mills (SYM)  with gauge group $SU(N)$ and 1-form symmetry $\Z_N^{(1)}$, which has also a $U(1)_R$ R-symmetry that acts by rotations on the supercharges, and is broken by the ABJ anomaly to $\Z_{2N}$. 

We want to gauge the $\mathbb{Z}_N^{(1)}$ to get to the $PSU(N)$ SYM theory. 
However there is already a non-trivial effect due to the mixed anomaly when we coupled the theory to background fields for the symmetries. Let $D_3^{(g)}$ be as usual the 0-form symmetry defects. Then coupling to  the background fields and performing a background gauge transformation, the generator $D_3^{(g)}$ is transformed by a phase 
\be
D_3^{(g)}(M_3) \rightarrow D_3^{(g)}(M_3) \exp \left(\int_{M_4} -{2 \pi i \over N}  {\mathfrak{P} (B_2)\over 2} \right)  \,,
\ee
for $\partial M_4 = M_3$. 
Gauging the 1-form symmetry is  not a consistent operation for this defect. This is a situation we discussed earlier, in the constext of non-universal, twisted theta defects. 
This can be remedied by stacking the defect $D_3^{(g)}$ with a 3d TQFT that has 
\begin{itemize}
\item 1-form symmetry 
\item  anomaly $\mathcal{A} = {2\pi i \over N} \int_{M_4} {\mathfrak{P}(B_2)\over 2}$.
\end{itemize}
There are many such TQFTs with boundaries that are not gapped. The minimal choice was determined in \cite{Hsin:2018vcg}: any 3d TQFT which has these properties can be factored into a TQFT, which has topological lines, that are neutral under the 1-form symmetry  and have no anomaly, times, the TQFT 
\be\label{AN1}
\mathcal{A}^{N, 1} = U(1)_N \,.
\ee
This has lines charged under the 1-form symmetry which also exhibit the anomaly required. 
In general define $\mathcal{A}^{N, p}$ to be the minimal 3d TQFTs, with $\Z_N^{(1)}$ and anomaly $p\cA$. 
For $p=1$ this is the 3d CS-theory with $U(1)$ gauge group and  CS-level $N$ (\ref{AN1}). The idea is to then stack this TQFT on the defects $D_3$ before gauging. The combined defect is then invariant under background gauge transformations and we can gauge the 1-form symmetry. 

Again, this is a stacking with TQFTs before gauging, however now with a 3d TQFT that has non-trivial topological order, and in fact does not admit gapped boundary conditions: if the three-manifold $M_3$ has a boundary $\partial M_3= \Sigma_2$, the theory on $\Sigma_2$ is a chiral boson with central charge 1. The existence of such defects is theory-dependent, and provide an example of {\bf twisted theta defect}. 

The twisted theta defect in this case is 
\be
\mathcal{D}_3^{(p)}(M_3) = D_3^{(p)}(M_3) \otimes \mathcal{A}^{N,p} (M_3) \,,
\ee
which is a simple defect after gauging the 1-form symmetry. To determine the fusion, we again need to compute the fusion of the TQFTs.  
The action for the TQFT $\mathcal{A}^{N,1}$ is known explicitly as the $U(1)_N$ CS-theory, coupled to a background field $b_2$ for the 1-form symmetry (written in terms of $U(1)$-valued fields)
\be
\mathcal{A}^{N,1} (b_2)= {i N \over 4 \pi } a_1  d a_1+ {i N\over 2 \pi} a_1  b_2 \,.
\ee
The combined defect is 
\be
\mathcal{D}_3^{(1)}(M_3) = \int Da_1\,  D_3^{(1)} (M_3, b_2)\,  \exp\left(
{iN \over 2 \pi} \int_{M_3} {1\over 2} a_1 da_1 + a db_2\right) \,.
\ee
The claim is that this defect has non-invertible fusion, due to the fact that the TQFT that we dressed it with have a non-invertible fusion. 
Let us focus on $N$ odd. The non-invertible fusion of the defects is 
\be\label{KOZDD}
\mathcal{D}_3^{(1)} (M_3)\otimes \mathcal{D}_{3}^{(1)}(M_3) = \mathcal{A}^{N, 2} \, \mathcal{D}_3^{(2)} (M_3) \,.
\ee
This is a non-invertible 0-form symmetry in the ${PSU}(N)$ 
SYM theory. 
Defining the conjugate 
\be
\mathcal{D}_3^{(1)\dagger} = D_3^{(-1)} \otimes \mathcal{A}^{N, -1}
\ee
results furthermore in the fusion 
\be\label{KOZfuss}
\ba
\mathcal{D}_3^{(1)} (M_3)\otimes \mathcal{D}_3^{(1)\dagger}  (M_3)
&=  \sum_{M_2 \in H_2(M_3, \mathbb{Z}_N) } { (-1)^{\chi(M_2)} D_2(M_2) \over  |H^{0}(M_3,\mathbb{Z}_N)| } D_3^{(0)} \cr 
& = C\left(D_3^{(0)},  D_2\right) \,,
\ea
\ee
which is the condensation defect of the 1-form symmetry on $M_3$ with generator $D_2(M_2)= e^{i2\pi\int_{M_2} b_2/N}$, where $b_2$ is the gauge field for the 1-form symmetry and $\chi$ the Euler characteristic of $M_2$\footnote{We omit writing the explicit 3-manifold dependence of the condensation defect.}.

\paragraph{Derivation of the fusion.}
Lets derive in detail the fusions in (\ref{KOZDD}) and (\ref{KOZfuss}). 

The fusion of the 3d TQFTs that we use to stack the defects with can be computed as follows: These are spin TQFTs, and their properties were studied in detail in \cite{Hsin:2018vcg, Choi:2022jqy}\footnote{The theories $\mathcal{A}^{N,p}$ have topological lines, which can braid, and only for $N$ and $p$ coprime is this braiding non-degenerate, i.e. only the trivial line has trivial braiding with all other lines.}  
\be
\mathcal{A}^{N,1} \otimes \mathcal{A}^{N,1} = 
\mathcal{A}^{N,2} \otimes \mathcal{A}^{N,2} \,.
\ee
From the sum of the actions on the left hand side, we can construct topological line operators $W_1^{\pm} = e^{i \int a_1 \pm a_1'}$. They satisfy $(W_1^{\pm})^N= \id$ and their spin is twice that of the spin of the lines in the original theory. 
Using the  group like fusion of the invertible defects $D_3^{(1)}\otimes D_3^{(1)} = D_3^{(2)}$  together with the fusion of the TQFT we get (\ref{KOZDD}).

Next we derive (\ref{KOZfuss}) as follows. First note that the invertible parts fuse as $D_3^{(1)}\otimes D_3^{(-1)} = D_3^{(0)}$, which is the identity. The interesting fusion arises from the dressing factor, as usual: 
\be
\ba
\mathcal{D}_3^{(1)} (M_3)\otimes \mathcal{D}_3^{(1)\dagger}  (M_3)
\sim
&\int D a_1 Da_1' e^{{i\over 2\pi} {N\over 2} \int_{M_3}( a_1da_1 - a_1'da_1') + {i\over 2\pi} N \int_{M_3} (a_1-a_1')b_2}  \, D_3^{(0)}(M_3)\cr 
 \stackrel{\hat{a} = a_1-a_1'}{=} 
 &
\int D a_1 D\hat{a}_1 e^{{i\over 2\pi} {N\over 2} (\int-2a_1d\hat{a}_1 +\hat{a}_1d\hat{a}_1 )+{i\over 2\pi} N \int \hat{a}_1b_2}  \, D_3^{(0)}(M_3)\,.
\ea
\ee
The equations of motion for $a_1$ imply that $\hat{a}_1$ is $N$-torsion. We thus want to rewrite the integral $\int D\hat{a}_1$ in terms of a sum over $\hat{a}_1 \in H^1 (M_3, \Z_N)$, using (\ref{boom}).
To simplify the expression further we need to use some more insights from algebraic topology. We discussed the Bockstein homomorphism in the context of 2-groups. Here we consider the map $\Bock: H^n(M, \Z_N) \to H^{n+1} (M, \Z_N)$, which we can used to transform $da_1/N$ into the cochain formulation $\Bock(a_1)$. 
We then find
\be
\ba
\mathcal{D}_3^{(1)} (M_3)\otimes \mathcal{D}_3^{(1)\dagger}  (M_3)
\sim & \sum_{\hat{a}_1\in H^1 (M_3, \Z_N)} (-1)^{\hat{a}_1\Bock(\hat{a}_1)} e^{i \int_{M_3} \hat{a}_1 b_2 } \, D_3^{(0)}(M_3)\,.
\ea
\ee
It is useful to replace the sum over  $a_1\in H^1(M_3, \Z_N)$ by the Poincar\'e dual in $M_3$, and sum over $M_2 \in H_2(M_3, \Z_N)$. The integral $\int_{M_3} \hat{a}_1 b_2$, becomes an integral of $b_2$ over the Poincar'e dual 2-cycle to $a_1$. The term $(-1)^{\int_{M_3}\hat{a}_1\Bock\hat{a}_1}$ measures $\int \hat{a}_1 \Bock\hat{a}_1$ mod 2, and we rewrite it as $(-1)^{Q(M_2)}$. 
Putting this all together we get (for the normalization, see \cite{Kaidi:2021xfk}) 
\be
\mathcal{D}_3^{(1)} (M_3)\otimes \mathcal{D}_3^{(1)\dagger}  (M_3)
= \sum_{M_2\in H_2 (M_3, \Z_N)}
 { (-1)^{Q(M_2)} e^{i \int_{M_2} b_2}  \over  |H^{0}(M_3,\mathbb{Z}_N)| }D_3^{(0)}(M_3)
\,.
\ee
Here $b_2$ is a $\Z_N$-valued field, and we can transform it back to the $U(1)$-formulation in (\ref{KOZfuss}). 

If $N=2$ we can replace $\Bock (\hat{a}_1)= \text{Sq}^1(\hat{a}_1) = \hat{a}_1\cup \hat{a}_1$, where $\text{Sq}^1$ is the first Steenrod square (see \cite{Hatcher}). Furthermore, $\int_{M_3} \hat{a}_1^3$ has a nice formulation in terms of the Euler characteristic of the Poincar\'e dual 2-cycle $M_2$ (which for oriented manifolds is even) in \cite{banchoff1974triple}, i.e. for $N=2$ we have $Q(M_2) = \chi(M_2)$.

What we have seen here is that theories with mixed anomalies can result in non-invertible symmetries. Roughly speaking, whenever a background field $B_{q+1}$ for a $q$-form symmetry appears linearly in a mixed 't Hooft anomaly, there is a gauged version of the theory, obtained by gauging all the other symmetries (not associated to $B_{q+1}$) that appear in the anomaly. The defect for the $q$-form symmetry is then  
 non-invertible in the gauged theory. 

\paragraph{Applications: ABJ Anomaly.}
The idea that stacking with TQFTs to cure anomalies, which then result in non-invertible symmetries is also applicable to ABJ-anomalies, as shown in 
\cite{Choi:2022jqy, Cordova:2022ieu}. In QED, i.e. $U(1)$ gauge theory with a charge one Dirac fermion $\psi$ the chiral symmetry 
\be
\psi \to e^{{i\theta\over 2} \gamma_5 } \psi \,,
\ee
has an ABJ anomaly. This means that the associated current is not conserved. Let $j^\mu = {1\over 2} \bar \psi \gamma^5 \gamma^\mu \psi$ then 
\be
d * j = F \wedge F\,.
\ee
Subsequently, the would-be Noether symmetry generator 
\be
D_3^{(\alpha)}= e^{i \alpha \int * j} 
\ee
is not invariant. Again this can be remedied by coupling this defect to a 3d TQFT. Let $\alpha= 2\pi /N$, then \cite{Choi:2022jqy, Cordova:2022ieu}
\be
\mathcal{D}_{3}^{(2 \pi/N)} = \int Da\,  e^{i \int_{M_3} {2 \pi\over N } * j + {N\over 4 \pi} ada + {1\over 2\pi} a d A } \,,
\ee
is gauge invariant, where $A$ is the gauge field of the 4d bulk QED, and $a$ is localized on the defect -- i.e. we again couple the system to a 3d $U(1)$ CS-theory, and thus consider a twisted theta-defect. The fusion of this defect with its conjugate can be similarly evaluated as in (\ref{KOZfuss}) and gives rise to the $\Z_N^{(1)}$ condensation defect.

\subsection{Non-invertible Symmetries in 4d from Outer-Automorphisms}

We already have seen that the symmetry of the $\Pin^+(4N)$ gauge theory in 3d is non-invertible, by gauging the $D_8^{(0)}$ form symmetry starting with $PSO(4N)$, see figure \ref{fig:miniweb}.
There is was argued using the theta-defects for the 0-form symmetry. An alternative way would be to gauge the outer automorphism of $\Spin(4N)$, which acts on the 1-form symmetry generators $\Z_2^S \times \Z_2^C$. 

More generally gauging non-normal subgroups, such as outer automorphisms, charge conjugations, yields non-invertible symmetries. This was developed in \cite{Bhardwaj:2022yxj}. 
Here we focus on two examples in 4d: the $\Pin^+(4N)$ pure gauge theory and the $O(2)$ gauge theory, which we already encountered before. This results in finite and continuous non-invertible symmetries, respectively. 

\subsubsection{Gauging Outer Automorphisms: 
4d $\Pin^+(4N)$ YM}

Let us consider the {4d $\Spin(4N)$ pure gauge theory}. This has 1-form symmetry given by the center
\be
G^{(1)} = \Z_2^{S} \times \Z_2^{C}\,,
\ee
and $\Z_2^{V}$ will denote the diagonal $\Z_2$. 
The line operators that are charged under these are the spinor and cospinor Wilson lines. 
In addition there is an outer automorphism $\Z_2^{(0)}$, which acts as in (\ref{SpinOut}). Gauging this outer automorphism maps the theory to $\Pin^+(4N)$ pure gauge theory, which we will show has non-invertible higher-categorical symmetries. More details and examples can be found in \cite{Bhardwaj:2022yxj}.

\begin{figure}
\centering
\begin{tikzpicture}
\draw [blue, fill=blue,opacity=0.1] (0,0) -- (0,4) -- (1.5,5) -- (1.5,1) -- (0,0);
 \draw [blue, thick] (0,0) -- (0,4) -- (1.5,5) -- (1.5,1) -- (0,0);
\begin{scope}[shift={(0,-0.5)}]
\draw [orange, fill= orange,opacity=0.1] (-2, 1.5) -- (2,1.5) -- (3.5, 2.5) -- (-0.5, 2.5) -- (-2, 1.5);
\draw [orange, thick] (-2,1.5) -- (2,1.5) -- (3.5,2.5) -- (-0.5,2.5) -- (-2,1.5);
\draw  (0,1.5) -- (1.5,2.5);
\node [orange] at (3,1.5) {$D_{2}^{(V)}$} ;
\end{scope}
\begin{scope}[shift={(0,1)}]
\draw [cyan, fill=cyan,opacity=0.2] (-2, 1.5) -- (2,1.5) -- (3.5, 2.5) -- (-0.5, 2.5) -- (-2, 1.5);
\draw [cyan, thick] 
(-2,1.5) -- (2,1.5) -- (3.5,2.5) -- (-0.5,2.5) -- (-2,1.5);
\draw [PineGreen, fill =PineGreen, opacity =0.3] (0,1.5) -- (1.5,2.5)  -- (3.5,2.5) --(2,1.5) -- (0,1.5);
\draw [PineGreen, thick] (0,1.5) -- (1.5,2.5)  -- (3.5,2.5) --(2,1.5) -- (0,1.5);
\node [PineGreen] at (3.5,2) {$D_{2}^{(C)}$} ;
\end{scope}
\node [cyan] at (-1.8, 3.2) {$D_2^{(S)}$} ;
\node [orange] at (-1.8,1.6) {$D_{2}^{(V)}$} ;
\node[blue] at (0.8,-0.5) {$D^{(g)}_{3}, \, g\in \Z_2^{(0)}$};
\end{tikzpicture}
\caption{The action of the 0-form outer automorphism $D_3^{(g)}$ on the topological surface defects $D_2^{(i)}$ which generate 
 $G^{(1)}$.\label{fig:OutSpin}}
\end{figure}
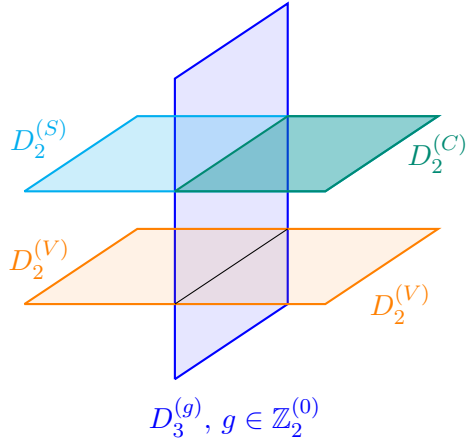

\paragraph{Symmetries of 4d $\Spin(4N)$ YM.}
In a 4d theory we have topological operators (top ops) of dimensions 0, 1, 2, 3: for $\Spin(4N)$ these are 
\be
\ba
\cC_{\Spin(4N)}^{\text{3d top ops}}& = \left\{ D_3^{(\id)} \,,\ D_3^{(-)} \right\}
\cr 
\cC_{\Spin(4N)}^{\text{2d top ops}}& = \left\{D_2^{(\id)}\,,\ 
D_2^{(S)}\,,\ 
D_2^{(C)}\,,\ 
D_2^{(V)} \right\} \,.
\ea\ee
There are no non-trivial topological lines between distinct topological surfaces, 
 but each surface has a topological line on it, which we denote as 
\be
\cC_{\Spin(4N)}^{\text{1d top ops}}=\left\{D_1^{(\id)},D_1^{(S)}, D_1^{(C)}, D_1^{(V)} \right\} \,.
\ee
Similarly we have 0d topological operators. 
These topological operators satisfy group-like fusion at all layers
\be
D_p^{(i)}\otimes D_p^{(j)}=D_p^{(ij)}\,.
\ee
In particular $S \otimes C = V$ and $S^2 = C^2 =V^2= \id$.
The 0-form symmetry generators, i.e. topological defects $D_3$, act as outer automorphism  on the 1-form symmetry generators, i.e. surface defects $D_2$ (and in fact also the spinor and cospinor Wilson lines), see figure \ref{fig:OutSpin}.
Concretely, the $G^{(0)}=\mathbb{Z}_2$ outer automorphism acts by 
\be
 D_i^{(S)} \longleftrightarrow D_i^{(C)} \,,
\ee
and leaves $D_i^{(V)}$ invariant. 
As in 3d, we should think of this as a 2-group $\mathbb{G}^{(2)}$ (of split type, i.e. similar to a semi-direct product of groups), which can be written as 
\be\label{Spin2gp}
\mathbb{G}^{(2)}= \left(\Z_2^{(S), (1)} \times \Z_2^{(C), (1)}\right) \rtimes \Z_2^{(0)} \,.
\ee
We will now gauge $G^{(0)}$ to pass to $\Pin^+(4N)$ using two, slightly different, approaches: 
\begin{itemize}
\item Theta-defects: We will stack $G^{(0)}$-TQFTs, however due to the presence of the 1-form symmetry, we need to now also account for the topological surfaces. This results instead of $2\Rep(\Z_2)$ in $2\Rep(\mathbb{G}^{(2)})$, i.e. the two-representations of the 2-group  (\ref{Spin2gp}), as discussed in \cite{Bhardwaj:2022lsg}.
\item A step-wise construction of the gauge-invariant defects following \cite{Barkeshli:2014cna, Bhardwaj:2022yxj}.
\end{itemize}

\paragraph{Gauging $\Z_2^{(0)}$ to $\Pin^+(4N)$: Take 1.}
Lets consider the theta-defect construction first. We need to determine the SPT-phases and SSB vacua, as well as SPT plus SSB for the split 2-group (\ref{Spin2gp}). Before embarking on this, let us guess what we expect: the 0-form symmetry acts non-trivially on some of the defects that generate the 1-form symmetry ($S$ and $C$ are exchanged), so we expect invariant combintations of these. In addition, there will also be the invariant 1-form symmetry generators $D_2^{(\id)}$ and $D_2^{(V)}$. In addition, gauging a 0-form symmetry in 4d, we expect there to be a dual 2-form symmetry generated by lines $D_1^{(-)}$
\be
G^{(0)} = \Z_2 =\left\{D_3^{(\id)}, D_3^{(-)}\right\} 
\quad \rightarrow \quad \widehat{G}^{(2)}= \Z_2 = 
\left\{D_1^{(\id)}, D_1^{(-)} \right\} \,.
\ee
We will see that in addition to the invariant surface defects we also get condensation defects for the lines $D_1^{(-)}$ that generate this dual symmetry. 

To start with, we need to discuss 2-group SPTs. 
As shown in \cite{Bhardwaj:2022lsg} split 2-group SPTs are given by 
\begin{enumerate}
\item 0-form SPTs for $G^{(0)}$, i.e. $\alpha \in H^2 (G^{(0)}, U(1))$
\item 1-form symmetry SPT, for the invariant part of the 1-form symmetry under $G^{(0)}$: $\beta \in \widehat{G^{(1)}}|_{G^{(0)}-\text{inv}}$. 
\end{enumerate}
So the TQFTs that we can use to stack before gauging are characterized by the following data, taking the SPTs and SSB to subgroups $H$ of $G^{(0)}$ into account: 
\be
\bm{T}_{H, \alpha_H, \beta} \,.
\ee
Lets apply this concretely in the example of the 2-group $\mathbb{G}^{(2)}$ in (\ref{Spin2gp}): 
the invariant SPTs are characters of $\Z_2 \times \Z_2$, i.e. pairs of phases $\pm 1$ which are invariant under exchange: 
\be
 \beta \in  \{ (++), (--)\} \,.
\ee
We denote these by 
\be
\bm{T}_{\Z_2}\,,\qquad \bm{T}_{\id}\,. 
\ee
These correspond to the surface defects 
\be
D_2^{(\id)}\,,\qquad D_2^{(V)} \,,
\ee
which are 1-form symmetry generators that are invariant under the $\Z_2^{(0)}$. 

Lets consider the spontaneous symmetry breaking to $H=\id$. These TQFTs have two vacua (which are mapped into each other by the broken 0-form symmetry): $v_i$, $i=1, 2$. In each vacuum we can have SPT-phases for the 1-form symmetry such that $\beta|_{v_1} = g \beta|_{v_2}$ where $g\in G^{(0)}$, i.e. the 1-form symmetry SPTs in the two vacua are related by the 0-form symmetry action. The  inequivalent configurations label three distinct TQFTs 
\be
\bm{T}_{\id, \beta}:\qquad \beta= (\text{SPT}_{v_1},\text{SPT}_{v_2} ) = (++, ++)\,, \ (--, --) \,,\ (+-, -+) \,.
\ee
They give rise to topological surface defects 
\be
D_2^{(\Z_2)}\,,\qquad D_2^{(V_{\Z_2})} \,,\qquad D_2^{(SC)} \,.
\ee
The latter can be thought of as the linear combination $D_2^{(S)}\oplus D_2^{(C)}$ in the pre-gauged theory. The remaining two are condensation defects for the dual symmetry to the 0-form symmetry 
\be
\ba
D_2^{(\Z_2)} &= C\left(D_2^{(\id)}, D_1^{(\id)} \oplus D_1^{(-)}\right) \cr 
D_2^{(V_{\Z_2})} &= C\left(D_2^{(V)}, D_1^{(\id)} \oplus D_1^{(-)}\right)\,.
\ea
\ee
The defect $D_2^{(\Z_2)}$ is precisely the defect that we have seen already in $2\Rep(\Z_2)$ as a condensation defect in section \ref{sec:Rep}. The defect $D_2^{(V_{\Z_2})}$ is the same condensation of lines, but now on the non-trivial surface defect $D_2^{(V)}$. 

Following similar reasoning as in section \ref{sec:Rep} regarding the fusion of TQFTs  the fusions for these defects is as follows:
\be
\ba
D_2^{(\id)} \otimes D_2^{(i)} &=D_2^{(i)} \cr 
D_2^{(V)} \otimes D_2^{(V)} & =D_2^{(\id)} \cr 
D_2^{(V)} \otimes D_2^{({\Z}_2)} & =D_2^{(V_{\Z_2})} \cr 
D_2^{(V)} \otimes D_2^{(V_{\Z_2})} & =D_2^{(\mathbb{Z}_2)} \cr 
\cr 
\cr 
\ea
\qquad 
\ba
D_2^{(\mathbb{Z}_2)} \otimes D_2^{(\mathbb{Z}_2)} & =2 D_2^{(\mathbb{Z}_2)} \cr 
D_2^{(V_{\mathbb{Z}_2})} \otimes D_2^{({\Z}_2)} & =2 D_2^{(V_{\mathbb{Z}_2})} \cr 
D_2^{(V_{\mathbb{Z}_2})} \otimes D_2^{(V_{{\Z}_2})} & =2 D_2^{(\mathbb{Z}_2)} \cr 
D_2^{(\mathbb{Z}_2)} \otimes D_2^{(S C)} & =2 D_2^{(S C)} \cr 
D_2^{(V_{Z_2})} \otimes D_2^{(S C)} & =2 D_2^{(S C)} \cr 
D_2^{(S C)} \otimes D_2^{(S C)} & =D_2^{({\Z}_2)} \oplus D_2^{(V_{\Z_2})}\,.
\ea
\ee
Whereas the left hand column seems invertible, the right hand column clearly is non-invertible. The first are the standard condensation defect fusion rules, whereas the last entry is the truly non-invertible fusion
\be\label{PinPlusFus}
D_2^{(S C)} \otimes D_2^{(S C)}  =D_2^{(\mathbb{Z}_2)} \oplus D_2^{(V_{\Z_2})} = C\left(D_2^{(\id)}, D_1^{(\id)} \oplus D_1^{(-)}\right) \oplus C\left(D_2^{(V)}, D_1^{(\id)} \oplus D_1^{(-)}\right) \,.
\ee
The right hand side modulo condensation is what we may have expected from the pre-gauged theory, where $D_2^{(SC)} = D_2^{(S)} \oplus D_2^{(C)}$ and we thus would derive $D_2^{(SC)} \otimes D_2^{(SC)} = 2 D_2^{(\id)} \oplus 2 D_2^{(V)}$, which is however {\it not} the correct fusion after gauging. 

Wrapped up into  a full-fledged categorical language, the symmetry of the 4d $\Pin^+(4N)$ gauge theory is the fusion 2-category
\be
2\Rep (\mathbb{G}^{(2)})= 2\Rep \left((\Z_2^{(1)} \times \Z_2^{(1)})\rtimes \Z_2^{(0)}\right)\,,
\ee
i.e. the 2-representation category of the 2-group that we started with \cite{Bhardwaj:2022yxj, Bhardwaj:2022maz, Bartsch:2022mpm}.

\paragraph{Gauging $\Z_2^{(0)}$ to $\Pin^+(4N)$: Take 2.}
We now repeat the previous gauging, but from a slightly more pedestrian way (without using the stacking with TQFTs). 
We now gauge the 0-form symmetry, which results in the $\Pin^+(4N)$ 4d gauge theory. We want to determine symmetries of this gauged theory. 

Let us start with the topological surfaces (after gauging  there are no more non-trivial topological 3d operators\footnote{This statement is only true modulo condensation defects, i.e. gauging of higher-form symmetries on 3d subspaces.}). The invariant combinations are 
\be
\cC_{\text{Pin}^+(4N)}^{\text{2d top ops}}=\left\{D_2^{(\id)},D_2^{(SC)},D_2^{(V)}\right\}\,,
\ee
where we define 
\be
D_2^{(SC)} = D_2^{(S)} \oplus D_2^{(C)} \,.
\ee
This is invariant under the 0-form symmetry and becomes a simple (irreducible) surface defect after gauging. 

Next, consider the topological lines in the gauged theory.
$G^{(0)}$ of $\Spin(4N)$ is an abelian group-like symmetry, so we get a dual 2-form symmetry, which is generated by a topological line 
\be
\widehat{G}^{(2)} = \Z_2 \,,\qquad \text{generated by } D_1^{(-)} \,.
\ee

The main principle for determining the lines is the following: we need to take gauge-invariant combinations. If a line is fixed by a subgroup $H$ of $G$, then we can in addition dress these lines by an irreducible representation of $H$. 

In this concrete example define the invariant combination $D_1^{(S)} \oplus D_1^{(C)} = D_1^{(SC)}$, on the surface $D_2^{(SC)}$. On $D_2^{(V)}$ the line  $D_1^{(V)}$ invariant (with stabilizer $\Z_2$), however there is now a second line, $D_1^{(V-)}$, which is obtained by dressing $D_1^{(V)}$ with the dual $\Z_2$ line (i.e. the non-trivial representation of $\Z_2$). In summary we have the lines 
\be
\cC_{\text{Pin}^+(4N)}^{\text{1d top ops}}=\left\{D_1^{(\id)},D_1^{(-)},D_1^{(SC)},D_1^{(V)},D_1^{(V_-)}\right\} \,.
\ee
Note that the fact that we cannot dress $D_1^{(SC)}$ with $D_1^{(-)}$, which in turn means that we can trivially absorb the line $D_1^{(-)}$ into the surface $D_2^{(SC)}$.

Let us consider the fusion of these defects. Naively, in the $\Spin(4N)$ theory we would compute the following fusion for the surfaces 
\be
\Spin(4N): \qquad D_2^{(SC)}\otimes D_2^{(SC)}=2 D_2^{(\id)}\oplus 2 D_2^{(V)} \,.
\ee
This is however not the correct fusion in $\Pin^+(4N)$!

\begin{figure}
\centering
\begin{tikzpicture}
\begin{scope}[shift={(0,0)}]
\draw [cyan, fill=cyan,opacity=0.1](0,1) -- (0,4) -- (1.5,5) -- (1.5,2) --  (0,1);
 \draw [cyan, thick]  (0,1) -- (0,4) -- (1.5,5) -- (1.5,2) --  (0,1);
\begin{scope}[shift={(0,1)}]
\draw [LimeGreen, fill =LimeGreen, opacity =0.1] (0,1.5) -- (1.5,2.5)  -- (3.5,2.5) --(2,1.5) -- (0,1.5);
\draw [LimeGreen, thick] (0,1.5) -- (1.5,2.5)  -- (3.5,2.5) --(2,1.5) -- (0,1.5);
\node [red, left] at (0,1.5) {$D_1^{(S, S) \to \id}$} ;
 \draw [red, ultra thick] (0,1.5) --(1.5,2.5) ;
\node [LimeGreen]at (3.5,2) {$D_{2}^{(\id)}$} ;
\end{scope}
\node [cyan] at (1, 3.8) {$D_2^{(S)}$} ;
\node [cyan] at (1,2) {$D_{2}^{(S)}$} ;
\end{scope}
\begin{scope}[shift={(6,0)}]
\draw [PineGreen, fill=PineGreen,opacity=0.1](0,1) -- (0,4) -- (1.5,5) -- (1.5,2) --  (0,1);
 \draw [PineGreen, thick]  (0,1) -- (0,4) -- (1.5,5) -- (1.5,2) --  (0,1);
\begin{scope}[shift={(0,1)}]
\draw [LimeGreen, fill =LimeGreen, opacity =0.1] (0,1.5) -- (1.5,2.5)  -- (3.5,2.5) --(2,1.5) -- (0,1.5);
\draw [LimeGreen, thick] (0,1.5) -- (1.5,2.5)  -- (3.5,2.5) --(2,1.5) -- (0,1.5);
\node [red, left] at (0,1.5) {$D_1^{(C, C) \to \id}$} ;
 \draw [red, ultra thick] (0,1.5) --(1.5,2.5) ;
\node [LimeGreen]at (3.5,2) {$D_{2}^{(\id)}$} ;
\end{scope}
\node [PineGreen] at (1, 3.8) {$D_2^{(C)}$} ;
\node [PineGreen] at (1,2) {$D_{2}^{(C)}$} ;
\end{scope}

\end{tikzpicture}
\caption{The junction (1-morphism) between $D_2^{(S)}$ and $D_2^{(S)}$ and $D_2^{(\id)}$, and $D_2^{(C)}\otimes D_2^{(C)}$ and $D_2^{(\id)}$. The two 1d junctions $D_1$ get exchanged under the action of the  outer automorphism $\Z_2^{(0)}$.\label{fig:SSid}}
\end{figure}
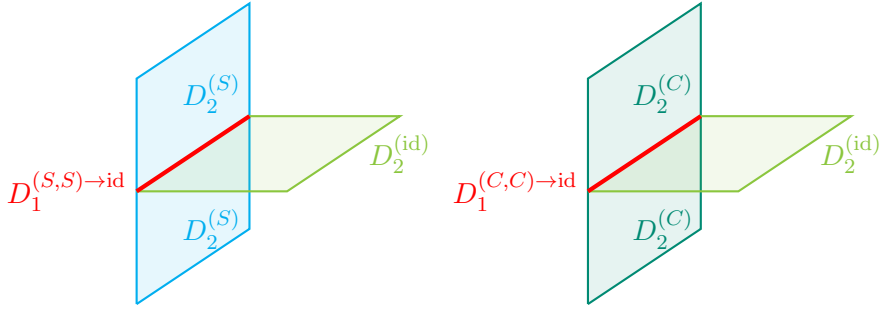

We need to be careful what the precise set of 1d interfaces (in categorical terms: morphisms) is. In the $\Spin(4N)$ theory
we have 1d interfaced from $D_2^{(SC)} \otimes D_2^{(SC)}$ to  $D_2^{(\id)}$ and $D_2^{(V)}$, respectively. Let us start with the identity surface defect and determine the morphisms: 
\be
D_2^{(SC)} \otimes D_2^{(SC)}  \to D_2^{(\id)} \,.
\ee
We can details lines that are junctions between $S, C,\id$ surfaces using the standard composition rules:
\be
\ba
D_1^{(S, S) \to \id} : &\qquad D_2^{(S)} \otimes D_2^{(S)} \to D_2^{(\id)} \cr 
D_1^{(C, C) \to \id} : &\qquad D_2^{(C)} \otimes D_2^{(C)} \to D_2^{(\id)} \,.
\ea
\ee
This is shown in figure \ref{fig:SSid}. 
However the $\Z_2$ outer automorphism acts by exchange of these
\be
\Z_2^{(0)}:\qquad D_1^{(S, S) \to \id} \longleftrightarrow D_1^{(C, C) \to \id} \,.
\ee
Likewise there are two junctions to $V$: 
\be
\ba
D_1^{(S, C) \to V} : &\qquad D_2^{(S)} \otimes D_2^{(C)} \to D_2^{(V)} \cr 
D_1^{(C, S) \to V} : &\qquad D_2^{(C)} \otimes D_2^{(S)} \to D_2^{(V)} \,.
\ea
\ee
However they are again exchanged under the outer automorphism. 
Thus, there is only one gauge-invariant morphism to $\id$ and to $V$, which means the fusion in the gauged theory is 
\be
\Pin^+(4N): \qquad D_2^{(SC)}\otimes D_2^{(SC)}= D_2^{(\id)}\oplus  D_2^{(V)}  \qquad \text{modulo condensation}\,.
\ee
Some remarks are in order. This is clearly a non-invertible fusion.  The fusion written above is what one could call ``local" fusion. Technically this means we consider the fusion modulo condensation (i.e. gauging of symmetries on the 2d surfaces). The full fusion including the condensation replaces the RHS with surfaces including condensation of the lines $D_1^{(\id)}\oplus D_1^{(-)}$ on $D_2^{(\id)}$ and $D_1^{(V)} \oplus D_1^{(V_-)}$ on $D_2^{(V)}$.

Similarly one can infer that the fusion of the other defects is 
\be
D_2^{(SC)} \otimes D_2^{(i)} = D_2^{(SC)}  \,,\qquad i=\id, V \,.
\ee
We also can compute the fusion of the lines, by again considering invariant local operators and find the non-invertible fusion
\be
D_1^{(SC)} \otimes D_1^{(SC)} =  D_1^{(\id)} \oplus D_1^{(-)}\oplus  D_1^{(V)} \oplus D_1^{(V_-)} \,.
\ee
This equation is to be interpreted as follows: the lines on the surface $D_2^{(SC)}$ fuse to lines in the identity surface $D_2^{(\id)}$, namely $D_1^{(\id)}$ and $D_1^{(-)}$, as well as to lines in the surface $D_2^{(V)}$, which are $D_1^{(V)}$ and $D_1^{(V-)}$. This results then in the same fusion as in (\ref{PinPlusFus}). 

\bigskip

Several applications of this gauging of outer automorphisms have appeared in the past year, in particular using the categorical formulation \cite{Antinucci:2022eat, Decoppet:2022dnz, Bhardwaj:2022kot, Bhardwaj:2022maz, GarciaEtxebarria:2022jky, Bartsch:2022ytj, Delcamp:2023kew, Kaidi:2023maf}.

\subsubsection{Gauging Outer Automorphisms 2: 4d $O(2)$ YM}
\label{sec:O2}

With these detailed discussion on outer automorphism gaugings, we can now return to the example of the $O(2)$ gauge theory, which we encountered in section \ref{sec:O2Take1}, and with our improved understanding of non-invertible symmetries, determine the correct symmetry structure. This is an example of a continious non-invertible symmetry. 

Recall that we start with the $U(1)$ theory that has topological (electric) 1-form symmetry generators $D_2^{(\theta)}$, $\theta\in [0, 2\pi]$, with  $\theta \sim  \theta + 2\pi$.
The $\Z_2^{(0)}$ charge conjugation leaves invariant $D_2^{(0)}$ and $D_2^{(\pi)}$, while exchanging for all other values 
\be
D_2^{(\theta)}\longleftrightarrow D_2^{(-\theta)}\,.
\ee
Gauging this, results in the $O(2)$ gauge theory, and we would like to determine the symmetry defects. 
In terms of surface defects we have  
 \be
\cC_{O(2)}^{\text{2d top ops}}=\left\{ D_2^{(0)}, D_2^{(\pi)}, D_2^{(\theta, +)}\, \ 0<\theta<\pi \right\} \,,
\ee
where
\be
D_2^{(\theta,+)}=\left( D_2^{(\theta)} \oplus D_2^{(-\theta)} \right) \,,
\ee
as a surface in the $U(1)$ theory (it becomes an irreducible, simple, object in $O(2)$). 
The fusions of surfaces can be determined completely analogously to the analysis we carried out in the $\Spin(4N)$ gauge theory in the last section, and results in the following fusions:
\be
\ba
D_2^{(\pi)} \otimes  D_2^{(\pi)} 
&= D_2^{(0)}\\
D_2^{(\pi)} \otimes  D_2^{(\theta,+)} 
&= D_2^{(\pi-\theta, +)}\\
D_2^{(\theta, +)}\otimes D_2^{(\theta',+)} 
&= D_2^{\left(\theta+\theta', +\right)}\oplus D_2^{\left(\theta-\theta', +\right)}\\
D_2^{(\theta, +)}\otimes D_2^{(\pi-\theta, +)} 
&=  {D_2^{(\pi)}\over \Z_2}\oplus D_2^{\left(2\theta-\pi, +\right)}\,, \quad\theta\neq \pi/2\\
D_2^{(\theta, +)}\otimes D_2^{(\theta, +)} 
&= {D_2^{(0)}\over \Z_2}\oplus D_2^{\left(2\theta, +\right)}\,, \quad\theta\neq \pi/2\\
D_2^{(\pi/2)}\otimes D_2^{(\pi/2)} &= {D_2^{(0)}\over \Z_2}\oplus {D_2^{(\pi)} \over \Z_2} \,,
\ea
\ee
where we have $\theta'\neq \pi -\theta$ and $\theta'\neq\theta$.
Here, we included the condensation defects denoting them by
$D_2^{(0)}/\Z_2$ and $D_2^{(\pi)}/\Z_2$, which corresponds to condensing the dual line $D_1^{-}$ and $D_1^{\pi, -}$ on the respective surfaces
\cite{Bhardwaj:2022yxj}.

Likewise the topological lines are 
\be
\cC_{O(2)}^{\text{1d top ops}}=\left\{ D_1^{(0)}, D_1^{(-)}, D_1^{(\pi)}, D_1^{(\pi,-)}, D_1^{(\theta)}\,\middle\vert\, 0<\theta<\pi \right\} \,,
\ee
where $D_1^{(0)}$, $D_1^{(\pi)}$ and $D_1^{(\theta)}$ are the lines on $D_2^{(0)}$, $D_2^{(\pi)}$ and $D_2^{(\theta)}$ respectively, and
\be
D_1^{(\theta, +)}= D_1^{(\theta)} \oplus D_1^{(-\theta)} \,,
\ee
which is the invariant combination, which becomes an irreducible line in $O(2)$. Since $D_1^{(0)}$ and $D_1^{(\pi)}$ are invariant under  $\Z_2$, we can attach $D_1^{(-)}$, to get the new lines $D_1^{(-)}$ in $D_2^{(\id)}$, and $D_1^{(\pi,-)}$ in  $D_2^{(\pi)}$. 

Again one can repeat the arguments similar to those we made in the $\Pin^+(4N)$ theory and obtain the fusion of lines: 
\be\label{O2Lines}
\ba
D_1^{(-)} \otimes  D_1^{(\theta, +)} &= D_1^{(\theta, +)}\\
D_1^{(\pi)} \otimes  D_1^{(\theta, +)} &= D_1^{(\pi -\theta, +)} \\
D_1^{(\pi,-)} \otimes D_1^{(\theta, +)} &= D_1^{(\pi-\theta, +)}\\
D_1^{(\theta, +)}\otimes D_1^{(\theta', +)} &= 
D_1^{(\theta+\theta', +)}\oplus D_1^{(\theta- \theta', +)}\\
D_1^{(\theta, +)}\otimes D_1^{(\pi -\theta, +)} &= D_1^{(\pi)}\oplus D_1^{(\pi,-)}\oplus D_1^{(2\theta-\pi, +)}
\,, \quad\theta\neq\pi/2\\
D_1^{(\theta, +)}\otimes D_1^{(\theta, +)} &= D_1^{(0)}\oplus D_1^{(-)}\oplus D_1^{(2\theta, +)}\,, \qquad\theta\neq \pi/2\\
D_1^{(\pi/2, +)}\otimes D_1^{(\pi/2, +)} &= D_1^{(0)}\oplus D_1^{(-)}\oplus D_1^{(\pi)}\oplus D_1^{(\pi,-)} \,,
\ea
\ee
where  again $\theta'\neq \pi -\theta$ and $\theta'\neq\theta$. Again the last line confirms that we need to include the condensation defects into the fusion of $D_2^{(\pi/2)}$ with itself, ad otherwise the fusion of  $D_1^{\pi/2, +}$ lines on these surfaces (last line in (\ref{O2Lines})) is inconsistent. 

We emphasize that the fusion is an operation that happens at each level of the symmetry category, i.e. within each fixed dimension of topological defects (rather than fusion of surfaces resulting in lines etc as initially proposed in \cite{Heidenreich:2021xpr}). In this sense, the structure of higher fusion categories becomes indispensable intertwined with generalized symmetries.

\subsection{Duality Defects}

A closely related set of defects are the duality defects \cite{Choi:2021kmx, Kaidi:2021xfk}. 
The main idea is extremely simple and elegant: Consider a theory $\cT$ with a $p$-form symmetry $G^{(p)}$ that does not have an 't Hooft anomaly. 
Also assume that $\cT$ is invariant under gauging this $p$-form symmetry
\be
\cT \cong \cT/G^{(p)} \,,\qquad \text{therefore } p={d-2\over 2} \,,
\ee
so that the dual symmetry is of the same type as the original symmetry.
Consider the theory, with spacetime split into two halves by a codimension 1 wall, and gauge $G^{(p)}$ only on one half. Impose Dirichlet boundary conditions on the codimension 1 subspace for the gauge field for $G^{(p)}$. This defines a topological defect $D_{d-1}$ -- the duality defect.  

This works by the self-duality constraint in 2d and 4d, in particular. Indeed, the main motivation  for this construction comes from the Kramers-Wannier duality in the 2d Ising model. Let us recap this briefly. 

\paragraph{Kramers-Wannier duality in the 2d Ising.}
Recall, that the Ising model  has three topological lines:
\be
D_1^{(\id)}\,,\quad D_1^{(-)} \,,\quad D_1^{(S)} 
\ee
with fusion 
\be\ba\label{Isingfu}
D_1^{(\id)} \otimes D_1^{(-)} &= D_1^{(-)} \,,\qquad 
D_1^{(-)} \otimes D_1^{(-)} = D_1^{(\id)} \cr 
D_1^{(S)} \otimes D_1^{(-)} &= D_1^{(S)} \otimes D_1^{(\id )} = D_1^{(S)} \cr
D_1^{(S)} \otimes D_1^{(S)} & = D_1^{(\id)} \oplus D_1^{(-)} \,.
\ea
\ee
We can think of this as a self-dual theory under the gauging of the 0-form symmetry $\Z_2^{(0)}$, generated by the line $D_1^{(-)}$. The duality defect is $D_1^{(S)}$. 
Gauging the 0-form symmetry on a half-space and requiring that the background $b_1=0$ (Dirichlet) on the 1d wall, implies that the fusion with the duality defect leaves that invariant.

\paragraph{Duality Defects in 4d.}
The duality defects in 4d arise when gauging a 1-form symmetry. Assume that the theory is invariant (note that we need to have invariance of the absolute theory here, so if there is a choice of polarization then one has to impose that this is also invariant) under the gauging of a 1-form symmetry. 
Given that 
\be
\cT \cong \cT/G^{(1)}\,,
\ee
we can repeat the construction of the defects. Let $D_2^{(g)}= e^{i \int b_2}$. Then the  1-form symmetry background has again Dirichlet boundary conditions $b_2=0$ at the defect and thus  $D_2^{(g)} \otimes D_3^{(\text{duality})} = D_3 ^{(\text{duality})}$. 
Denote by $\overline{D_3}^{(\text{duality})}$ the conjugate obtained by orientation reversal (the two defects separate a slab of $G^{(1)}$ gauged space with a the rest where the symmetry is not gauged). 
Let us consider for simplicity $G^{(1)} = \Z_N$. 
Then 
as in the 2d case 
\be\ba
D_3^{(\text{duality})}(M_3) \otimes \overline{D_3}^{(\text{duality})} (M_3)
=&{1\over N} \sum_{M_2 \in H_2 (M_3, G^{(1)})}  D_2^{(1)} (M_2 ) D_3^{(\id)}  \cr
=& C(M_3, \Z_N^{(1)})  \,.
\ea\ee
This is of course very reminiscent of the defects in the section on gauging with mixed anomalies. Indeed, when the theory is self-dual and has a self-dual polarization, then this is the same construction. Otherwise the duality defects are not of the type that they can be related by gauging to an invertible symmetry. Such symmetries were (regrettably) coined {\it (non-) intrinsically non-invertible}. 
Many examples of this type have appeared by now \cite{Kaidi:2022cpf, Bashmakov:2022uek, Antinucci:2022vyk, Heckman:2022xgu}.


\section{From Representations to Higher-Charges}
\label{sec:Charges}

So far we have been focused on the structure of the symmetries -- i.e. the topological operators. The generalizations from standard 0-form symmetries to higher-form, and higher-group, but also non-invertibles is vast. 

We now address the question: what are the ``generalized charges", i.e. the operators (usually extended) that carry charge under generalized symmetries -- largely following the recent works \cite{Bhardwaj:2023wzd, Bhardwaj:2023ayw} (see also \cite{Bartsch:2023pzl}). 

\paragraph{Group-Like Symmetries and Representations.}
For group-like symmetries, e.g. 0-from, the charges transform in representations of the group. This is fairly standard and well-established. Also for $p$-form symmetry groups $G^{(p)}$, the charges are naturally $p$-dimensional extended operators and transform in representations of $G^{(p)}$. For higher-form symmetries, recall this is due to the fact that the codimension $p+1$ topological operators 
\be
D_{d-p-1}^{(g)} \,,\quad g\in G^{(p)}\,,
\ee
link with $p$-dimensional defects $\cO_p$. 
We expect the charges to be 
\be
D_{d-p-1}^{(g)} \cO_p = \bm{R}_g(\cO_p) \cO_p  \,.
\ee
Here $\bm{R}_g$ is a representation of $G^{(p)}$ evaluated on the element $g$. E.g. for abelian groups (which most higher-form symmetries are), this is a phase.

We will answer the following question: 
{\it What is the most general set of charges of a $p$-form symmetry?}
The curious and surprising insight is that the above ``$p$-form symmetry has $p$-dimensional charges" mantra is only a subset of generalized charges. 

Let first fix some terminology and define:
\be
\text{Generalized charges given by $p$-dimensional operators $\cO_p$ will be called $p$-charges.}
\ee
So the standard statement about $p$-dimensional operators being charged under $p$-form symmetries is:
\be
\text{The $p$-charges of $p$-form symmetries $G^{(p)}$ are representations of $G^{(p)}$. }
\ee
We will see that there are $q$-charges for $p$-form symmetries, with $q\not=p$. In short the statement is \cite{Bhardwaj:2023wzd}
\begin{tcolorbox}[colback=white, colframe=black!50, rounded corners]
\centering
{$q$-charges of a $p$-form symmetry $G^{(p)}$ form $(q+1)$-representations of $G^{(p)}$.}
\end{tcolorbox}
In these lectures we will focus only on invertible 0-form symmetries and genuine charges. An extension to other invertible higher-form symmetries or higher-groups and their $q$-charges (twisted or untwisted) can be found in \cite{Bhardwaj:2023wzd} and to non-invertibles in \cite{Bhardwaj:2023ayw}.

\subsection{Higher-Charges: 0-Form Symmetry}

We have already seen examples of generalized charges for 0-form symmetries. When the theory has a 0-form outer automorphism or charge conjugation, this does not only act on local operators but also line operators. 

E.g. for the $U(1)$ gauge theory in 4d, see section \ref{sec:O2}, we considered the action of the charge conjugation, which maps 
\be
A \to - A \,,
\ee
and thus acts on the 1-form symmetry generators as 
\be
D_2^{(\theta, i)} \to D_2^{(-\theta, i)} \,,
\ee
where $i= E, M$, but also it act on the Wilson line operators 
\be
W_\alpha= P e^{i \alpha \int A}  \to W_{-\alpha} \,.
\ee
Similarly, it acts on the 't Hooft lines $H_a$ (replacing $A$ with $A^D$).

Clearly with the earlier terminology that we agreed upon we have: 
\be
\ba
F= dA:\qquad & \text{0-charge of $G^{(0)}= \Z_2^{(0)}$}\cr
W_\alpha :\qquad &\text{1-charges of $G^{(0)}= \Z_2^{(0)}$}\cr
H_a :\qquad &\text{1-charges of $G^{(0)}= \Z_2^{(0)}$}\cr 
D_2^{(\theta, i)} : \qquad & \text{2-charges of $G^{(0)}= \Z_2^{(0)}$} \,.
\ea
\ee
Another example are the Wilson line operators in $\Spin(4N)$ gauge theories, in the spinor and co-spinor representations, which are a 1-charge of the outer automorphism.

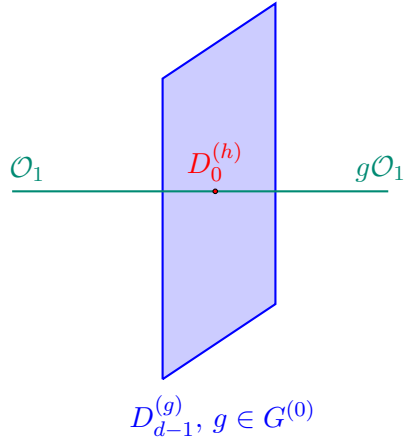
\begin{figure}
\centering
\begin{tikzpicture}
\draw [blue, fill=blue,opacity=0.2] 
(0,0) -- (0,4) -- (1.5,5) -- (1.5,1) -- (0,0);
 \draw [blue, thick] 
 (0,0) -- (0,4) -- (1.5,5) -- (1.5,1) -- (0,0);
 \draw [PineGreen, thick] (0.7,2.5) -- (3, 2.5) ;
 \draw [PineGreen, thick] (-2,2.5) -- (0.7, 2.5) ;
\node [PineGreen] at (-1.8, 2.8) {$\cO_1$} ;
\node[blue] at (0.8,-0.5) {$D^{(g)}_{d-1}, \, g\in G^{(0)}$};
\node [PineGreen] at (2.9, 2.8) {$g \cO_1$} ;
 \draw [black,fill=red] (0.7,2.5) ellipse (0.03 and 0.03);
 \node[above, red] at (0.7, 2.5) {$D_0^{(h)}$};
\end{tikzpicture}
\caption{The action of $g\in G^{(0)}$ on a $1$-dimensional operator $\cO_1$, transforming it into $g \cO_1$. 
The junction $D_0^{(h)}$ are labeled by $h\in H_{\cO_1}$, the stabilizer group of $\cO_1$.\label{fig:GonO}}
\end{figure}

\begin{figure}
\centering
\begin{tikzpicture}
\begin{scope}[shift={(0,0)}]
\draw [PineGreen, thick] (0,0) -- (5, 0) ;
\draw [blue, thick] (1.5,2.5) -- (1.5, -2.5) ;
\draw [blue, thick] (3.5,2.5) -- (3.5, -2.5) ;
\node[PineGreen] at (-0.5, 0) {$\cO_1$};
\node[blue] at (3, 2.5) {$D_{d-1}^{(h_2)}$};
\node[blue] at (1, 2.5) {$D_{d-1}^{(h_1)}$};
\node [red] at (2, 0.4) {$D_{0}^{(h_1)}$} ;
\node [red] at (4, 0.4) {$D_{0}^{(h_2)}$} ;
\draw [thick,red,fill=red] (1.5, 0) ellipse (0.05 and 0.05);
\draw [thick,red,fill=red] (3.5, 0) ellipse (0.05 and 0.05);
\end{scope}
\begin{scope}[shift={(8,0)}]
\node [] at (-1,0) {$= \ \sigma (h_1, h_2)~\times$};
\draw [PineGreen, thick] (0.5,0) -- (5, 0) ;
\draw [blue, thick] (2.5,2.5) -- (2.5, -2.5) ;
\node[PineGreen] at (5.5, 0) {$\cO_1$};
\node[blue] at (1.9, 2.5) {$D_{d-1}^{(h_1 h_2)}$};
\node [red] at (3.2, 0.4) {$D_{0}^{(h_1 h_2)}$} ;
\draw [thick,red,fill=red] (2.5, 0) ellipse (0.05 and 0.05);
\end{scope}
\end{tikzpicture} 
\caption{The induced 0-form symmetry $H$ has fusion, which is modified by the cocycle $\sigma \in H^2(H, U(1))$.  \label{fig:cacao}}
\end{figure}
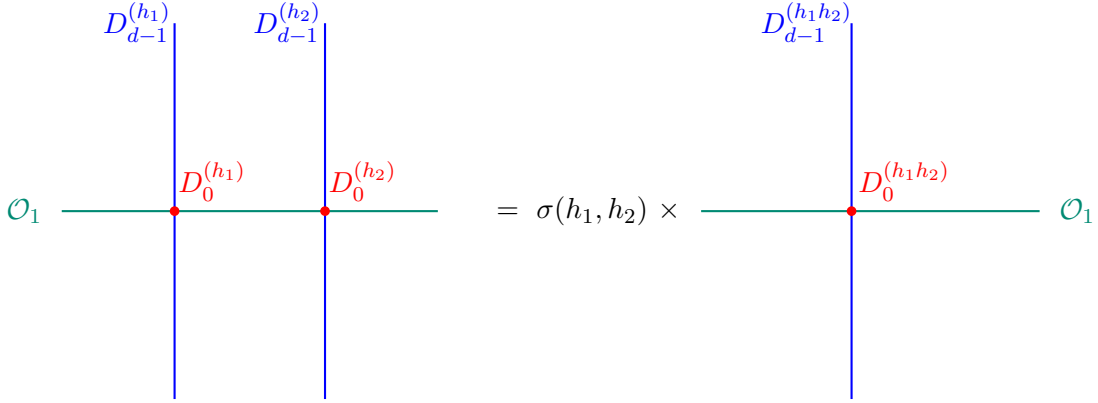

\paragraph{1-charges for $G^{(0)}$.} We can give a precise characterization of the 1-charges of 0-form symmetries. Consider a line operator $\cO_1$. A 0-form symmetry generator $D_{d-p-1}^{(g)}$ acts on this by sending it possibly to another operator $g \cO_1$. 
Let us denote all operators that can be reached by $G^{(0)}$-action in this way from one operator $\cO_1$ as $M_{\cO_1}$. These orbits have a stabilizer group  $H_{\cO_1}$, i.e. 
\be
h \cO_1 =\cO_1 \,,\qquad \text{ for all } h\in H_{\cO_1} \,.
\ee
The intersection of the line $\cO_1$ and the 0-form symmetry generator gives rise to a 0-form generator $D_0^{(h)}$ on the line. These topologial operators do not have to fuse according to the group law, but can have an anomaly: 
\be
D_0^{(h)} \otimes  D_0^{(h')} = \sigma (h, h') D_0^{(hh')}\,, 
\ee
where $\sigma \in H^2 (H_{\cO_1}, U(1))$. This is depicted in figure \ref{fig:cacao}.

As was shown in \cite{Bhardwaj:2023wzd} this is all we need to uniquely characterize the  1-charges for $G^{(0)}$: 
\be
(H, \sigma)\,,\qquad H \text{ subgroup of }G\,, \ \sigma \in H^2 (H, U(1))\,.
\ee
Mathematically, (as we already have seen in section \ref{sec:Rep}) this is what characterizes a 2-representation of a group! 
We will call this a dimension $n$ 2-representation, where 
$n$ is the number of cosets $G/H$ in $G$. 
So we have seen: 
\be
\text{1-charges of 0-form symmetries $G^{(0)}$ are 2-representations $\brho^{(2)}$ of $G^{(0)}$.}
\ee
We indicate $k$-representation by $\brho^{(k)}$.

\paragraph{Example: $2\Rep (\Z_2)$ reloaded.} 
We studied the 2-representations of $\Z_2$ in the context of theta defects in section \ref{sec:Rep}. Let us recap: 
$H= 1, \Z_2$ and there is no cocycle in either case. So there are two 2-representations: 
\be
\brho^{(2)}_1= (H=1, \sigma=0) \,,\qquad 
\brho^{(2)}_{\Z_2}= (H =\Z_2, \sigma=0) \,.
\ee
So we should identify these e.g. in the field theories with 0-form outer automorphism. 

For 4d $U(1)$ with $\Z_2$ the charge conjugation, $W_\alpha$ and $W_{-\alpha}$ are in the same orbit. The stabilizer is only trivial for $\alpha=0, \pi$, which are in $\rho^{(2)}_{\Z_2}$. For all other $\alpha$ we have $W_{\alpha} \oplus W_{-\alpha}$ in the non-trivial 2-representation $\rho^{(2)}_{1}$. 

Likewise for $\Spin(4N)$ the spinor and co-spinor Wilson lines form the 2-representation $\rho^{(2)}_{1}$, whereas for the vector representation Wilson line, $W_V$, we have the 2-representation $\rho^{(2)}_{\Z_2}$ under the outer automorphism $\mathbb{Z}_2^{(0)}$, see (\ref{SpinOut}).

\paragraph{$q$-Charges are $(q+1)$-representations of $G^{(0)}$.}
These observations generalize to $q$-charges of 0-form symmetries in the following way: 
Consider $q$-dimensional operators $\cO_q$, then again a 0-form symmetry will map this to another operator $g \cO_q$. The orbits of this type will have a stabilizer group $H_{\cO_q}$.
The $q$-charges for $G^{(0)}$ are 
\be
\brho^{(q+1)}= (H, H^{q+1} (H, U(1))) \,,
\ee
which are again the stabilizer groups $H$ of orbits of $q$-dimensional operators $\cO_q$, and cocycles, and not surprisingly these are called $(q+1)$-representations of the group $G^{(0)}$.
We should think of $H$ as the 0-form symmetry that is induced on the defects $\cO_q$ (at the intersection with the 0-form symmetry generators).
 A word of caution however in that these are not all $q$-charges in general -- only the ones of group cohomology type.

Extending this analysis to the most general higher form symmetry we can in fact show that 
\be
\text{$q$-charges of higher-form $G^{(p)}$ are $(q+1)$-representations $\brho^{(q+1)}$ of $G^{(p)}$.}
\ee
Extensions to higher-groups can be found in \cite{Bhardwaj:2023wzd}, as well as the important discussion of non-genuine and twisted charges, i.e. charges, which are boundaries of higher-dimensional defects.

\subsection{Symmetry Fractionalization}

So far we considered topological defects of the (bulk) QFT, e.g. for a $G^{(0)}$ 0-form symmetry the $D_{d-1}^{(g)}$ generators, which give rise to induced symmetries on the defects $\cO_q$. This is what we would call a cohomology type $q$-charge.
 However, for $q>1$ there can be in addition localized symmetries on the defects themselves\footnote{For $q=1$, assuming that the line operator is simple (``irreducible"), there cannot be a non-trivial operator $D_0$.}. We will discuss the simplest instance which is surface operators in a theory with 0-form symmetry, i.e. $2$-charges. 

This effect is called symmetry fractionalization, which means  that a symmetry -- here $G^{(0)}$ --  of the theory gets enlarged, when restricted to an operator $\cO_q$. The larger group, $G^{(0), \text{frac}}$ arises because the defect itself may have topological lines that are not arising from restrictions of topological defects of the bulk (they are localized symmetries). In the context of condensed matter physics this was already observed in \cite{Barkeshli:2014cna}. 
 
\begin{figure}
\centering
\begin{tikzpicture}
\draw [blue, fill=blue,opacity=0.1] (0,0) -- (0,4) -- (1.5,5) -- (1.5,1) -- (0,0);
 \draw [blue, thick] (0,0) -- (0,4) -- (1.5,5) -- (1.5,1) -- (0,0);
 \draw [Green, thick] (0,3) -- (1.5, 4) ;
\draw [orange, fill=orange,opacity=0.1] (-2, 1.5) -- (2,1.5) -- (3.5, 2.5) -- (-0.5, 2.5) -- (-2, 1.5);
\draw [orange, thick] (-2, 1.5) -- (2,1.5) -- (3.5, 2.5) -- (-0.5, 2.5) -- (-2, 1.5);
\draw [purple, thick] (0, 1.5) -- (1.5, 2.5) ;
\node[blue] at (0.8,-0.5) {$\mathcal{O}_2$};
\node [orange] at (3,1.5) {$D_{d-1}^{(g)}$} ;
\node [Green] at (0.8,3.9) {$D_{1}^{\text{loc}}$} ;
\node [purple] at (2,3) {$D_{1}^{\text{ind}}$} ;
\end{tikzpicture}
\caption{Localized $D_1^{\text{loc}}$ (green) and induced $D_1^{\text{ind}}$ (purple) symmetries on the 2d operator $\mathcal{O}_2$. Localized symmetries exist on the surface defect, whereas induced ones arise as topological lines that are the intersection with the 0-form symmetry generators $D_{d-1}^{(g)}$. \label{fig:indu}}
\end{figure}
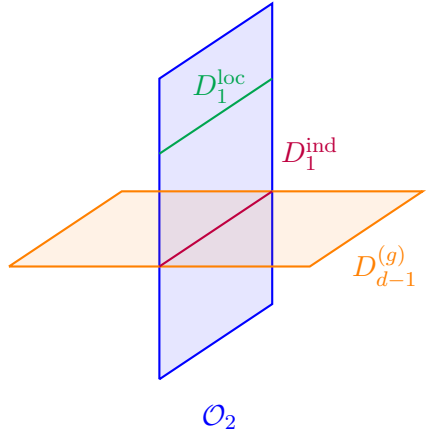

The distinction between induced and localized symmetries is illustrated in figure \ref{fig:indu}:
\begin{itemize}
\item Localized: topological defects that are localized on the operator $\cO_2$ 
\item Induced: topological defects that arise as intersections of $\cO_2$ with bulk 0-form symmetry generators $D_{d-1}^{(g)}$.
\end{itemize}

A $2$-charge $\cO_2$ for a 0-form symmetry $G^{(0)}$ that admits such symmetry fractionalization is clearly beyond the 2-representations that we studied earlier. These generalizations are specified by the following data: 
\begin{itemize}
\item Bulk 0-form symetry $G^{(0)}$, and resulting $D_1^{(g)}$ induced line operators on the 2-charge 
\item Symmetry $\cC^{\text{frac}}$ that is the enhanced (fractionalized) symmetry, which has a map to $G^{(0)}$. If this is group like, then we need a sequence 
\be
1 \to G^{(0), \text{loc}} \to G^{(0) ,\text{ind}} \stackrel{\pi}{\to} G^{(0)}  \to 1 \,,
\ee
where $G^{(0), \text{loc}}$ is the localized symmetry, and $G^{(0) ,\text{ind}}$ the induced one. 

However we will see that the fractionalized symmetry need not be group like, but can be non-invertible, e.g. the symmetry can fractionalize to a fusion category symmetry. 

\item An anomaly, which in the group case is  $\omega \in H^{3} (G^{(0), \text{ind}}, U(1))$. 
\end{itemize}
It is natural to ask what the relation between these higher charges and higher-representations and categories is. The statement (which is substantiated in \cite{Bhardwaj:2023wzd}) is that $q$-charges are $(q+1)$-representations with associated $G^{(0)}$-graded fusion $(q-1)$-category $(q-1)\Vec_{G^{(0),\text{ind}}}^\omega$. 
An extension to different dimensions and higher-form symmetries/higher groups is also found there. 

Let us finish this with two examples: one which is a group-like fractionalization, the other which fractionalizes to a non-invertible symmetry!

\begin{figure}
\centering
\begin{tikzpicture}
\draw [blue, fill=blue,opacity=0.1] (0,0) -- (0,4) -- (1.5,5) -- (1.5,1) -- (0,0);
 \draw [blue, thick] (0,0) -- (0,4) -- (1.5,5) -- (1.5,1) -- (0,0);
\begin{scope}[shift={(0,-0.5)}]
\draw [orange, fill=orange,opacity=0.1] (-2, 1.5) -- (2,1.5) -- (3.5, 2.5) -- (-0.5, 2.5) -- (-2, 1.5);
\draw [orange, thick] (-2,1.5) -- (2,1.5) -- (3.5,2.5) -- (-0.5,2.5) -- (-2,1.5);
\draw [purple, thick] (0,1.5) -- (1.5,2.5) ;
\node [purple] at (-0.5,1) {$D^{(i)}_{1}$} ;
\node [orange] at (3,1.5) {$D_{d-1}^{(-)}$} ;
\end{scope}
\node[blue] at (0.8,-0.5) {$\mathcal{O}_2$};
\begin{scope}[shift={(0,1)}]
\draw [orange, fill=orange,opacity=0.1] (-2, 1.5) -- (2,1.5) -- (3.5, 2.5) -- (-0.5, 2.5) -- (-2, 1.5);
\draw [orange, thick] (-2,1.5) -- (2,1.5) -- (3.5,2.5) -- (-0.5,2.5) -- (-2,1.5);
\draw [purple, thick] (0,1.5) -- (1.5,2.5) ;
\node [purple] at (2,3) {$D^{(i)}_{1}$} ;
\node [orange] at (3.5,2) {$D_{d-1}^{(-)}$} ;
\end{scope}
\node at (5,2.5) {=};
\begin{scope}[shift={(6.5,0)}]
\draw [blue, fill=blue,opacity=0.1] (0,0) -- (0,4) -- (1.5,5) -- (1.5,1) -- (0,0);
 \draw [blue, thick] (0,0) -- (0,4) -- (1.5,5) -- (1.5,1) -- (0,0);
 \node[blue] at (0.8,-0.5) {$\mathcal{O}_2$};
 \node [Green] at (2,3) {$D^{(-)}_{1}$} ;
\end{scope}
\draw [thick,Green](6.5,2) -- (8,3);
\end{tikzpicture}
\caption{Symmetry fractionalization from a bulk $G^{(0)}=\Z_2$ 0-form to $G^{(0), \text{frac}}=\Z_4$ on the 2-charge $\cO_2$. The induced symmetries $D_1^{(i)}$ arise from the intersection of $\cO_2$ with the bulk 0-form symmetry $D_{d-1}^{(-)}$. In the bulk these fuse as 
$D_{d-1}^{(-)} \otimes D_{d-1}^{(-)} = D_{d-1}^{(\text{id})}$. However due to the presence of the non-trivial localized symmetry on $\cO_2$, the induced lines $D_1^{(i)}$ fuse as $D_{1}^{(i)}\otimes D_{1}^{(i)}=D_{1}^{(-)}$ on the surface, where $D_1^{(-)}$ generates the localized $\Z_2$ 0-form symmetry. 
\label{fig:Z2Z4}}
\end{figure}
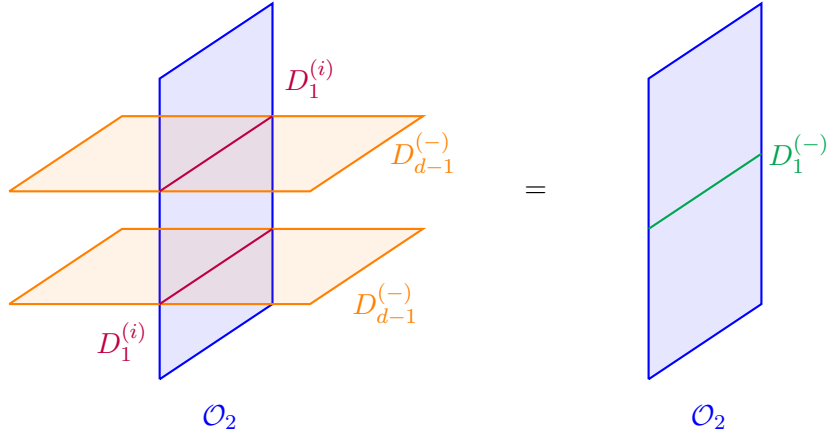

\paragraph{Example: Fractionalization of $\Z_2$ to $\Z_4$.}
Consider a $G^{(0)}=\Z_2^{(0)}$ 0-form symmetry, which fractionalizes on a 2-charge $\cO_2$ to $G^{(0), \text{frac}}= \Z_4$, see figure \ref{fig:Z2Z4}. This occurs e.g. in the 3d gauge theory with $\Z_4^{(0)}$ form symmetry, where we gauge a $\Z_2^{(0)}$ (see \cite{Bhardwaj:2022maz}). 
For instance 3d gauge theory with gauge group $PSU(4)$, where we gauge a $\Z_2^{(0)} \subset \Z_4^{(0)}$ 
\be
  \begin{tikzcd}
  PSU(4) \quad    \rar["\text{gauge } \Z_2^{(0)}"]&  \quad 
     SU(4)/\Z_2   \end{tikzcd}
    \,.
\ee
The resulting theory has $\Z_2^{(0)}$ and $\Z_2^{(1)}$ 0- and 1-form symmetries, but the fact that we started with $\Z_4^{(0)}$ (as opposed to $\Z_2 \times \Z_2$, i.e. there is a non-trivial extension class in $H^2(\Z_2 , \Z_2)$) means, that the gauged theory $SU(4)/\Z_2$ has a mixed 't Hooft anomaly 
\be
\mathcal{A} = e^{i \pi \int  B_2 \cup A_1^2} \,.
\ee
From the theta-defect construction, we know that there is a surface defect $D_2^{(\Z_2)}$, which is the condensation defect for the 1-form symmetry $\Z_2^{(1)}$. This is the 2-charge that carries non-trivial symmetry fractionalization: 
intersections with 0-form $\Z_2$ gives induced lines, which fuse to the localized symmetry on $D_2^{\Z_2}$, which is dual to the gauged symmetry on the condensation defect.

\paragraph{Fractionalization to the Ising category.}
Group-like induced symmetries can also fractionalize to non-invertible symmetries. E.g. this occurs when considering a theory with a $\Z_2$ 0-form symmetry $D_{d-1}^{(g)}$and a surface operator $\cO_2$, which has a localized $\Z_2$ 0-form symmetry as well. Denote the generator of the  localized symmetry by $D_1^{(-)}$ and the induced one $D_1^{(S)}$. 
This can then have the fusion 
\be
D_1^{(S)}\otimes D_1^{(S)}=D_1^{(\id)}\oplus D_1^{(-)}\,,
\ee
which is precisely the non-invertible fusion of the Kramers-Wannier duality defect (\ref{Isingfu}), which is the fractionalized symmetry on a 2-charge in a theory with only $\Z_2^{(0)}$!


\section{The Symmetry TFT and the Sandwich}
\label{sec:SymTFT}

We have discussed gauging of symmetries as a central tool to  the
construction of non-invertible symmetries. Even starting with a 0-form symmetry $G^{(0)}$, we find that there are many ways to gauge -- resulting in an intricate web of symmetry structures, an example is shown in figure \ref{fig:miniweb}. It would be first of all useful to obtain a unified description of these gauge-related symmetry structures. 
The Symmetry TFT provides precisely that. It also has direct connections with string and holographic setups, which naturally give rise to closely related TFTs. 

However, perhaps equally importantly, it was noted recently \cite{Bhardwaj:2023ayw}, that the SymTFT also encodes the generalized charges, and thus has a fundamental role to play in constraining theories with given symmetries. 

The SymTFT has many points of origin -- it has existed prior to its naming in \cite{Apruzzi:2021nmk}, in works of Freed and Teleman and \cite{Gaiotto:2020iye} and 2d theories as the Turaev-Viro theories, but the theoretical framework was laid out in full in \cite{Freed:2022qnc}. Many applications have been put forward since then \cite{Apruzzi:2021nmk, Apruzzi:2022rei, Kaidi:2022cpf, vanBeest:2022fss, Kaidi:2023maf, Antinucci:2022vyk}.
The connection to generalized charges was established in \cite{Bhardwaj:2023ayw}.


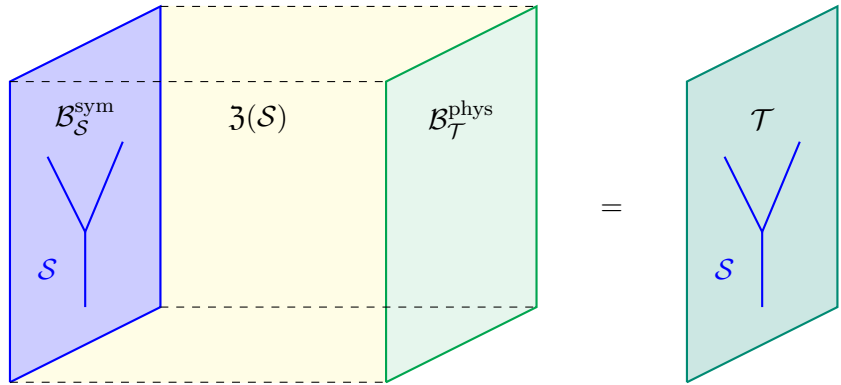
\begin{figure}
\centering
\begin{tikzpicture}
\draw [yellow, fill= yellow, opacity =0.1]
(0,0) -- (0,4) -- (2,5) -- (7,5) -- (7,1) -- (5,0)--(0,0);
\begin{scope}[shift={(0,0)}]
\draw [white, thick, fill=white,opacity=1]
(0,0) -- (0,4) -- (2,5) -- (2,1) -- (0,0);
\end{scope}
\begin{scope}[shift={(5,0)}]
\draw [white, thick, fill=white,opacity=1]
(0,0) -- (0,4) -- (2,5) -- (2,1) -- (0,0);
\end{scope}
\begin{scope}[shift={(5,0)}]
\draw [Green, thick, fill=Green,opacity=0.1] 
(0,0) -- (0, 4) -- (2, 5) -- (2,1) -- (0,0);
\draw [Green, thick]
(0,0) -- (0, 4) -- (2, 5) -- (2,1) -- (0,0);
\node at (1,3.5) {$\mathcal{B}^{\text{phys}}_{\mathcal{T}}$};
\end{scope}
\begin{scope}[shift={(0,0)}]
\draw [blue, thick, fill=blue,opacity=0.2]
(0,0) -- (0, 4) -- (2, 5) -- (2,1) -- (0,0);
\draw [blue, thick]
(0,0) -- (0, 4) -- (2, 5) -- (2,1) -- (0,0);
\node at (1,3.5) {$\mathcal{B}^{\text{sym}}_\cS$};
\end{scope}
\node at (3.3, 3.5) {$\mathfrak{Z}({\mathcal{S}})$};
\draw[dashed] (0,0) -- (5,0);
\draw[dashed] (0,4) -- (5,4);
\draw[dashed] (2,5) -- (7,5);
\draw[dashed] (2,1) -- (7,1);
    \draw [blue, thick] (1, 1) -- (1, 2) ; 
    \draw [blue, thick] (0.5, 3) -- (1, 2) -- (1.5, 3.2) ; 
    \node[blue] at (0.5, 1.5) {$\cS$};
\begin{scope}[shift={(9,0)}]
\node at (-1,2.3) {$=$} ;
\draw [PineGreen, thick, fill=PineGreen,opacity=0.2] 
(0,0) -- (0, 4) -- (2, 5) -- (2,1) -- (0,0);
\draw [PineGreen, thick]
(0,0) -- (0, 4) -- (2, 5) -- (2,1) -- (0,0);
\node at (1,3.5) {$\cT$};    
    \draw [blue, thick] (1, 1) -- (1, 2) ; 
    \draw [blue, thick] (0.5, 3) -- (1, 2) -- (1.5, 3.2) ; 
    \node[blue] at (0.5, 1.5) {$\cS$};
\end{scope}
\end{tikzpicture}
\caption{The SymTFT Sandwich:
The SymTFT $\mathfrak{Z}(\cS)$ of a symmetry $\cS$ of a $d$-dimensional theory $\cT$ is a $(d+1)$-dimensional TQFT. It has two boundaries (bread):  the topological boundary condition $\mathcal{B}^{\text{sym}}_\cS$ and the physical (not gapped) boundary condition $\mathcal{B}^{\text{phys}}_\cS$. 
After the interval compactification, the topological defects (drawn as blue lines) of the topological boundary condition $\mathcal{B}^{\text{sym}}_\cS$ become the topological defects of the QFT $\cT$  that generate the $\cS$ symmetry of $\cT$. \label{fig:SymTFT}}
\end{figure}

\subsection{General Philosophy}

Consider a $d$-dimensional theory $\cT$ with a symmetry $\cS$ (usually a collection of topological defects that form a fusion higher-category). The SymTFT $\mathfrak{Z}(\cS)$ is then defined as a $(d+1)$-dimensional topological field theory, 
which admits a $d$-dimensional topological boundary condition (where the symmetry $\cS$ is realized) and a physical boundary condition. See figure \ref{fig:SymTFT}. After compactifying this interval, one obtains back the theory $\cT$ with the symmetry $\cS$.

In the simplest instances, e.g. when the symmetry $\cS$ is an invertible group-like symmetry (or higher group) with maybe an 't Hooft anomaly, the SymTFT is  a BF-theory or Dijkgraaf-Witten theory with twist. We will discuss this case in detail below. 
One can think of the SymTFT as a $d+1$-dimensional gauging of the symmetry $\cS$, which is in fact precisely what it is in the abelian higher-form symmetry case. 

For $\cS$ a categorical (non-invertible) symmetry given by higher-fusion categories $\cS$, we conjecture that the SymTFT has topological operators that are given by the Drinfeld center of $\cS$. 

The most salient features of this setup are as follows, which often goes under the name of the sandwich construction:
\begin{enumerate}
\item The SymTFT $\mathfrak{Z}(\cS)$ has a topological boundary conditions, $\mathcal{B}_{\cS}^{\text{sym}}$, whose topological defects form the higher fusion category $\cS$.
\item It also has a physical (not necessarily topological) boundary condition $\mathcal{B}_{\mathcal{T}}^{\text{phys}}$, as shown on the right hand side of figure \ref{fig:SymTFT}.
\item Upon compactification of the SymTFT along the interval in figure \ref{fig:SymTFT}, results in the original theory $\cT$ with symmetry $\cS$. 
\item The topological defects of the SymTFT $\mathfrak{Z}(\cS)$ form the Drinfeld center $\mathcal{Z}(\cS)$. A topological defect $\bm{Q}_{q+1}\in \mathcal{Z}(\cS)$ can have Dirichlet boundary conditions, i.e. it ends and gives rise to a $q$-dimensional operator; or it has Neumann boundary conditions, and becomes a topological defect in the boundary. There are other configurations which e.g. give rise to twisted sector charges (see \cite{Bhardwaj:2023ayw} for a comprehensive discussion). 

\item Gauging: changing the symmetry boundary conditions, $\mathcal{B}^{\text{sym}}_{\cS}$ to $\mathcal{B}^{\text{sym}}_{\cS'}$, means bulk topological defects that were terminating could now end and vice versa. The condition for what is allowed is in general a choice of Lagrangian algebra. We will discuss this in the context of abelian symmetries in detail below.

\item Physically perhaps most interesting is the observation, that the  generalized charges are precisely the topological operators of the SymTFT  \cite{Bhardwaj:2023ayw}. 

\end{enumerate}

\subsection{Examples of SymTFTs and Generalized Charges}

We now discuss two simple examples, for SymTFTs for abelian symmetries: in 2d, abelian 0-form symmetries, and in any dimensions, abelian higher-form symmetries, and discuss how in this special context the generalized charges emerge from the SymTFT. We conclude with some general observations about the general case of symmetries that are not necessarily abelian, and discuss the relation of the Drinfeld center to the generalized charges. 

\paragraph{2d Theories.}
We will focus here on a simple class of SymTFT and refer to \cite{Bhardwaj:2023ayw} for the case of a general symmetry fusion or higher-fusion category. 
Let us consider again for simplicity the case of an abelian group $G$. 
We can gauge a 0-form symmetry $G$ in 2d, with symmetry $\Vec_G$ and get the symmetry $\Rep(G)$. There are variants on this, where e.g. we only gauge a subgroup $H\in G$. And in addition we can include 't Hooft anomalies $\omega \in H^3 (G, U(1))$, which can obstruct the gauging. 
It is useful to separate the symmetry structure from the actual physical theory, which is what the SymTFT achieves. 
Start with a 2d theory $\cT$  with 0-form symmetry $G = \Z_N$. The SymTFT is given in terms of a 3d theory 
\be\label{BFU1}
S_{\SymTFT} = {2\pi \over N}\int_{M_3} b_1 \cup \delta c_1 +  \omega (b_1) \,,
\ee
where $b_1, c_1 \in C^1 (M_3, \Z_N)$ are now dynamical gauge fields in a Dijkgraaf-Witten theory with twist (given by the anomaly $\omega$). This type of action is written in terms of the cochains with finite coefficients. 
In terms of $U(1)$-valued fields which roughly are identified with $b^{\text{continuum}} = {2\pi \over N} b^{\text{discrete}}$ we can also write the BF term as 
\be
S_{BF} = {N\over 2\pi} \int_{M_3} b_1 \wedge d c_1   \,,
\ee
where $b_1, c_1$ are now a $U(1)$-valued gauge field, whose equations of motion dictate that it is $\Z_N$ valued.

The 3d-dimensional SymTFT has two boundary conditions: 
\begin{itemize}
\item Symmetry boundary condition $\mathcal{B}^{\text{sym}}_\cS$: this is  gapped (topological) boundary condition, e.g. Neumann or Dirichlet for the 1-form fields. 
\item Physical boundary condition $\mathcal{B}^{\text{phys}}_\cT$.
\end{itemize}
The physical theory $\cT$ is  obtained by interval compactification, see figure \ref{fig:SymTFT}. 
What are these topological boundary conditions? In the specific examples above there are two canonical boundary conditions:
\be
\ba
\mathcal{B}_{\cS_1}^{\text{sym}}:\  & \quad  b_1 \text{ Dirichlet,} \ c_1\text{ Neumann} \cr 
\mathcal{B}_{\cS_2}^{\text{sym}}:\  & \quad  c_1 \text{ Dirichlet,} \ b_1\text{ Neumann} \,.
\ea
\ee
Dirichlet means that the field does not fluctuate in the boundary theory and thus becomes a background field, i.e. this corresponds to inserting the delta function 
\be
\text{Dirichlet:} \qquad \delta (b-B) \,,
\ee
where $B$ is now a non-dynamical background field. 
Whereas Neumann indicates that the resulting field remains dynamical. 
We can form the topological operators in the 3d bulk theory 
\be\label{D1bc}
D_1^{b}(M_1) = e^{i  \int_{M_1} b_1} \,,\qquad 
D_1^{c}(M_1) = e^{i \int_{M_1} c_1} \,,
\ee
which due to the BF-coupling $b\delta c$ have non-trivial exchange 
\be\label{BFlink}
D_1^b (M_1) D_1^c (N_1) = e^{2 \pi i L(M_1, N_1)\over N} D_1^c (N_1) D_1^b (M_1)  \,.
\ee
This follows from the canonical quantization of the BF-coupling, in terms of the Heisenberg algebra, thinking of $b$ and $c$ as canonically conjugate variables (see e.g. \cite{Witten:1998wy}). 
Here $L(M_1, N_1)$ is the linking of the 1-manifolds. 

Upon imposing boundary conditions, these topological operators will 
determine the symmetry generators and generalized charges. E.g. in the theory with 
symmetry boundary condition $\mathcal{B}_{\cS_1}$, the $D_1^{b}$ defects can end, and thus become point-like operators in the theory, whereas $D_1^{c}$ cannot end and remains a topological operator on the boundary. 
After interval compactification, these are the charges and symmetry generators, respectively in the theory with 0-form symmetry $\Z_N$, generated by $D_1^{c}$. 
The non-trivial linking (\ref{BFlink})  states the fact that the local operators carry charge under the topological lines. 
Changing the b.c. to $\mathcal{B}_{\cS_2}$ reverses the roles of $b$ and $c$: now the $c$-lines are endable and give rise to charged operators, where as the $b$-lines are topological symmetry generators, generating the dual symmetry (which in this case is again $\mathbb{Z}_N$). 

Changing from boundary conditions $\mathcal{B}_{\cS_1}$ to $\mathcal{B}_{\cS_2}$ corresponds to gauging of the 0-form symmetry. 
Indeed, we can see this as a change of boundary conditions from Neumann to Dirichlet and vice versa. 
 
In the presence of a  't Hooft anomaly $\omega \in H^3 (G, U(1))$ not all boundary conditions may be allowed (thus indicating the obstruction to gauging). Let us consider the simple instance when there is an abelian group-symmetry $\Z_2$ with anomaly 
\be
\omega (b_1) = b_1^3\,.
\ee
We can see that the equations of motion for $b_1$ are 
\be
\delta c_1 + b_1^2 =0 \,.
\ee
Fixing $b_1$ to be Dirichlet and $c_1$ Neumann is consistent, but $b_1$ Neumann and $c_1$ Dirichlet is not. 

\paragraph{Symmetry TFTs for Abelian Symmetries: General Dimension.}


The SymTFT is a much more general concept and its role in studying generalized symmetries, generalized charges and relation among symmetries via gaugings, is only being explored in higher dimensions as we write this -- see \cite{Apruzzi:2021nmk, Freed:2022qnc, vanBeest:2022fss, Apruzzi:2022rei, Kaidi:2022cpf, Bhardwaj:2023wzd,  Inamura:2023qzl, Bhardwaj:2023ayw}. However this very closely interlinks with holography and string theory constructions of QFTs, and thus may be useful to review in the simplest instance of abelian higher-form symmetries. 
In this case we introduce background fields for the higher-form symmetries $G^{(p_i)}$ and their duals. Consider a product of higher-form symmetries
\be
\cS= \prod_{i} G^{(p_i)} \,,
\ee
and  let $b_{p+1} \in C^{p+1}(M_{d+1}, G^{(p)})$ be the gauge field for a $G^{(p)}$, and $\widehat{b}_{d-p-1}$ the one for its dual symmetry $\widehat{G}^{(d-p-2)}$. Then the action for the SymTFT is a Dijkgraaf-Witten or BF-theory with ``twist" $\mathcal{A}$ (which is determined by the 't Hooft anomalies of the theory, and a polynomial in the background fields $b_{p+1}$)
\be
S_{\text{SymTFT}} = \int_{M_{d+1}}  \sum_{i} 
{2\pi \over n_i} b_{p_i+1}  \cup \delta \widehat{b}_{d-p_i-1} + \mathcal{A}(\{b_{p_i+1}\}) \,.
\ee
Here the groups $G^{(p_i)}$ are cyclic of order $n_i$.

Of course such BF-terms are ubiquituous in holographic setups. For instance as already noted in \cite{Witten:1998wy}, in AdS$_5\times S^5$ there is a topological coupling after integrating over the sphere with $N$ units of 5-form flux, which is a special instance of the above 
\be
S_{\text{SymTFT}} = N\int_{\text{AdS}_5} b_2  \wedge  dc_2 \,,
\ee
where $b_2$ and $c_2$ correspond to the NSNS and RR 2-form potentials in Type IIB. 
The SymTFT in this case is that of a $G^{(1)}=\Z_N$ 1-form symmetry, which is consistent with the dual CFT, which is 4d $\mathcal{N}=4$ Super-Yang-Mills with gauge algebra $\mathfrak{su}(N)$. Again we can specify the global form of the group, and thus its higher-form symmetries by considering the topological operators 
\be
D_2^{b}(M_2) = e^{i  \int_{M_2} b_2} \,,\qquad 
D_2^{c}(M_2) = e^{i \int_{M_2} c_2} \,,\qquad 
\ee
Again due to flux non-commutativity, these have non-trivial exchange relations in the bulk
\be\label{nonloc}
D_2^b (M_2) D_2^c (N_2) =  e^{2\pi i L(M_2, N_2) \over N} D_2^c (N_2)D_2^b (M_2) \,.
\ee
As in the 2d case we have choices of boundary conditions for the fields $b_2$ and $c_2$. In terms of the SymTFT this corresponds to letting the topological defects $D_2$ in the bulk end or terminate\footnote{In holography the distinction between the topological and physical boundary are somewhat less clearly distinguished. We will here focus on the discussion in a full-fledged SymTFT.}.


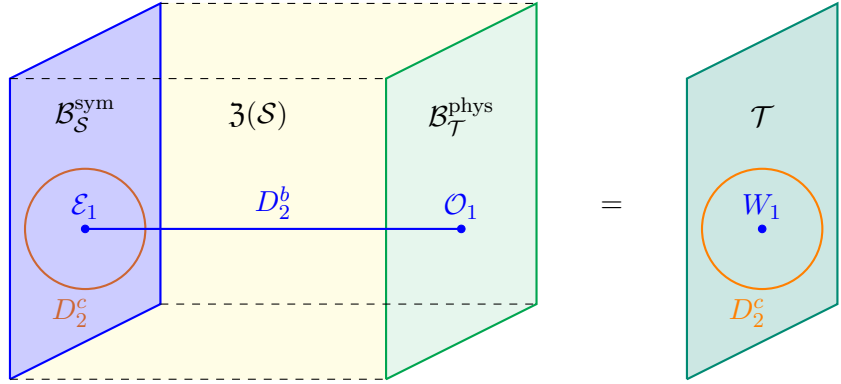
\begin{figure}
\centering
\begin{tikzpicture}
\draw [yellow, fill= yellow, opacity =0.1]
(0,0) -- (0,4) -- (2,5) -- (7,5) -- (7,1) -- (5,0)--(0,0);
\begin{scope}[shift={(0,0)}]
\draw [white, thick, fill=white,opacity=1]
(0,0) -- (0,4) -- (2,5) -- (2,1) -- (0,0);
\end{scope}
\begin{scope}[shift={(5,0)}]
\draw [white, thick, fill=white,opacity=1]
(0,0) -- (0,4) -- (2,5) -- (2,1) -- (0,0);
\end{scope}
\begin{scope}[shift={(5,0)}]
\draw [Green, thick, fill=Green,opacity=0.1] 
(0,0) -- (0, 4) -- (2, 5) -- (2,1) -- (0,0);
\draw [Green, thick]
(0,0) -- (0, 4) -- (2, 5) -- (2,1) -- (0,0);
\node at (1,3.5) {$\mathcal{B}^{\text{phys}}_{\mathcal{T}}$};
\end{scope}
\draw [orange, thick] (1,2) ellipse (0.8 and 0.8);
\node[below, orange] at  (0.8,1.2)  {$D_2^c$};
\draw[thick, blue]  (1,2) --(6,2) ;
\draw [blue,fill=blue] (6,2) ellipse (0.05 and 0.05);
\draw [blue,fill=blue] (1,2) ellipse (0.05 and 0.05);
\node[above, blue] at  (6, 2) {$\cO_1$};
\node[above, blue] at  (1, 2) {$\mathcal{E}_1$};
\node[above, blue] at  (3.5, 2) {$D_2^{b}$}; 
\begin{scope}[shift={(0,0)}]
\draw [blue, thick, fill=blue,opacity=0.2]
(0,0) -- (0, 4) -- (2, 5) -- (2,1) -- (0,0);
\draw [blue, thick]
(0,0) -- (0, 4) -- (2, 5) -- (2,1) -- (0,0);
\node at (1,3.5) {$\mathcal{B}^{\text{sym}}_\cS$};
\end{scope}
\node at (3.3, 3.5) {$\mathfrak{Z}({\mathcal{S}})$};
\draw[dashed] (0,0) -- (5,0);
\draw[dashed] (0,4) -- (5,4);
\draw[dashed] (2,5) -- (7,5);
\draw[dashed] (2,1) -- (7,1);
\begin{scope}[shift={(9,0)}]
\node at (-1,2.3) {$=$} ;
\draw [PineGreen, thick, fill=PineGreen,opacity=0.2] 
(0,0) -- (0, 4) -- (2, 5) -- (2,1) -- (0,0);
\draw [PineGreen, thick]
(0,0) -- (0, 4) -- (2, 5) -- (2,1) -- (0,0);
\node at (1,3.5) {$\cT$};    
\draw [orange, thick] (1,2) ellipse (0.8 and 0.8);
\node[below, orange] at  (0.8,1.2) {$D_2^c$};
\draw [blue,fill=blue] (1,2) ellipse (0.05 and 0.05);
\node[above, blue] at  (1, 2) {$W_1$};
\end{scope}
\end{tikzpicture}
\caption{5d SymTFT $\mathfrak{Z}(\cS)$ for a 4d theory $\cT$: on the left hand side, $D_2^b$ is  a bulk topological operator which ends on both boundaries. $D_2^c$ is a bulk topological operator that projects to a symmetry defect on $\mathcal{B}_\cS^{\text{sym}}$. 
 The non-trivial linking in the bulk translates to a non-trivial linking on the symmetry boundary.  After interval compactification we obtain precisely the picture on the right:
A Wilson line operator that is charged under the 1-form symmetry $\Z_N$ generated by topological surfaces $D_2^c$.
 \label{fig:charges}}
\end{figure}

Let again 
\be
\ba
\mathcal{B}_{SU(N)}^{\text{sym}}: &\qquad b_2 \text{ Dirichlet}, \ c_2 \text{ Neumann} \cr 
\mathcal{B}_{PSU(N)}^{\text{sym}}:& \qquad c_2 \text{ Dirichlet}, \ b_2 \text{ Neumann} \,.
\ea
\ee
In figure  \ref{fig:charges} we have shown the SymTFT setup with a 1-charge that is a Wilson line, arising as the end-point of the $D_2^{b}$ surface operator (which is a bulk topological operator with Dirichlet b.c. on the symmetry boundary). The defects $D_2^c$ on the other hand have Neumann boundary conditions, and restrict on the symmetry boundary to the generators of $\cS$. 
The non-trivial charge under these topological defects arises from (\ref{nonloc}).

This is the picture for the invertible symmetries. Of course once we include anomalies into the SymTFT, in particular mixed anomalies, then non-invertible symmetries have a unified description within this framework as well. A  simple example can be obtained again from physical theories such as 4d gauge theories, this time 4d $\mathcal{N}=1$ super-Yang-Mills with $SU(N)$ gauge group as discussed in section \ref{sec:TwistTheta}. Holographically this was realized in \cite{Apruzzi:2021phx, Apruzzi:2022rei}
Its 0-form $\Z_{2N}$ and 1-form $\Z_N$ have a mixed anomaly (\ref{mixedano}). The SymTFT is 
\be
S_{\text{SymTFT}} ={2 \pi \over N} \int_{M_5} b_2  \delta c_2 + a_1 \delta c_3 -\int_{M_5} {\pi \over N} a_1 b_2^2 \,.
\ee
The topological defects of the SymTFT are the defects $D_2^b$ and $D_2^c$ as in the analysis above, but in addition we also have the 0-form symmetry generator
\be
D_3(M_3)= e^{i\int_{M_3} c_3} e^{i\int_{M_4} {b_2^2/ N }} \,.
\ee
Here $\partial M_4 = M_3$. 
When $b_2$ has Dirichlet  boundary conditions, the second term  becomes a c-number and can be neglected. However, if we pick Neumann boundary conditions, then term is non-negligible, and results in the non-invertible twisted theta defects we discussed in section \ref{sec:TwistTheta}.

\paragraph{SymTFT and the Drinfeld Center.}

We only discussed the SymTFT $\mathfrak{Z}(\cS)$ for abelian fields, but mathematically its topological defects have a precise relation to the higher-fusion category symmetry $\cS$. 
It is the Drinfeld center of the category $\cS$: $\mathcal{Z}(\cS)$. 

In the abelian examples, we have seen that the topological defects of the SymTFT give rise to the generalized charges, and project to symmetry defects on the gapped boundary. 
This is precisely the part that generalizes to more intricate symmetries $\cS$. 
The topological defects of the SymTFT $\mathfrak{Z}(\cS)$ form the Drinfeld center $\mathcal{Z}(\cS)$ of the symmetry category $\cS$ and following \cite{Bhardwaj:2023ayw}
\begin{tcolorbox}[colback=white, colframe=black!50, rounded corners]
\centering{The generalized charges of a symmetry $\cS$ are elements of the Drinfeld center $\mathcal{Z}(\cS)$.}
\end{tcolorbox}
The specific gapped symmetry boundary conditions on $\mathcal{B}_\cS^{\text{sym}}$ only reassign which charges are genuine/non-genuine (or twisted sector) charges in the theory after the interval compactification. But the Drinfeld center is invariant. Mathematically this is the statement that gauging does not change the Drinfeld center (i.e. Morita equivalent higher fusion categories have the same Drinfeld center). 
In this sense, the charges (flavor) of the sandwich is indeed in the center. 

Understanding higher fusion categories from the perspective of Morita equivalence and the relation to the Drinfeld center has been studied in the math literature for 2-categories \cite{kong2015boundary, Kong:2019brm, Johnson-Freyd:2020usu, decoppet2022drinfeld, Zhao:2022yaw}. In this context the SymTFTs are known as Douglas-Reutter-Walker TFTs \cite{DouglasReutter, Walker} or in special cases, predating this, the Crane-Jetter \cite{Crane:1993if} TFTs.

The utility of the SymTFT and the Drinfeld center in terms of the generalized charges \cite{Bhardwaj:2023ayw} makes this a central object in the study of generalized symmetries. The above mathematics papers also provide a classification of fusion higher-categories, which will have important implications for the types of generalized symmetries that exist in QFTs. 

The SymTFT is also a point of contact with condensed matter physics, where similar concepts were introduced e.g. in
\cite{Ji:2019jhk, Kong:2020cie}, (for the hep-th community, slightly confusingly) calling it ``categorical symmetries". Clearly these connections, both to lattice models, continuum limits and the mathematics of fusion higher-categories are an exciting area of research.


\section{Outlook and Open Problems}
\label{sec:Outlook}

The main purpose of these lectures is to give an introduction to the topic generalized symmetries, including invertible, but mostly non-invertible symmetries, and hopefully get the readers hooked on this  exciting topic. The now hopefully highly motivated reader should be able  delve into the subject more deeply following various literature references.  

One of the corollaries of these recent developments in symmetries is the inevitable emergence of categorical structures: both in terms of the characterization of symmetries (the multi-layered structure of topological defects is manifesting the structure of fusion higher-categories), but also the generalized charges, which even for invertible symmetries (e.g. higher-form symmetry groups) have an inherent higher-categorical structure! 
So one goal of these lectures is to remove the fear of categories and put at least higher fusion-categories firmly into the realm of the standard vocabulary in theoretical physics. 

Obviously the field of generalized, non-invertible or categorical symmetries is still very much in development, and there are numerous future directions and open problems to tackle. Here are some that spring to mind: 

\paragraph{IR Constraints.}
The most pressing question is whether we can derive genuinely new properties of QFTs from non-invertible symmetries. Some interesting applications have been put forward in the context of the Standard Model, but they are still not quite as compelling as we would like them to be (e.g. the constraints on pion-decays can be derived without the use of non-invertible symmetries), see  \cite{Wang:2021vki, Choi:2022rfe, Choi:2022fgx, Cordova:2022ieu, Cordova:2022fhg, Cordova:2022qtz, Putrov:2023jqi} for a selection of such applications. 
In 2d many powerful applications of fusion category symmetries are known \cite{Chang:2018iay, Komargodski:2020mxz} and one important advance would be to derive similarly powerful statements in higher-dimensions. One obvious application is in the context of confinement/deconfinement, with some results in \cite{Apruzzi:2022rei}. 

\paragraph{Symmetries in String Theory/Holography.}
Some QFTs will not admit any weakly coupled Lagrangian descriptions at all, and require other methods to determine their global symmetries and anomalies. Examples are 5d and 6d superconformal field theories, but also 4d SCFTs for instance of class S type. In the past few years many results on symmetries of these theories have been obtained, using geometric or brane engineering both in the invertible  \cite{Tachikawa:2013hya, DelZotto:2015isa, GarciaEtxebarria:2019caf, Eckhard:2019jgg, Morrison:2020ool, Albertini:2020mdx, Gukov:2020btk, Bah:2020uev, Closset:2020scj, DelZotto:2020esg, Apruzzi:2020zot, Bhardwaj:2020phs, DelZotto:2020sop, Closset:2020afy, Bhardwaj:2021pfz, Bhardwaj:2021ojs, Apruzzi:2021vcu, Hosseini:2021ged, Cvetic:2021sxm, Bhardwaj:2021zrt, Closset:2021lhd, Bhardwaj:2021wif, Apruzzi:2021mlh, Tian:2021cif, Closset:2021lwy, Bhardwaj:2021mzl, Apruzzi:2021nmk, DelZotto:2022fnw, Bhardwaj:2022ekc, DelZotto:2022ohj, Cvetic:2022imb, DelZotto:2022joo, Apruzzi:2022dlm, Bhardwaj:2022scy, Hubner:2022kxr, DelZotto:2022ras, vanBeest:2022fss} and non-invertible world \cite{Bhardwaj:2022yxj, Bashmakov:2022jtl, Apruzzi:2022rei, GarciaEtxebarria:2022vzq, Heckman:2022muc,Bashmakov:2022uek, Heckman:2022xgu,Acharya:2023bth, Carta:2023bqn, Dierigl:2023jdp,Sacchi:2023omn,Chen:2023qnv,Bashmakov:2023kwo}. 
In the context of string theory realizations of QFTs and holographic descriptions of strongly coupled QFTs, an interesting conceptual question is how the categorical structures of symmetries arise from the UV. 
The stringy/holographic realization of SymTFTs (albeit only for abelian symmetries) is a very promising starting point \cite{Apruzzi:2021nmk, Apruzzi:2022dlm, Apruzzi:2022rei, Kaidi:2022cpf, Bashmakov:2022uek, vanBeest:2022fss, Antinucci:2022vyk}.
Complementing this, there is a proposals that symmetries  are realized in terms of branes (in a limit when they are topological) \cite{Apruzzi:2022rei, GarciaEtxebarria:2022vzq, Heckman:2022muc}. Some aspects of generalized charges also have been put into this language \cite{Apruzzi:2022rei, Heckman:2022xgu}, relying mostly on the Hanany-Witten effect, and fusion of topological defects seems to be manifested in tachyon condensation. 
However the ultimate dream in this context would be to derive the fusion higher-categories more manifestly from a string compactification or in the gravitational/string theoretic side of the holographic dual. From the identifications of branes as symmetry generators (or even topological defects of the SymTFTs), it would seem natural that a formulation of branes in terms of brane-anti-brane pairs with a tachyon profile, will be key to this endeavour, and thus e.g. for Calabi-Yau compactifications, closely related to the derived category of coherent sheaves.

\paragraph{Quantum Gravity Constraints.}
Within quantum gravity, the absence of global symmetries led to the initial proposal of non-invertible symmetries \cite{Heidenreich:2021xpr}. With the vast generalizations of non-invertible symmetries in QFTs, a return to these quantum gravity (or swampland) constraints is another obvious avenue. Global symmetries in quantum gravity are conjectured to be either gauged or broken. The mechanisms of this in turn will imply insightful constraints on quantum gravity theories. For reviews on this topic, see e.g. \cite{Palti:2019pca, vanBeest:2021lhn}.

\begin{figure}
\centering
\includegraphics[height=7cm]{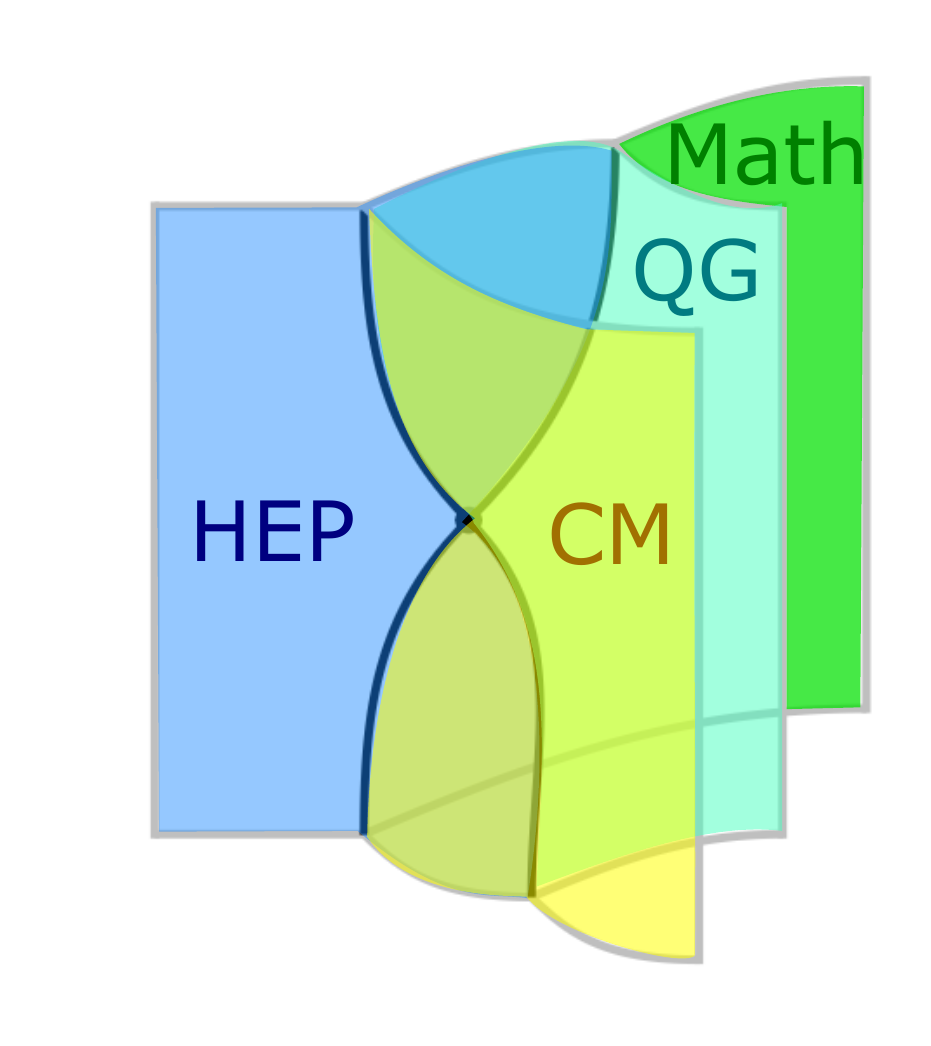}
\caption{Logo for the 2025 KITP program ``Generalized Symmetries in QFT: High-Energy Physics, Condensed Matter and Quantum Gravity". \label{fig:KITP}}
\end{figure}

\paragraph{Relation to Condensed Matter.} 
The research in the last 10 years in condensed matter has developed many of the ideas that are now entering high-energy physics. The main distinction here is that the categorical structures characterize topological order -- i.e. the focus is on the TQFTs that are associated to the categories. In high-energy physics, the perspective that we have taken in these lectures is that the categories act on physical (non-topological) QFTs. 
In the realm of cond-mat, the SymTFT has been discussed in \cite{Ji:2019jhk, Kong:2020cie}. Much of the gauging in 3d has appeared in the works \cite{Barkeshli:2014cna, Teo:2015xla}. The categorical framework was put forward in various papers, we refer to the review \cite{Kong:2022cpy} for an overview. But much remains to be explored at the juncture of cond-mat and hep-th, in particular connections between lattice and continuum models, and topological order versus symmetries. This promises to be a very fruitful future avenue for research interactions accross physics and math disciplines, for instance in the ``Symmetry Seminar" series \cite{SymSem} in the 2025 KITP program gensym25, see figure \ref{fig:KITP}.

\paragraph{Mathematical Question.} 
Fusion higher-categories are within mathematics also a topic that is in development, and many questions remain open, that would be important to address in order to further our understanding of generalized symmetries. Much progress on fusion higher-categories has been made in recent years \cite{DouglasReutter} and topological order in higher dimensions e.g. \cite{Johnson-Freyd:2020twl, Kong:2020jne, Johnson-Freyd:2021tbq}.
Examples of questions that would be on a wish-list are: what are the higher-structures (the analogs of $F$-symbols in fusion categories) in fusion higher-categories, and can they be encoded into simple algebraic constraints? What is the complete classification of fusion $n$-categories, to include non-group-like symmetries? Another important question is the  construction of the centers of fusion higher-categories, and thereby aspects of the SymTFTs and the associated higher-charges of the symmetries.

\subsection*{Acknowledgements}

I am indebted to 
Lakshya Bhardwaj, Fabio Apruzzi, Ibou Bah, Federico Bonetti, Lea Bottini, Dewi Gould, Apoorv Tiwari and Jinxiang Wu, for collaboration on this topic of non-invertible symmetries!
Thank you in particular to Lea Bottini, Fabio Apruzzi, Lakshya Bhardwaj, Dewi Gould, Mark Mezei, Daniel Pajer and Pavel Putrov for comments on an earlier draft, and Thibault D\'ecoppet and Theo Johnson-Freyd for discussions on fusion higher-categories. 
I would also like to thank Brian Willett, who first sparked my interest in the topic of generalized symmetries way back in 2018.
A great thanks goes to the organizers of the ICTP spring school and the participants, for giving me the opportunity to lecture on this topic and all the lively interactions during the school. 
My research on generalized symmetries has been supported by the  European Union's Horizon 2020 Framework through the ERC grant 682608, the Simons Foundation Collaboration on ``Special Holonomy in Geometry, Analysis, and Physics", Award ID: 724073, Schafer-Nameki, and most recently, the EPSRC Open Fellowship EP/X01276X/1.



\bibliography{Symbib}

\providecommand{\href}[2]{#2}\begingroup\raggedright\begin{thebibliography}{100}

\bibitem{Gaiotto:2014kfa}
D.~Gaiotto, A.~Kapustin, N.~Seiberg, and B.~Willett, ``{Generalized Global
  Symmetries},'' \href{http://dx.doi.org/10.1007/JHEP02(2015)172}{{\em JHEP}
  {\bfseries 02} (2015) 172}, \href{http://arxiv.org/abs/1412.5148}{{\ttfamily
  arXiv:1412.5148 [hep-th]}}.

\bibitem{EGNO}
P.~Etingof, S.~Gelaki, D.~Nikshych, and V.~Ostrik,
  \href{http://dx.doi.org/10.1090/surv/205}{{\em Tensor categories}}, vol.~205
  of {\em Mathematical Surveys and Monographs}.
\newblock American Mathematical Society, Providence, RI, 2015.
\newblock \url{https://doi.org/10.1090/surv/205}.

\bibitem{Frohlich:2004ef}
J.~Frohlich, J.~Fuchs, I.~Runkel, and C.~Schweigert, ``{Kramers-Wannier duality
  from conformal defects},''
  \href{http://dx.doi.org/10.1103/PhysRevLett.93.070601}{{\em Phys. Rev. Lett.}
  {\bfseries 93} (2004) 070601},
  \href{http://arxiv.org/abs/cond-mat/0404051}{{\ttfamily
  arXiv:cond-mat/0404051}}.

\bibitem{Fuchs:2007tx}
J.~Fuchs, M.~R. Gaberdiel, I.~Runkel, and C.~Schweigert, ``{Topological defects
  for the free boson CFT},''
  \href{http://dx.doi.org/10.1088/1751-8113/40/37/016}{{\em J. Phys. A}
  {\bfseries 40} (2007) 11403},
  \href{http://arxiv.org/abs/0705.3129}{{\ttfamily arXiv:0705.3129 [hep-th]}}.

\bibitem{Frohlich:2009gb}
J.~Frohlich, J.~Fuchs, I.~Runkel, and C.~Schweigert,
  \href{http://dx.doi.org/10.1142/9789814304634_0056}{``{Defect lines,
  dualities, and generalised orbifolds},''} in {\em {16th International
  Congress on Mathematical Physics}}.
\newblock 9, 2009.
\newblock \href{http://arxiv.org/abs/0909.5013}{{\ttfamily arXiv:0909.5013
  [math-ph]}}.

\bibitem{Bhardwaj:2017xup}
L.~Bhardwaj and Y.~Tachikawa, ``{On finite symmetries and their gauging in two
  dimensions},'' \href{http://dx.doi.org/10.1007/JHEP03(2018)189}{{\em JHEP}
  {\bfseries 03} (2018) 189}, \href{http://arxiv.org/abs/1704.02330}{{\ttfamily
  arXiv:1704.02330 [hep-th]}}.

\bibitem{Chang:2018iay}
C.-M. Chang, Y.-H. Lin, S.-H. Shao, Y.~Wang, and X.~Yin, ``{Topological Defect
  Lines and Renormalization Group Flows in Two Dimensions},''
  \href{http://dx.doi.org/10.1007/JHEP01(2019)026}{{\em JHEP} {\bfseries 01}
  (2019) 026}, \href{http://arxiv.org/abs/1802.04445}{{\ttfamily
  arXiv:1802.04445 [hep-th]}}.

\bibitem{Thorngren:2019iar}
R.~Thorngren and Y.~Wang, ``{Fusion Category Symmetry I: Anomaly In-Flow and
  Gapped Phases},'' \href{http://arxiv.org/abs/1912.02817}{{\ttfamily
  arXiv:1912.02817 [hep-th]}}.

\bibitem{Komargodski:2020mxz}
Z.~Komargodski, K.~Ohmori, K.~Roumpedakis, and S.~Seifnashri, ``{Symmetries and
  strings of adjoint QCD$_{2}$},''
  \href{http://dx.doi.org/10.1007/JHEP03(2021)103}{{\em JHEP} {\bfseries 03}
  (2021) 103}, \href{http://arxiv.org/abs/2008.07567}{{\ttfamily
  arXiv:2008.07567 [hep-th]}}.

\bibitem{Thorngren:2021yso}
R.~Thorngren and Y.~Wang, ``{Fusion Category Symmetry II: Categoriosities at
  $c$ = 1 and Beyond},'' \href{http://arxiv.org/abs/2106.12577}{{\ttfamily
  arXiv:2106.12577 [hep-th]}}.

\bibitem{Witten:1979ey}
E.~Witten, ``{Dyons of Charge e theta/2 pi},''
  \href{http://dx.doi.org/10.1016/0370-2693(79)90838-4}{{\em Phys. Lett. B}
  {\bfseries 86} (1979) 283--287}.

\bibitem{Heidenreich:2021xpr}
B.~Heidenreich, J.~McNamara, M.~Montero, M.~Reece, T.~Rudelius, and
  I.~Valenzuela, ``{Non-invertible global symmetries and completeness of the
  spectrum},'' \href{http://dx.doi.org/10.1007/JHEP09(2021)203}{{\em JHEP}
  {\bfseries 09} (2021) 203}, \href{http://arxiv.org/abs/2104.07036}{{\ttfamily
  arXiv:2104.07036 [hep-th]}}.

\bibitem{Kaidi:2021xfk}
J.~Kaidi, K.~Ohmori, and Y.~Zheng, ``{Kramers-Wannier-like Duality Defects in
  (3+1)D Gauge Theories},''
  \href{http://dx.doi.org/10.1103/PhysRevLett.128.111601}{{\em Phys. Rev.
  Lett.} {\bfseries 128} no.~11, (2022) 111601},
  \href{http://arxiv.org/abs/2111.01141}{{\ttfamily arXiv:2111.01141
  [hep-th]}}.

\bibitem{Choi:2021kmx}
Y.~Choi, C.~Cordova, P.-S. Hsin, H.~T. Lam, and S.-H. Shao, ``{Noninvertible
  duality defects in 3+1 dimensions},''
  \href{http://dx.doi.org/10.1103/PhysRevD.105.125016}{{\em Phys. Rev. D}
  {\bfseries 105} no.~12, (2022) 125016},
  \href{http://arxiv.org/abs/2111.01139}{{\ttfamily arXiv:2111.01139
  [hep-th]}}.

\bibitem{Roumpedakis:2022aik}
K.~Roumpedakis, S.~Seifnashri, and S.-H. Shao, ``{Higher Gauging and
  Non-invertible Condensation Defects},''
  \href{http://arxiv.org/abs/2204.02407}{{\ttfamily arXiv:2204.02407
  [hep-th]}}.

\bibitem{Bhardwaj:2022yxj}
L.~Bhardwaj, L.~E. Bottini, S.~Schafer-Nameki, and A.~Tiwari, ``{Non-invertible
  higher-categorical symmetries},''
  \href{http://dx.doi.org/10.21468/SciPostPhys.14.1.007}{{\em SciPost Phys.}
  {\bfseries 14} no.~1, (2023) 007},
  \href{http://arxiv.org/abs/2204.06564}{{\ttfamily arXiv:2204.06564
  [hep-th]}}.

\bibitem{Choi:2022zal}
Y.~Choi, C.~Cordova, P.-S. Hsin, H.~T. Lam, and S.-H. Shao, ``{Non-invertible
  Condensation, Duality, and Triality Defects in 3+1 Dimensions},''
  \href{http://arxiv.org/abs/2204.09025}{{\ttfamily arXiv:2204.09025
  [hep-th]}}.

\bibitem{Cordova:2022ieu}
C.~Cordova and K.~Ohmori, ``{Noninvertible Chiral Symmetry and Exponential
  Hierarchies},'' \href{http://dx.doi.org/10.1103/PhysRevX.13.011034}{{\em
  Phys. Rev. X} {\bfseries 13} no.~1, (2023) 011034},
  \href{http://arxiv.org/abs/2205.06243}{{\ttfamily arXiv:2205.06243
  [hep-th]}}.

\bibitem{Choi:2022jqy}
Y.~Choi, H.~T. Lam, and S.-H. Shao, ``{Noninvertible Global Symmetries in the
  Standard Model},''
  \href{http://dx.doi.org/10.1103/PhysRevLett.129.161601}{{\em Phys. Rev.
  Lett.} {\bfseries 129} no.~16, (2022) 161601},
  \href{http://arxiv.org/abs/2205.05086}{{\ttfamily arXiv:2205.05086
  [hep-th]}}.

\bibitem{Kaidi:2022uux}
J.~Kaidi, G.~Zafrir, and Y.~Zheng, ``{Non-invertible symmetries of $
  \mathcal{N} $ = 4 SYM and twisted compactification},''
  \href{http://dx.doi.org/10.1007/JHEP08(2022)053}{{\em JHEP} {\bfseries 08}
  (2022) 053}, \href{http://arxiv.org/abs/2205.01104}{{\ttfamily
  arXiv:2205.01104 [hep-th]}}.

\bibitem{Antinucci:2022eat}
A.~Antinucci, G.~Galati, and G.~Rizi, ``{On continuous 2-category symmetries
  and Yang-Mills theory},''
  \href{http://dx.doi.org/10.1007/JHEP12(2022)061}{{\em JHEP} {\bfseries 12}
  (2022) 061}, \href{http://arxiv.org/abs/2206.05646}{{\ttfamily
  arXiv:2206.05646 [hep-th]}}.

\bibitem{Bashmakov:2022jtl}
V.~Bashmakov, M.~Del~Zotto, and A.~Hasan, ``{On the 6d Origin of Non-invertible
  Symmetries in 4d},'' \href{http://arxiv.org/abs/2206.07073}{{\ttfamily
  arXiv:2206.07073 [hep-th]}}.

\bibitem{Damia:2022bcd}
J.~A. Damia, R.~Argurio, and E.~Garcia-Valdecasas, ``{Non-Invertible Defects in
  5d, Boundaries and Holography},''
  \href{http://dx.doi.org/10.21468/SciPostPhys.14.4.067}{{\em SciPost Phys.}
  {\bfseries 14} (2023) 067}, \href{http://arxiv.org/abs/2207.02831}{{\ttfamily
  arXiv:2207.02831 [hep-th]}}.

\bibitem{Choi:2022rfe}
Y.~Choi, H.~T. Lam, and S.-H. Shao, ``{Noninvertible Time-Reversal Symmetry},''
  \href{http://dx.doi.org/10.1103/PhysRevLett.130.131602}{{\em Phys. Rev.
  Lett.} {\bfseries 130} no.~13, (2023) 131602},
  \href{http://arxiv.org/abs/2208.04331}{{\ttfamily arXiv:2208.04331
  [hep-th]}}.

\bibitem{Bhardwaj:2022lsg}
L.~Bhardwaj, S.~Schafer-Nameki, and J.~Wu, ``{Universal Non-Invertible
  Symmetries},'' \href{http://dx.doi.org/10.1002/prop.202200143}{{\em Fortsch.
  Phys.} {\bfseries 70} no.~11, (2022) 2200143},
  \href{http://arxiv.org/abs/2208.05973}{{\ttfamily arXiv:2208.05973
  [hep-th]}}.

\bibitem{Bartsch:2022mpm}
T.~Bartsch, M.~Bullimore, A.~E.~V. Ferrari, and J.~Pearson, ``{Non-invertible
  Symmetries and Higher Representation Theory I},''
  \href{http://arxiv.org/abs/2208.05993}{{\ttfamily arXiv:2208.05993
  [hep-th]}}.

\bibitem{Damia:2022rxw}
J.~A. Damia, R.~Argurio, and L.~Tizzano, ``{Continuous Generalized Symmetries
  in Three Dimensions},'' \href{http://arxiv.org/abs/2206.14093}{{\ttfamily
  arXiv:2206.14093 [hep-th]}}.

\bibitem{Apruzzi:2022rei}
F.~Apruzzi, I.~Bah, F.~Bonetti, and S.~Schafer-Nameki, ``{Noninvertible
  Symmetries from Holography and Branes},''
  \href{http://dx.doi.org/10.1103/PhysRevLett.130.121601}{{\em Phys. Rev.
  Lett.} {\bfseries 130} no.~12, (2023) 121601},
  \href{http://arxiv.org/abs/2208.07373}{{\ttfamily arXiv:2208.07373
  [hep-th]}}.

\bibitem{Lin:2022xod}
L.~Lin, D.~G. Robbins, and E.~Sharpe, ``{Decomposition, Condensation Defects,
  and Fusion},'' \href{http://dx.doi.org/10.1002/prop.202200130}{{\em Fortsch.
  Phys.} {\bfseries 70} no.~11, (2022) 2200130},
  \href{http://arxiv.org/abs/2208.05982}{{\ttfamily arXiv:2208.05982
  [hep-th]}}.

\bibitem{GarciaEtxebarria:2022vzq}
I.~n. Garc\'\i{}a~Etxebarria, ``{Branes and Non-Invertible Symmetries},''
  \href{http://dx.doi.org/10.1002/prop.202200154}{{\em Fortsch. Phys.}
  {\bfseries 70} no.~11, (2022) 2200154},
  \href{http://arxiv.org/abs/2208.07508}{{\ttfamily arXiv:2208.07508
  [hep-th]}}.

\bibitem{Heckman:2022muc}
J.~J. Heckman, M.~H\"ubner, E.~Torres, and H.~Y. Zhang, ``{The Branes Behind
  Generalized Symmetry Operators},''
  \href{http://dx.doi.org/10.1002/prop.202200180}{{\em Fortsch. Phys.}
  {\bfseries 71} no.~1, (2023) 2200180},
  \href{http://arxiv.org/abs/2209.03343}{{\ttfamily arXiv:2209.03343
  [hep-th]}}.

\bibitem{Niro:2022ctq}
P.~Niro, K.~Roumpedakis, and O.~Sela, ``{Exploring non-invertible symmetries in
  free theories},'' \href{http://dx.doi.org/10.1007/JHEP03(2023)005}{{\em JHEP}
  {\bfseries 03} (2023) 005}, \href{http://arxiv.org/abs/2209.11166}{{\ttfamily
  arXiv:2209.11166 [hep-th]}}.

\bibitem{Kaidi:2022cpf}
J.~Kaidi, K.~Ohmori, and Y.~Zheng, ``{Symmetry TFTs for Non-Invertible
  Defects},'' \href{http://arxiv.org/abs/2209.11062}{{\ttfamily
  arXiv:2209.11062 [hep-th]}}.

\bibitem{Antinucci:2022vyk}
A.~Antinucci, F.~Benini, C.~Copetti, G.~Galati, and G.~Rizi, ``{The holography
  of non-invertible self-duality symmetries},''
  \href{http://arxiv.org/abs/2210.09146}{{\ttfamily arXiv:2210.09146
  [hep-th]}}.

\bibitem{Chen:2022cyw}
S.~Chen and Y.~Tanizaki, ``{Solitonic symmetry beyond homotopy: invertibility
  from bordism and non-invertibility from TQFT},''
  \href{http://arxiv.org/abs/2210.13780}{{\ttfamily arXiv:2210.13780
  [hep-th]}}.

\bibitem{Lin:2022dhv}
Y.-H. Lin, M.~Okada, S.~Seifnashri, and Y.~Tachikawa, ``{Asymptotic density of
  states in 2d CFTs with non-invertible symmetries},''
  \href{http://dx.doi.org/10.1007/JHEP03(2023)094}{{\em JHEP} {\bfseries 03}
  (2023) 094}, \href{http://arxiv.org/abs/2208.05495}{{\ttfamily
  arXiv:2208.05495 [hep-th]}}.

\bibitem{Bashmakov:2022uek}
V.~Bashmakov, M.~Del~Zotto, A.~Hasan, and J.~Kaidi, ``{Non-invertible
  Symmetries of Class $\mathcal{S}$ Theories},''
  \href{http://arxiv.org/abs/2211.05138}{{\ttfamily arXiv:2211.05138
  [hep-th]}}.

\bibitem{Karasik:2022kkq}
A.~Karasik, ``{On anomalies and gauging of U(1) non-invertible symmetries in 4d
  QED},'' \href{http://arxiv.org/abs/2211.05802}{{\ttfamily arXiv:2211.05802
  [hep-th]}}.

\bibitem{Cordova:2022fhg}
C.~Cordova, S.~Hong, S.~Koren, and K.~Ohmori, ``{Neutrino Masses from
  Generalized Symmetry Breaking},''
  \href{http://arxiv.org/abs/2211.07639}{{\ttfamily arXiv:2211.07639
  [hep-ph]}}.

\bibitem{GarciaEtxebarria:2022jky}
I.~n. Garc\'\i{}a~Etxebarria and N.~Iqbal, ``{A Goldstone theorem for
  continuous non-invertible symmetries},''
  \href{http://arxiv.org/abs/2211.09570}{{\ttfamily arXiv:2211.09570
  [hep-th]}}.

\bibitem{Decoppet:2022dnz}
T.~D. D\'ecoppet and M.~Yu, ``{Gauging noninvertible defects: a 2-categorical
  perspective},'' \href{http://dx.doi.org/10.1007/s11005-023-01655-1}{{\em
  Lett. Math. Phys.} {\bfseries 113} no.~2, (2023) 36},
  \href{http://arxiv.org/abs/2211.08436}{{\ttfamily arXiv:2211.08436
  [math.CT]}}.

\bibitem{Moradi:2022lqp}
H.~Moradi, S.~F. Moosavian, and A.~Tiwari, ``{Topological Holography: Towards a
  Unification of Landau and Beyond-Landau Physics},''
  \href{http://arxiv.org/abs/2207.10712}{{\ttfamily arXiv:2207.10712
  [cond-mat.str-el]}}.

\bibitem{Runkel:2022fzi}
I.~Runkel, L.~Szegedy, and G.~M.~T. Watts, ``{Parity and Spin CFT with
  boundaries and defects},'' \href{http://arxiv.org/abs/2210.01057}{{\ttfamily
  arXiv:2210.01057 [hep-th]}}.

\bibitem{Choi:2022fgx}
Y.~Choi, H.~T. Lam, and S.-H. Shao, ``{Non-invertible Gauss Law and Axions},''
  \href{http://arxiv.org/abs/2212.04499}{{\ttfamily arXiv:2212.04499
  [hep-th]}}.

\bibitem{Bhardwaj:2022kot}
L.~Bhardwaj, S.~Schafer-Nameki, and A.~Tiwari, ``{Unifying Constructions of
  Non-Invertible Symmetries},''
  \href{http://arxiv.org/abs/2212.06159}{{\ttfamily arXiv:2212.06159
  [hep-th]}}.

\bibitem{Bhardwaj:2022maz}
L.~Bhardwaj, L.~E. Bottini, S.~Schafer-Nameki, and A.~Tiwari, ``{Non-Invertible
  Symmetry Webs},'' \href{http://arxiv.org/abs/2212.06842}{{\ttfamily
  arXiv:2212.06842 [hep-th]}}.

\bibitem{Bartsch:2022ytj}
T.~Bartsch, M.~Bullimore, A.~E.~V. Ferrari, and J.~Pearson, ``{Non-invertible
  Symmetries and Higher Representation Theory II},''
  \href{http://arxiv.org/abs/2212.07393}{{\ttfamily arXiv:2212.07393
  [hep-th]}}.

\bibitem{Heckman:2022xgu}
J.~J. Heckman, M.~Hubner, E.~Torres, X.~Yu, and H.~Y. Zhang, ``{Top Down
  Approach to Topological Duality Defects},''
  \href{http://arxiv.org/abs/2212.09743}{{\ttfamily arXiv:2212.09743
  [hep-th]}}.

\bibitem{Antinucci:2022cdi}
A.~Antinucci, C.~Copetti, G.~Galati, and G.~Rizi, ``{''Zoology'' of
  non-invertible duality defects: the view from class $\mathcal{S}$},''
  \href{http://arxiv.org/abs/2212.09549}{{\ttfamily arXiv:2212.09549
  [hep-th]}}.

\bibitem{Apte:2022xtu}
A.~Apte, C.~Cordova, and H.~T. Lam, ``{Obstructions to Gapped Phases from
  Non-Invertible Symmetries},''
  \href{http://arxiv.org/abs/2212.14605}{{\ttfamily arXiv:2212.14605
  [hep-th]}}.

\bibitem{Delcamp:2023kew}
C.~Delcamp and A.~Tiwari, ``{Higher categorical symmetries and gauging in
  two-dimensional spin systems},''
  \href{http://arxiv.org/abs/2301.01259}{{\ttfamily arXiv:2301.01259
  [hep-th]}}.

\bibitem{Kaidi:2023maf}
J.~Kaidi, E.~Nardoni, G.~Zafrir, and Y.~Zheng, ``{Symmetry TFTs and Anomalies
  of Non-Invertible Symmetries},''
  \href{http://arxiv.org/abs/2301.07112}{{\ttfamily arXiv:2301.07112
  [hep-th]}}.

\bibitem{Li:2023mmw}
L.~Li, M.~Oshikawa, and Y.~Zheng, ``{Non-Invertible Duality Transformation
  Between SPT and SSB Phases},''
  \href{http://arxiv.org/abs/2301.07899}{{\ttfamily arXiv:2301.07899
  [cond-mat.str-el]}}.

\bibitem{Brennan:2023kpw}
T.~D. Brennan, S.~Hong, and L.-T. Wang, ``{Coupling a Cosmic String to a
  TQFT},'' \href{http://arxiv.org/abs/2302.00777}{{\ttfamily arXiv:2302.00777
  [hep-ph]}}.

\bibitem{Etheredge:2023ler}
M.~Etheredge, I.~Garcia~Etxebarria, B.~Heidenreich, and S.~Rauch, ``{Branes and
  symmetries for $\mathcal N=3$ S-folds},''
  \href{http://arxiv.org/abs/2302.14068}{{\ttfamily arXiv:2302.14068
  [hep-th]}}.

\bibitem{Lin:2023uvm}
Y.-H. Lin and S.-H. Shao, ``{Bootstrapping Non-invertible Symmetries},''
  \href{http://arxiv.org/abs/2302.13900}{{\ttfamily arXiv:2302.13900
  [hep-th]}}.

\bibitem{Putrov:2023jqi}
P.~Putrov and J.~Wang, ``{Categorical Symmetry of the Standard Model from
  Gravitational Anomaly},'' \href{http://arxiv.org/abs/2302.14862}{{\ttfamily
  arXiv:2302.14862 [hep-th]}}.

\bibitem{Carta:2023bqn}
F.~Carta, S.~Giacomelli, N.~Mekareeya, and A.~Mininno, ``{Comments on
  Non-invertible Symmetries in Argyres-Douglas Theories},''
  \href{http://arxiv.org/abs/2303.16216}{{\ttfamily arXiv:2303.16216
  [hep-th]}}.

\bibitem{Koide:2023rqd}
M.~Koide, Y.~Nagoya, and S.~Yamaguchi, ``{Non-invertible symmetries and
  boundaries in four dimensions},''
  \href{http://arxiv.org/abs/2304.01550}{{\ttfamily arXiv:2304.01550
  [hep-th]}}.

\bibitem{Zhang:2023wlu}
C.~Zhang and C.~C\'ordova, ``{Anomalies of $(1+1)D$ categorical symmetries},''
  \href{http://arxiv.org/abs/2304.01262}{{\ttfamily arXiv:2304.01262
  [cond-mat.str-el]}}.

\bibitem{Cao:2023doz}
W.~Cao, L.~Li, M.~Yamazaki, and Y.~Zheng, ``{Subsystem Non-Invertible Symmetry
  Operators and Defects},'' \href{http://arxiv.org/abs/2304.09886}{{\ttfamily
  arXiv:2304.09886 [cond-mat.str-el]}}.

\bibitem{Dierigl:2023jdp}
M.~Dierigl, J.~J. Heckman, M.~Montero, and E.~Torres, ``{R7-Branes as Charge
  Conjugation Operators},'' \href{http://arxiv.org/abs/2305.05689}{{\ttfamily
  arXiv:2305.05689 [hep-th]}}.

\bibitem{Inamura:2023qzl}
K.~Inamura and K.~Ohmori, ``{Fusion Surface Models: 2+1d Lattice Models from
  Fusion 2-Categories},'' \href{http://arxiv.org/abs/2305.05774}{{\ttfamily
  arXiv:2305.05774 [cond-mat.str-el]}}.

\bibitem{Chen:2023qnv}
J.~Chen, W.~Cui, B.~Haghighat, and Y.-N. Wang, ``{SymTFTs and Duality Defects
  from 6d SCFTs on 4-manifolds},''
  \href{http://arxiv.org/abs/2305.09734}{{\ttfamily arXiv:2305.09734
  [hep-th]}}.

\bibitem{Bashmakov:2023kwo}
V.~Bashmakov, M.~Del~Zotto, and A.~Hasan, ``{Four-manifolds and Symmetry
  Categories of 2d CFTs},'' \href{http://arxiv.org/abs/2305.10422}{{\ttfamily
  arXiv:2305.10422 [hep-th]}}.

\bibitem{Choi:2023xjw}
Y.~Choi, B.~C. Rayhaun, Y.~Sanghavi, and S.-H. Shao, ``{Comments on Boundaries,
  Anomalies, and Non-Invertible Symmetries},''
  \href{http://arxiv.org/abs/2305.09713}{{\ttfamily arXiv:2305.09713
  [hep-th]}}.

\bibitem{Gaiotto:2019xmp}
D.~Gaiotto and T.~Johnson-Freyd, ``{Condensations in higher categories},''
  \href{http://arxiv.org/abs/1905.09566}{{\ttfamily arXiv:1905.09566
  [math.CT]}}.

\bibitem{DouglasReutter}
C.~L. Douglas and D.~J. Reutter, ``Fusion 2-categories and a state-sum
  invariant for 4-manifolds,''
  \href{http://arxiv.org/abs/1812.11933}{{\ttfamily arXiv:1812.11933}}.

\bibitem{Hatcher}
A.~Hatcher, {\em {Algebraic topology}}.
\newblock Cambridge Univ. Press, Cambridge, 2000.
\newblock \url{https://cds.cern.ch/record/478079}.

\bibitem{Kapustin:2014gua}
A.~Kapustin and N.~Seiberg, ``{Coupling a QFT to a TQFT and Duality},''
  \href{http://dx.doi.org/10.1007/JHEP04(2014)001}{{\em JHEP} {\bfseries 04}
  (2014) 001}, \href{http://arxiv.org/abs/1401.0740}{{\ttfamily arXiv:1401.0740
  [hep-th]}}.

\bibitem{Benini:2018reh}
F.~Benini, C.~C\'ordova, and P.-S. Hsin, ``{On 2-Group Global Symmetries and
  their Anomalies},'' \href{http://dx.doi.org/10.1007/JHEP03(2019)118}{{\em
  JHEP} {\bfseries 03} (2019) 118},
  \href{http://arxiv.org/abs/1803.09336}{{\ttfamily arXiv:1803.09336
  [hep-th]}}.

\bibitem{Hsin:2020nts}
P.-S. Hsin and H.~T. Lam, ``{Discrete theta angles, symmetries and
  anomalies},'' \href{http://dx.doi.org/10.21468/SciPostPhys.10.2.032}{{\em
  SciPost Phys.} {\bfseries 10} no.~2, (2021) 032},
  \href{http://arxiv.org/abs/2007.05915}{{\ttfamily arXiv:2007.05915
  [hep-th]}}.

\bibitem{FultonHarris}
W.~Fulton and J.~Harris, {\em Representation Theory}.
\newblock Springer-Verlag, 1991.

\bibitem{Cordova:2022ruw}
C.~Cordova, T.~T. Dumitrescu, K.~Intriligator, and S.-H. Shao, ``{Snowmass
  White Paper: Generalized Symmetries in Quantum Field Theory and Beyond},'' in
  {\em {Snowmass 2021}}.
\newblock 5, 2022.
\newblock \href{http://arxiv.org/abs/2205.09545}{{\ttfamily arXiv:2205.09545
  [hep-th]}}.

\bibitem{Gomes:2023ahz}
P.~R.~S. Gomes, ``{An Introduction to Higher-Form Symmetries},''
  \href{http://arxiv.org/abs/2303.01817}{{\ttfamily arXiv:2303.01817
  [hep-th]}}.

\bibitem{Bhardwaj:2023wzd}
L.~Bhardwaj and S.~Schafer-Nameki, ``{Generalized Charges, Part I: Invertible
  Symmetries and Higher Representations},''
  \href{http://arxiv.org/abs/2304.02660}{{\ttfamily arXiv:2304.02660
  [hep-th]}}.

\bibitem{Bartsch:2023pzl}
T.~Bartsch, M.~Bullimore, and A.~Grigoletto, ``{Higher representations for
  extended operators},'' \href{http://arxiv.org/abs/2304.03789}{{\ttfamily
  arXiv:2304.03789 [hep-th]}}.

\bibitem{Gaiotto:2020iye}
D.~Gaiotto and J.~Kulp, ``{Orbifold groupoids},''
  \href{http://dx.doi.org/10.1007/JHEP02(2021)132}{{\em JHEP} {\bfseries 02}
  (2021) 132}, \href{http://arxiv.org/abs/2008.05960}{{\ttfamily
  arXiv:2008.05960 [hep-th]}}.

\bibitem{Apruzzi:2021nmk}
F.~Apruzzi, F.~Bonetti, I.~n.~G. Etxebarria, S.~S. Hosseini, and
  S.~Schafer-Nameki, ``{Symmetry TFTs from String Theory},''
  \href{http://arxiv.org/abs/2112.02092}{{\ttfamily arXiv:2112.02092
  [hep-th]}}.

\bibitem{Freed:2022qnc}
D.~S. Freed, G.~W. Moore, and C.~Teleman, ``{Topological symmetry in quantum
  field theory},'' \href{http://arxiv.org/abs/2209.07471}{{\ttfamily
  arXiv:2209.07471 [hep-th]}}.

\bibitem{Bhardwaj:2023ayw}
L.~Bhardwaj and S.~Schafer-Nameki, ``{Generalized Charges, Part II:
  Non-Invertible Symmetries and the Symmetry TFT},''
  \href{http://arxiv.org/abs/2305.17159}{{\ttfamily arXiv:2305.17159
  [hep-th]}}.

\bibitem{TongGauge}
D.~Tong, ``{Part III Lecture Notes: Gauge Theory}.''
  \url{https://www.damtp.cam.ac.uk/user/tong/gaugetheory.html}.

\bibitem{Tachikawa:2017gyf}
Y.~Tachikawa, ``{On gauging finite subgroups},''
  \href{http://dx.doi.org/10.21468/SciPostPhys.8.1.015}{{\em SciPost Phys.}
  {\bfseries 8} no.~1, (2020) 015},
  \href{http://arxiv.org/abs/1712.09542}{{\ttfamily arXiv:1712.09542
  [hep-th]}}.

\bibitem{horowitz1990quantum}
G.~T. Horowitz and M.~Srednicki, ``A quantum field theoretic description of
  linking numbers and their generalization,'' {\em Communications in
  Mathematical Physics} {\bfseries 130} (1990) 83--94.

\bibitem{Kapustin:2005py}
A.~Kapustin, ``{Wilson-'t Hooft operators in four-dimensional gauge theories
  and S-duality},'' \href{http://dx.doi.org/10.1103/PhysRevD.74.025005}{{\em
  Phys. Rev. D} {\bfseries 74} (2006) 025005},
  \href{http://arxiv.org/abs/hep-th/0501015}{{\ttfamily arXiv:hep-th/0501015}}.

\bibitem{Aharony:2013hda}
O.~Aharony, N.~Seiberg, and Y.~Tachikawa, ``{Reading between the lines of
  four-dimensional gauge theories},''
  \href{http://dx.doi.org/10.1007/JHEP08(2013)115}{{\em JHEP} {\bfseries 08}
  (2013) 115}, \href{http://arxiv.org/abs/1305.0318}{{\ttfamily arXiv:1305.0318
  [hep-th]}}.

\bibitem{Goddard:1976qe}
P.~Goddard, J.~Nuyts, and D.~I. Olive, ``{Gauge Theories and Magnetic
  Charge},'' \href{http://dx.doi.org/10.1016/0550-3213(77)90221-8}{{\em Nucl.
  Phys. B} {\bfseries 125} (1977) 1--28}.

\bibitem{Bhardwaj:2021wif}
L.~Bhardwaj, ``{2-Group Symmetries in Class S},''
  \href{http://arxiv.org/abs/2107.06816}{{\ttfamily arXiv:2107.06816
  [hep-th]}}.

\bibitem{Apruzzi:2021vcu}
F.~Apruzzi, S.~Schafer-Nameki, L.~Bhardwaj, and J.~Oh, ``{The Global Form of
  Flavor Symmetries and 2-Group Symmetries in 5d SCFTs},''
  \href{http://dx.doi.org/10.21468/SciPostPhys.13.2.024}{{\em SciPost Phys.}
  {\bfseries 13} no.~2, (2022) 024},
  \href{http://arxiv.org/abs/2105.08724}{{\ttfamily arXiv:2105.08724
  [hep-th]}}.

\bibitem{Bhardwaj:2022dyt}
L.~Bhardwaj, M.~Bullimore, A.~E.~V. Ferrari, and S.~Schafer-Nameki,
  ``{Anomalies of Generalized Symmetries from Solitonic Defects},''
  \href{http://arxiv.org/abs/2205.15330}{{\ttfamily arXiv:2205.15330
  [hep-th]}}.

\bibitem{Sharpe:2015mja}
E.~Sharpe, ``{Notes on generalized global symmetries in QFT},''
  \href{http://dx.doi.org/10.1002/prop.201500048}{{\em Fortsch. Phys.}
  {\bfseries 63} (2015) 659--682},
  \href{http://arxiv.org/abs/1508.04770}{{\ttfamily arXiv:1508.04770
  [hep-th]}}.

\bibitem{Cordova:2018cvg}
C.~C\'ordova, T.~T. Dumitrescu, and K.~Intriligator, ``{Exploring 2-Group
  Global Symmetries},'' \href{http://dx.doi.org/10.1007/JHEP02(2019)184}{{\em
  JHEP} {\bfseries 02} (2019) 184},
  \href{http://arxiv.org/abs/1802.04790}{{\ttfamily arXiv:1802.04790
  [hep-th]}}.

\bibitem{Apruzzi:2021mlh}
F.~Apruzzi, L.~Bhardwaj, D.~S.~W. Gould, and S.~Schafer-Nameki, ``{2-Group
  symmetries and their classification in 6d},''
  \href{http://dx.doi.org/10.21468/SciPostPhys.12.3.098}{{\em SciPost Phys.}
  {\bfseries 12} no.~3, (2022) 098},
  \href{http://arxiv.org/abs/2110.14647}{{\ttfamily arXiv:2110.14647
  [hep-th]}}.

\bibitem{Lee:2021crt}
Y.~Lee, K.~Ohmori, and Y.~Tachikawa, ``{Matching higher symmetries across
  Intriligator-Seiberg duality},''
  \href{http://dx.doi.org/10.1007/JHEP10(2021)114}{{\em JHEP} {\bfseries 10}
  (2021) 114}, \href{http://arxiv.org/abs/2108.05369}{{\ttfamily
  arXiv:2108.05369 [hep-th]}}.

\bibitem{Bourget:2018ond}
A.~Bourget, A.~Pini, and D.~Rodr\'\i{}guez-G\'omez, ``{Gauge theories from
  principally extended disconnected gauge groups},''
  \href{http://dx.doi.org/10.1016/j.nuclphysb.2019.02.004}{{\em Nucl. Phys. B}
  {\bfseries 940} (2019) 351--376},
  \href{http://arxiv.org/abs/1804.01108}{{\ttfamily arXiv:1804.01108
  [hep-th]}}.

\bibitem{Kapustin:2013uxa}
A.~Kapustin and R.~Thorngren, ``{Higher symmetry and gapped phases of gauge
  theories},'' \href{http://arxiv.org/abs/1309.4721}{{\ttfamily arXiv:1309.4721
  [hep-th]}}.

\bibitem{Vishwanath:2012tq}
A.~Vishwanath and T.~Senthil, ``{Physics of three dimensional bosonic
  topological insulators: Surface Deconfined Criticality and Quantized
  Magnetoelectric Effect},''
  \href{http://dx.doi.org/10.1103/PhysRevX.3.011016}{{\em Phys. Rev. X}
  {\bfseries 3} no.~1, (2013) 011016},
  \href{http://arxiv.org/abs/1209.3058}{{\ttfamily arXiv:1209.3058
  [cond-mat.str-el]}}.

\bibitem{Gaiotto:2017zba}
D.~Gaiotto and T.~Johnson-Freyd, ``{Symmetry Protected Topological phases and
  Generalized Cohomology},''
  \href{http://dx.doi.org/10.1007/JHEP05(2019)007}{{\em JHEP} {\bfseries 05}
  (2019) 007}, \href{http://arxiv.org/abs/1712.07950}{{\ttfamily
  arXiv:1712.07950 [hep-th]}}.

\bibitem{Chen:2011pg}
X.~Chen, Z.-C. Gu, Z.-X. Liu, and X.-G. Wen, ``{Symmetry protected topological
  orders and the group cohomology of their symmetry group},''
  \href{http://dx.doi.org/10.1103/PhysRevB.87.155114}{{\em Phys. Rev. B}
  {\bfseries 87} no.~15, (2013) 155114},
  \href{http://arxiv.org/abs/1106.4772}{{\ttfamily arXiv:1106.4772
  [cond-mat.str-el]}}.

\bibitem{Kong:2019brm}
L.~Kong, Y.~Tian, and S.~Zhou, ``{The center of monoidal 2-categories in 3+1D
  Dijkgraaf-Witten theory},''
  \href{http://dx.doi.org/10.1016/j.aim.2019.106928}{{\em Adv. Math.}
  {\bfseries 360} (2020) 106928},
  \href{http://arxiv.org/abs/1905.04644}{{\ttfamily arXiv:1905.04644
  [math.QA]}}.

\bibitem{Johnson-Freyd:2020ivj}
T.~Johnson-Freyd and M.~Yu, ``{Fusion 2-categories With no Line Operators are
  Grouplike},'' \href{http://dx.doi.org/10.1017/S0004972721000095}{{\em Bull.
  Austral. Math. Soc.} {\bfseries 104} no.~3, (2021) 434--442},
  \href{http://arxiv.org/abs/2010.07950}{{\ttfamily arXiv:2010.07950
  [math.QA]}}.

\bibitem{decoppet2022drinfeld}
T.~D. D{\'e}coppet, ``On the drinfeld centers of fusion 2-categories,'' {\em
  arXiv preprint arXiv:2211.04917} (2022) .

\bibitem{Hsin:2018vcg}
P.-S. Hsin, H.~T. Lam, and N.~Seiberg, ``{Comments on One-Form Global
  Symmetries and Their Gauging in 3d and 4d},''
  \href{http://dx.doi.org/10.21468/SciPostPhys.6.3.039}{{\em SciPost Phys.}
  {\bfseries 6} no.~3, (2019) 039},
  \href{http://arxiv.org/abs/1812.04716}{{\ttfamily arXiv:1812.04716
  [hep-th]}}.

\bibitem{banchoff1974triple}
T.~Banchoff, ``Triple points and singularities of projections of smoothly
  immersed surfaces,'' {\em Proceedings of the American Mathematical Society}
  {\bfseries 46} no.~3, (1974) 402--406.

\bibitem{Barkeshli:2014cna}
M.~Barkeshli, P.~Bonderson, M.~Cheng, and Z.~Wang, ``{Symmetry
  Fractionalization, Defects, and Gauging of Topological Phases},''
  \href{http://dx.doi.org/10.1103/PhysRevB.100.115147}{{\em Phys. Rev. B}
  {\bfseries 100} no.~11, (2019) 115147},
  \href{http://arxiv.org/abs/1410.4540}{{\ttfamily arXiv:1410.4540
  [cond-mat.str-el]}}.

\bibitem{vanBeest:2022fss}
M.~van Beest, D.~S.~W. Gould, S.~Schafer-Nameki, and Y.-N. Wang, ``{Symmetry
  TFTs for 3d QFTs from M-theory},''
  \href{http://dx.doi.org/10.1007/JHEP02(2023)226}{{\em JHEP} {\bfseries 02}
  (2023) 226}, \href{http://arxiv.org/abs/2210.03703}{{\ttfamily
  arXiv:2210.03703 [hep-th]}}.

\bibitem{Witten:1998wy}
E.~Witten, ``{AdS / CFT correspondence and topological field theory},''
  \href{http://dx.doi.org/10.1088/1126-6708/1998/12/012}{{\em JHEP} {\bfseries
  12} (1998) 012}, \href{http://arxiv.org/abs/hep-th/9812012}{{\ttfamily
  arXiv:hep-th/9812012}}.

\bibitem{Apruzzi:2021phx}
F.~Apruzzi, M.~van Beest, D.~S.~W. Gould, and S.~Sch\"afer-Nameki,
  ``{Holography, 1-form symmetries, and confinement},''
  \href{http://dx.doi.org/10.1103/PhysRevD.104.066005}{{\em Phys. Rev. D}
  {\bfseries 104} no.~6, (2021) 066005},
  \href{http://arxiv.org/abs/2104.12764}{{\ttfamily arXiv:2104.12764
  [hep-th]}}.

\bibitem{kong2015boundary}
L.~Kong, X.-G. Wen, and H.~Zheng, ``Boundary-bulk relation for topological
  orders as the functor mapping higher categories to their centers,'' {\em
  arXiv preprint arXiv:1502.01690} (2015) .

\bibitem{Johnson-Freyd:2020usu}
T.~Johnson-Freyd, ``{On the Classification of Topological Orders},''
  \href{http://dx.doi.org/10.1007/s00220-022-04380-3}{{\em Commun. Math. Phys.}
  {\bfseries 393} no.~2, (2022) 989--1033},
  \href{http://arxiv.org/abs/2003.06663}{{\ttfamily arXiv:2003.06663
  [math.CT]}}.

\bibitem{Zhao:2022yaw}
J.~Zhao, J.-Q. Lou, Z.-H. Zhang, L.-Y. Hung, L.~Kong, and Y.~Tian, ``{String
  Condensations in 3+1D and Lagrangian Algebras},''
  \href{http://arxiv.org/abs/2208.07865}{{\ttfamily arXiv:2208.07865
  [cond-mat.str-el]}}.

\bibitem{Walker}
K.~{Walker}, ``{A universal state sum},''
  \href{http://dx.doi.org/10.48550/arXiv.2104.02101}{{\em arXiv e-prints}
  (Apr., 2021) arXiv:2104.02101},
  \href{http://arxiv.org/abs/2104.02101}{{\ttfamily arXiv:2104.02101
  [math.QA]}}.

\bibitem{Crane:1993if}
L.~Crane and D.~Yetter, ``{A Categorical construction of 4-D topological
  quantum field theories},''
\newblock 3, 1993.
\newblock \href{http://arxiv.org/abs/hep-th/9301062}{{\ttfamily
  arXiv:hep-th/9301062}}.

\bibitem{Ji:2019jhk}
W.~Ji and X.-G. Wen, ``{Categorical symmetry and noninvertible anomaly in
  symmetry-breaking and topological phase transitions},''
  \href{http://dx.doi.org/10.1103/PhysRevResearch.2.033417}{{\em Phys. Rev.
  Res.} {\bfseries 2} no.~3, (2020) 033417},
  \href{http://arxiv.org/abs/1912.13492}{{\ttfamily arXiv:1912.13492
  [cond-mat.str-el]}}.

\bibitem{Kong:2020cie}
L.~Kong, T.~Lan, X.-G. Wen, Z.-H. Zhang, and H.~Zheng, ``{Algebraic higher
  symmetry and categorical symmetry -- a holographic and entanglement view of
  symmetry},'' \href{http://dx.doi.org/10.1103/PhysRevResearch.2.043086}{{\em
  Phys. Rev. Res.} {\bfseries 2} no.~4, (2020) 043086},
  \href{http://arxiv.org/abs/2005.14178}{{\ttfamily arXiv:2005.14178
  [cond-mat.str-el]}}.

\bibitem{Wang:2021vki}
J.~Wang and Y.-Z. You, ``{Gauge Enhanced Quantum Criticality Between Grand
  Unifications: Categorical Higher Symmetry Retraction},''
  \href{http://arxiv.org/abs/2111.10369}{{\ttfamily arXiv:2111.10369
  [hep-th]}}.

\bibitem{Cordova:2022qtz}
C.~Cordova and S.~Koren, ``{Higher Flavor Symmetries in the Standard Model},''
  \href{http://arxiv.org/abs/2212.13193}{{\ttfamily arXiv:2212.13193
  [hep-ph]}}.

\bibitem{Tachikawa:2013hya}
Y.~Tachikawa, ``{On the 6d origin of discrete additional data of 4d gauge
  theories},'' \href{http://dx.doi.org/10.1007/JHEP05(2014)020}{{\em JHEP}
  {\bfseries 05} (2014) 020}, \href{http://arxiv.org/abs/1309.0697}{{\ttfamily
  arXiv:1309.0697 [hep-th]}}.

\bibitem{DelZotto:2015isa}
M.~Del~Zotto, J.~J. Heckman, D.~S. Park, and T.~Rudelius, ``{On the Defect
  Group of a 6D SCFT},''
  \href{http://dx.doi.org/10.1007/s11005-016-0839-5}{{\em Lett. Math. Phys.}
  {\bfseries 106} no.~6, (2016) 765--786},
  \href{http://arxiv.org/abs/1503.04806}{{\ttfamily arXiv:1503.04806
  [hep-th]}}.

\bibitem{GarciaEtxebarria:2019caf}
I.~n. Garc\'\i{}a~Etxebarria, B.~Heidenreich, and D.~Regalado, ``{IIB flux
  non-commutativity and the global structure of field theories},''
  \href{http://dx.doi.org/10.1007/JHEP10(2019)169}{{\em JHEP} {\bfseries 10}
  (2019) 169}, \href{http://arxiv.org/abs/1908.08027}{{\ttfamily
  arXiv:1908.08027 [hep-th]}}.

\bibitem{Eckhard:2019jgg}
J.~Eckhard, H.~Kim, S.~Schafer-Nameki, and B.~Willett, ``{Higher-Form
  Symmetries, Bethe Vacua, and the 3d-3d Correspondence},''
  \href{http://dx.doi.org/10.1007/JHEP01(2020)101}{{\em JHEP} {\bfseries 01}
  (2020) 101}, \href{http://arxiv.org/abs/1910.14086}{{\ttfamily
  arXiv:1910.14086 [hep-th]}}.

\bibitem{Morrison:2020ool}
D.~R. Morrison, S.~Schafer-Nameki, and B.~Willett, ``{Higher-Form Symmetries in
  5d},'' \href{http://dx.doi.org/10.1007/JHEP09(2020)024}{{\em JHEP} {\bfseries
  09} (2020) 024}, \href{http://arxiv.org/abs/2005.12296}{{\ttfamily
  arXiv:2005.12296 [hep-th]}}.

\bibitem{Albertini:2020mdx}
F.~Albertini, M.~Del~Zotto, I.~n. Garc\'\i{}a~Etxebarria, and S.~S. Hosseini,
  ``{Higher Form Symmetries and M-theory},''
  \href{http://dx.doi.org/10.1007/JHEP12(2020)203}{{\em JHEP} {\bfseries 12}
  (2020) 203}, \href{http://arxiv.org/abs/2005.12831}{{\ttfamily
  arXiv:2005.12831 [hep-th]}}.

\bibitem{Gukov:2020btk}
S.~Gukov, P.-S. Hsin, and D.~Pei, ``{Generalized global symmetries of $T[M]$
  theories. Part I},'' \href{http://dx.doi.org/10.1007/JHEP04(2021)232}{{\em
  JHEP} {\bfseries 04} (2021) 232},
  \href{http://arxiv.org/abs/2010.15890}{{\ttfamily arXiv:2010.15890
  [hep-th]}}.

\bibitem{Bah:2020uev}
I.~Bah, F.~Bonetti, and R.~Minasian, ``{Discrete and higher-form symmetries in
  SCFTs from wrapped M5-branes},''
  \href{http://dx.doi.org/10.1007/JHEP03(2021)196}{{\em JHEP} {\bfseries 03}
  (2021) 196}, \href{http://arxiv.org/abs/2007.15003}{{\ttfamily
  arXiv:2007.15003 [hep-th]}}.

\bibitem{Closset:2020scj}
C.~Closset, S.~Schafer-Nameki, and Y.-N. Wang, ``{Coulomb and Higgs Branches
  from Canonical Singularities: Part 0},''
  \href{http://dx.doi.org/10.1007/JHEP02(2021)003}{{\em JHEP} {\bfseries 02}
  (2021) 003}, \href{http://arxiv.org/abs/2007.15600}{{\ttfamily
  arXiv:2007.15600 [hep-th]}}.

\bibitem{DelZotto:2020esg}
M.~Del~Zotto, I.~n. Garc\'\i{}a~Etxebarria, and S.~S. Hosseini, ``{Higher form
  symmetries of Argyres-Douglas theories},''
  \href{http://dx.doi.org/10.1007/JHEP10(2020)056}{{\em JHEP} {\bfseries 10}
  (2020) 056}, \href{http://arxiv.org/abs/2007.15603}{{\ttfamily
  arXiv:2007.15603 [hep-th]}}.

\bibitem{Apruzzi:2020zot}
F.~Apruzzi, M.~Dierigl, and L.~Lin, ``{The fate of discrete 1-form symmetries
  in 6d},'' \href{http://dx.doi.org/10.21468/SciPostPhys.12.2.047}{{\em SciPost
  Phys.} {\bfseries 12} no.~2, (2022) 047},
  \href{http://arxiv.org/abs/2008.09117}{{\ttfamily arXiv:2008.09117
  [hep-th]}}.

\bibitem{Bhardwaj:2020phs}
L.~Bhardwaj and S.~Sch\"afer-Nameki, ``{Higher-form symmetries of 6d and 5d
  theories},'' \href{http://dx.doi.org/10.1007/JHEP02(2021)159}{{\em JHEP}
  {\bfseries 02} (2021) 159}, \href{http://arxiv.org/abs/2008.09600}{{\ttfamily
  arXiv:2008.09600 [hep-th]}}.

\bibitem{DelZotto:2020sop}
M.~Del~Zotto and K.~Ohmori, ``{2-Group Symmetries of 6D Little String Theories
  and T-Duality},'' \href{http://dx.doi.org/10.1007/s00023-021-01018-3}{{\em
  Annales Henri Poincare} {\bfseries 22} no.~7, (2021) 2451--2474},
  \href{http://arxiv.org/abs/2009.03489}{{\ttfamily arXiv:2009.03489
  [hep-th]}}.

\bibitem{Closset:2020afy}
C.~Closset, S.~Giacomelli, S.~Schafer-Nameki, and Y.-N. Wang, ``{5d and 4d
  SCFTs: Canonical Singularities, Trinions and S-Dualities},''
  \href{http://dx.doi.org/10.1007/JHEP05(2021)274}{{\em JHEP} {\bfseries 05}
  (2021) 274}, \href{http://arxiv.org/abs/2012.12827}{{\ttfamily
  arXiv:2012.12827 [hep-th]}}.

\bibitem{Bhardwaj:2021pfz}
L.~Bhardwaj, M.~Hubner, and S.~Schafer-Nameki, ``{1-form Symmetries of 4d N=2
  Class S Theories},''
  \href{http://dx.doi.org/10.21468/SciPostPhys.11.5.096}{{\em SciPost Phys.}
  {\bfseries 11} (2021) 096}, \href{http://arxiv.org/abs/2102.01693}{{\ttfamily
  arXiv:2102.01693 [hep-th]}}.

\bibitem{Bhardwaj:2021ojs}
L.~Bhardwaj, ``{Global form of flavor symmetry groups in 4d N=2 theories of
  class S},'' \href{http://dx.doi.org/10.21468/SciPostPhys.12.6.183}{{\em
  SciPost Phys.} {\bfseries 12} no.~6, (2022) 183},
  \href{http://arxiv.org/abs/2105.08730}{{\ttfamily arXiv:2105.08730
  [hep-th]}}.

\bibitem{Hosseini:2021ged}
S.~S. Hosseini and R.~Moscrop, ``{Maruyoshi-Song flows and defect groups of $
  {\mathrm{D}}_{\mathrm{p}}^{\mathrm{b}} $(G) theories},''
  \href{http://dx.doi.org/10.1007/JHEP10(2021)119}{{\em JHEP} {\bfseries 10}
  (2021) 119}, \href{http://arxiv.org/abs/2106.03878}{{\ttfamily
  arXiv:2106.03878 [hep-th]}}.

\bibitem{Cvetic:2021sxm}
M.~Cvetic, M.~Dierigl, L.~Lin, and H.~Y. Zhang, ``{Higher-form symmetries and
  their anomalies in M-/F-theory duality},''
  \href{http://dx.doi.org/10.1103/PhysRevD.104.126019}{{\em Phys. Rev. D}
  {\bfseries 104} no.~12, (2021) 126019},
  \href{http://arxiv.org/abs/2106.07654}{{\ttfamily arXiv:2106.07654
  [hep-th]}}.

\bibitem{Bhardwaj:2021zrt}
L.~Bhardwaj, M.~Hubner, and S.~Schafer-Nameki, ``{Liberating confinement from
  Lagrangians: 1-form symmetries and lines in 4d N=1 from 6d N=(2,0)},''
  \href{http://dx.doi.org/10.21468/SciPostPhys.12.1.040}{{\em SciPost Phys.}
  {\bfseries 12} no.~1, (2022) 040},
  \href{http://arxiv.org/abs/2106.10265}{{\ttfamily arXiv:2106.10265
  [hep-th]}}.

\bibitem{Closset:2021lhd}
C.~Closset and H.~Magureanu, ``{The $U$-plane of rank-one 4d $\mathcal{N}=2$ KK
  theories},'' \href{http://dx.doi.org/10.21468/SciPostPhys.12.2.065}{{\em
  SciPost Phys.} {\bfseries 12} no.~2, (2022) 065},
  \href{http://arxiv.org/abs/2107.03509}{{\ttfamily arXiv:2107.03509
  [hep-th]}}.

\bibitem{Tian:2021cif}
J.~Tian and Y.-N. Wang, ``{5D and 6D SCFTs from $\mathbb{C}^3$ orbifolds},''
  \href{http://dx.doi.org/10.21468/SciPostPhys.12.4.127}{{\em SciPost Phys.}
  {\bfseries 12} no.~4, (2022) 127},
  \href{http://arxiv.org/abs/2110.15129}{{\ttfamily arXiv:2110.15129
  [hep-th]}}.

\bibitem{Closset:2021lwy}
C.~Closset, S.~Sch\"afer-Nameki, and Y.-N. Wang, ``{Coulomb and Higgs branches
  from canonical singularities. Part I. Hypersurfaces with smooth Calabi-Yau
  resolutions},'' \href{http://dx.doi.org/10.1007/JHEP04(2022)061}{{\em JHEP}
  {\bfseries 04} (2022) 061}, \href{http://arxiv.org/abs/2111.13564}{{\ttfamily
  arXiv:2111.13564 [hep-th]}}.

\bibitem{Bhardwaj:2021mzl}
L.~Bhardwaj, S.~Giacomelli, M.~H\"ubner, and S.~Sch\"afer-Nameki, ``{Relative
  defects in relative theories: Trapped higher-form symmetries and irregular
  punctures in class S},''
  \href{http://dx.doi.org/10.21468/SciPostPhys.13.4.101}{{\em SciPost Phys.}
  {\bfseries 13} no.~4, (2022) 101},
  \href{http://arxiv.org/abs/2201.00018}{{\ttfamily arXiv:2201.00018
  [hep-th]}}.

\bibitem{DelZotto:2022fnw}
M.~Del~Zotto, J.~J. Heckman, S.~N. Meynet, R.~Moscrop, and H.~Y. Zhang,
  ``{Higher symmetries of 5D orbifold SCFTs},''
  \href{http://dx.doi.org/10.1103/PhysRevD.106.046010}{{\em Phys. Rev. D}
  {\bfseries 106} no.~4, (2022) 046010},
  \href{http://arxiv.org/abs/2201.08372}{{\ttfamily arXiv:2201.08372
  [hep-th]}}.

\bibitem{Bhardwaj:2022ekc}
L.~Bhardwaj, ``{Discovering T-Dualities of Little String Theories},''
  \href{http://arxiv.org/abs/2209.10548}{{\ttfamily arXiv:2209.10548
  [hep-th]}}.

\bibitem{DelZotto:2022ohj}
M.~Del~Zotto, M.~Liu, and P.-K. Oehlmann, ``{Back to heterotic strings on ALE
  spaces. Part I. Instantons, 2-groups and T-duality},''
  \href{http://dx.doi.org/10.1007/JHEP01(2023)176}{{\em JHEP} {\bfseries 01}
  (2023) 176}, \href{http://arxiv.org/abs/2209.10551}{{\ttfamily
  arXiv:2209.10551 [hep-th]}}.

\bibitem{Cvetic:2022imb}
M.~Cveti\v{c}, J.~J. Heckman, M.~H\"ubner, and E.~Torres, ``{0-form, 1-form,
  and 2-group symmetries via cutting and gluing of orbifolds},''
  \href{http://dx.doi.org/10.1103/PhysRevD.106.106003}{{\em Phys. Rev. D}
  {\bfseries 106} no.~10, (2022) 106003},
  \href{http://arxiv.org/abs/2203.10102}{{\ttfamily arXiv:2203.10102
  [hep-th]}}.

\bibitem{DelZotto:2022joo}
M.~Del~Zotto, I.~n. Garc\'\i{}a~Etxebarria, and S.~Schafer-Nameki, ``{2-Group
  Symmetries and M-Theory},''
  \href{http://dx.doi.org/10.21468/SciPostPhys.13.5.105}{{\em SciPost Phys.}
  {\bfseries 13} (2022) 105}, \href{http://arxiv.org/abs/2203.10097}{{\ttfamily
  arXiv:2203.10097 [hep-th]}}.

\bibitem{Apruzzi:2022dlm}
F.~Apruzzi, ``{Higher form symmetries TFT in 6d},''
  \href{http://dx.doi.org/10.1007/JHEP11(2022)050}{{\em JHEP} {\bfseries 11}
  (2022) 050}, \href{http://arxiv.org/abs/2203.10063}{{\ttfamily
  arXiv:2203.10063 [hep-th]}}.

\bibitem{Bhardwaj:2022scy}
L.~Bhardwaj and D.~S.~W. Gould, ``{Disconnected 0-Form and 2-Group
  Symmetries},'' \href{http://arxiv.org/abs/2206.01287}{{\ttfamily
  arXiv:2206.01287 [hep-th]}}.

\bibitem{Hubner:2022kxr}
M.~Hubner, D.~R. Morrison, S.~Schafer-Nameki, and Y.-N. Wang, ``{Generalized
  Symmetries in F-theory and the Topology of Elliptic Fibrations},''
  \href{http://dx.doi.org/10.21468/SciPostPhys.13.2.030}{{\em SciPost Phys.}
  {\bfseries 13} no.~2, (2022) 030},
  \href{http://arxiv.org/abs/2203.10022}{{\ttfamily arXiv:2203.10022
  [hep-th]}}.

\bibitem{DelZotto:2022ras}
M.~Del~Zotto and I.~n. Garc\'\i{}a~Etxebarria, ``{Global Structures from the
  Infrared},'' \href{http://arxiv.org/abs/2204.06495}{{\ttfamily
  arXiv:2204.06495 [hep-th]}}.

\bibitem{Acharya:2023bth}
B.~S. Acharya, M.~Del~Zotto, J.~J. Heckman, M.~Hubner, and E.~Torres,
  ``{Junctions, Edge Modes, and $G_2$-Holonomy Orbifolds},''
  \href{http://arxiv.org/abs/2304.03300}{{\ttfamily arXiv:2304.03300
  [hep-th]}}.

\bibitem{Sacchi:2023omn}
M.~Sacchi, O.~Sela, and G.~Zafrir, ``{5d to 3d compactifications and discrete
  anomalies},'' \href{http://arxiv.org/abs/2305.08185}{{\ttfamily
  arXiv:2305.08185 [hep-th]}}.

\bibitem{Palti:2019pca}
E.~Palti, ``{The Swampland: Introduction and Review},''
  \href{http://dx.doi.org/10.1002/prop.201900037}{{\em Fortsch. Phys.}
  {\bfseries 67} no.~6, (2019) 1900037},
  \href{http://arxiv.org/abs/1903.06239}{{\ttfamily arXiv:1903.06239
  [hep-th]}}.

\bibitem{vanBeest:2021lhn}
M.~van Beest, J.~Calder\'on-Infante, D.~Mirfendereski, and I.~Valenzuela,
  ``{Lectures on the Swampland Program in String Compactifications},''
  \href{http://dx.doi.org/10.1016/j.physrep.2022.09.002}{{\em Phys. Rept.}
  {\bfseries 989} (2022) 1--50},
  \href{http://arxiv.org/abs/2102.01111}{{\ttfamily arXiv:2102.01111
  [hep-th]}}.

\bibitem{Teo:2015xla}
J.~C.~Y. Teo, T.~L. Hughes, and E.~Fradkin, ``{Theory of Twist Liquids: Gauging
  an Anyonic Symmetry},''
  \href{http://dx.doi.org/10.1016/j.aop.2015.05.012}{{\em Annals Phys.}
  {\bfseries 360} (2015) 349--445},
  \href{http://arxiv.org/abs/1503.06812}{{\ttfamily arXiv:1503.06812
  [cond-mat.str-el]}}.

\bibitem{Kong:2022cpy}
L.~Kong and Z.-H. Zhang, ``{An invitation to topological orders and category
  theory},'' \href{http://arxiv.org/abs/2205.05565}{{\ttfamily arXiv:2205.05565
  [cond-mat.str-el]}}.

\bibitem{SymSem}
L.~Bhardwaj, M.~Bullimore, S.~Schafer-Nameki, and A.~Tiwari, ``{Symmetry
  Seminar},''. \url{{https://sites.google.com/view/symmetryseminar/home}}.

\bibitem{Johnson-Freyd:2020twl}
T.~Johnson-Freyd, ``{(3+1)D topological orders with only a
  $\mathbb{Z}_2$-charged particle},''
  \href{http://arxiv.org/abs/2011.11165}{{\ttfamily arXiv:2011.11165
  [math.QA]}}.

\bibitem{Kong:2020jne}
L.~Kong, T.~Lan, X.-G. Wen, Z.-H. Zhang, and H.~Zheng, ``{Classification of
  topological phases with finite internal symmetries in all dimensions},''
  \href{http://dx.doi.org/10.1007/JHEP09(2020)093}{{\em JHEP} {\bfseries 09}
  (2020) 093}, \href{http://arxiv.org/abs/2003.08898}{{\ttfamily
  arXiv:2003.08898 [math-ph]}}.

\bibitem{Johnson-Freyd:2021tbq}
T.~Johnson-Freyd and M.~Yu, ``{Topological Orders in (4+1)-Dimensions},''
  \href{http://dx.doi.org/10.21468/SciPostPhys.13.3.068}{{\em SciPost Phys.}
  {\bfseries 13} no.~3, (2022) 068},
  \href{http://arxiv.org/abs/2104.04534}{{\ttfamily arXiv:2104.04534
  [hep-th]}}.

\end{thebibliography}\endgroup
\bibliographystyle{ytphys}


\end{document}